\documentclass[a4paper,twocolumn,11pt,accepted=2026-04-25]{quantumarticle}
\pdfoutput=1
\usepackage[utf8]{inputenc}
\usepackage[english]{babel}
\usepackage[T1]{fontenc}
\usepackage{amsmath,amssymb}
\usepackage{hyperref}

\usepackage{tikz}
\usepackage{lipsum}

\usepackage[numbers]{natbib}
\usepackage{graphicx}
\usepackage{subcaption}

\newtheorem{theorem}{Theorem}
\newtheorem{corollary}{Corollary}
\newtheorem{remark}{Remark}

\begin{document}

\title{Quantum Optimal Control for Coherent Spin Dynamics of Radical Pairs via Pontryagin Maximum Principle}

\author{U.~G.~Abdulla$^{*}$}
\affiliation{Okinawa Institute of Science and Technology 1919-1 Tancha, Onna-son, Kunigami-gun, Okinawa, Japan 904-0495}
\orcid{0000-0002-4908-1660}
\email{Ugur.Abdulla@oist.jp}

\author{J.~H.~Rodrigues}
\orcid{0000-0002-9896-5761}
\affiliation{Okinawa Institute of Science and Technology 1919-1 Tancha, Onna-son, Kunigami-gun, Okinawa, Japan 904-0495}

\author{J.~-J.~Slotine}
\orcid{0000-0002-7161-7812}
\affiliation{Nonlinear Systems Laboratory, Department of Mechanical Engineering, Massachusetts Institute of Technology, Cambridge, Massachusetts, USA}

\maketitle


\begin{abstract}
This paper aims to devise the shape of the external electromagnetic field that drives the spin dynamics of radical pairs to a quantum coherent state through maximization of the triplet-born singlet yield in biochemical reactions. The model is a Schr\"{o}dinger system with spin Hamiltonians given by the sum of Zeeman interaction and hyperfine coupling interaction terms. We introduce a one-parameter family of optimal control problems by coupling the Schr\"{o}dinger system to a control field through filtering equations for the electromagnetic field. Fr\'echet differentiability and the Pontryagin Maximum Principle in Hilbert space are proved, and the bang-bang structure of the optimal control is established. A new iterative Pontryagin Maximum Principle (IPMP) method for the identification of the bang-bang optimal control is developed. Numerical simulations based on IPMP and the gradient projection method (GPM) in Hilbert spaces are pursued, and the convergence, stability, and the regularization effect are demonstrated. Comparative analysis of filtering with regular optimal electromagnetic field versus non-filtering with bang-bang optimal field ({\it Abdulla et al, Quantum Sci. Technol., {\bf9}, 4, 2024}) demonstrates that the change of the maxima of the singlet yield is less than 1\%. The results open a venue for a potential experimental work on magnetoreception as a manifestation of quantum biological phenomena.
\end{abstract}


\section{\label{intronew}Introduction}
One of the great challenges of modern science is to bridge the gap between atomic and cellular level phenomena that determine the condition and outcomes of our cells.
Quantum biology is an emerging field with the aim to comprehend quantum effects in biochemical mechanisms and biological function \cite{BUC2010,BAL2011,LAM2013}. The field of quantum biology currently focuses on several areas of research: photosynthesis, and magnetoreception \cite{RIT2009,RIT2004,NIE2013,USS2016,CAI2012,CAI2013,CIN2003}; 
olfaction \cite{TUR1996,TUR2014}; enzymatic activity, fruit fly navigation, ATP production, and vision \cite{MOH2014,SIA2014,RIE1998,TIN2016,ZUE2019,DOD2019,PAL2014}.
However, it should be mentioned that there is no clear consensus on the presence of genuine quantum mechanical effects in biology, e.g., in avian magnetoreception and photosynthesis, there are other proposed mechanisms, such as navigation using a magnetite-based mechanism \cite{WIL2013,SHA2015}. In \cite{USS2014}, quantum coherence effects that control the biological production of specific reactive oxygen species (ROS) in mammalian cells are presented. ROS balance is known to affect cellular signaling, proliferation, and redox homeostasis \cite{THA2000,DIE2009,FIN2011,RAY2012,BIG2014}. Increased understanding of bio-physical and bio-molecular mechanisms that control specific ROS product channels, at the point of generation, may open new therapeutic avenues, particularly in neural-heightened medicine, enhanced wound healing, and improved human performance.

The spin-correlated radical-pair mechanism (SRPM) offers the best understood and established explanation of how magnetic fields might influence biochemical reactions \cite{STE1989,HAG22}. Magnetic field effects are possible during the singlet-triplet spin evolution of two radicals \cite{SCH1978b}, which are in the same “cage” that form a radical pair. Applied magnetic fields can affect two main internal interactions of the radical pair whose time-dependences affect the temporal evolution of the singlet-triplet mixing, and thereby modulate relative reaction rates: electron-nuclear hyperfine interactions (HFI) and radical Zeeman ($\Delta g$) energies; both interactions have energies (E $10^8 - 10^9$ rad/sec) much less than kT thermal energy ($10^{13}$ rad/sec). SRPM is typically an adiabatic process, uncoupled to the thermal bath; non-thermal effects of magnetic fields in biological systems are expected through these magnetic field and resonant frequency regimes without explicit influence of temperature. The effects of the HFI and Zeeman resonances are dependent on the frequency (resonant), strength, and orientation (relative to the molecular frame) of magnetic fields; these parameters can thus influence the relative reaction rates and product distributions of reactions of radical pairs. It is of great importance to identify magnetic field parameters to modulate quantum coherences in a radical pair reaction \cite{CAI2012,CAI2013,plos23,qst24,Cho2024}. In this context, the quantum coherence refers to the phase-coherent superposition of singlet and triplet states in a pair of free radicals, allowing them to exist simultaneously in both states. The quantum coherence plays a crucial role for radical pairs to sense a weak magnetic field by oscillating between states before reacting, and directly influencing reaction yields. 

In a broader context, the goal of this paper is to reveal the major mathematical principle of quantum optimal control theory (QOCT) and develop a roadmap leading straight to the experimental applications where a new generation of quantum technology is created on physical or biological systems serving as hardware platforms.  The description and understanding of the mathematical principles of the QOCT is of fundamental importance for the development of a new generation of quantum technologies \cite{KOC2015,KOC2022,RAB2001,RAB2010,BOS2021}. 
Some of the recent examples of successful application of QOCT methods in experiments for the development of a new generation of quantum technology are \cite{ofek2016,wern2021,larrouy2020,omran2019,borselli2021,figg2019,magrini2021}. 

In \cite{plos23}, an optimization method that combines sensitivity analysis with Tikhonov regularization is implemented \cite{ABD1995a,ABD1995b,ABD2018,ABD2019}. The numerical results of \cite{plos23} demonstrated that the quantum singlet-triplet yield of the radical pair system can be significantly reduced if optimization is simultaneously pursued for both external magnetic fields and internal hyperfine parameters. However, the method implemented in \cite{plos23} is limited to constant-in-time magnetic field parameters.

In a recent paper \cite{qst24}, optimal control of the external electromagnetic field that maximizes the quantum triplet-born singlet yield of simplified radical pairs modeled by Schr\"{o}dinger system with spin Hamiltonians given by the sum of Zeeman interaction and hyperfine coupling interaction terms is analyzed. 
Fr\'echet differentiability and Pontryagin Maximum Principle in Hilbert space setting is proved, and bang-bang structure of the optimal control is established. A closed optimality system of nonlinear differential equations for the identification of the bang-bang optimal control is revealed, and an effective numerical method for the identification of the bang-bang optimal magnetic field intensity for achieving a quantum coherence for the radical pair system is developed.

Remarkably, the optimal magnetic pulse driving the complex quantum system to a coherent state has a simple bang-bang structure. 
The Pontryagin Maximum Principle - a fundamental mathematical principle for the optimality of complex dynamical systems - turns out to be a fundamental principle for quantum coherence as well. 
The Pontryagin Maximum Principle implies that, along the optimal state and adjoint trajectories, almost every time instance, the Hamilton-Pontryagin function attains its maximum with respect to a finite-dimensional control parameter precisely at the optimal control parameter. Due to the structure of the Hamilton-Pontryagin function, which inherited the structure of the Hamiltonian, its maximum with respect to each component of the control parameter is always achieved at the extreme values. Hence, the complex quantum system is driven to a coherent state through banging of the components of the external magnetic pulse between its extreme values in specific time intervals. 

It should be noted that primarily due to the structure of the Zeeman interaction term of the Hamiltonian, the optimal control problem is non-convex. Therefore, in general, there is a non-uniqueness of the optimal control, which causes numerical instability of the algorithms for the identification of the bang-bang optimal control. The goal of this paper is to introduce a novel regularization of the quantum optimal control problem for the identification of the external electromagnetic field that drives the spin dynamics of radical pairs to a coherent state via maximization of the triplet-born singlet yield in biochemical reactions. We aim to develop new regularized methods with enhanced regularity of the optimal electromagnetic field input, and improved numerical stability of the algorithms based on Fr\'echet differentiability and the Pontryagin Maximum Principle in Hilbert space.
Despite its great importance and potential for experimental applications in magnetoreception, the result of \cite{qst24} raised a major technical question about how practical it will be to generate a bang-bang electromagnetic wave on a short time scale. In this paper, we aim to answer the following questions:

{\it What would be the simplest continuous-in-time electromagnetic wave input that is almost optimal, in the sense that it provides a quantum triplet-singlet yield with minimal loss relative to the optimal bang-bang electromagnetic field input? How to improve the numerical instability of algorithms caused by the non-uniqueness of the bang-bang optimal control?}\\

To answer this question, we introduce an idea of coupling the Schr\"{o}dinger system to a control field through the first-order filtering ordinary differential equation (ODE) for the electromagnetic field, and the new control vector is taken as a linear input of the filtering ODE (see Section~\ref{sct:model}, \eqref{schrodinger},\eqref{filter}). This transformation has a crucial effect on the qualitative nature of the optimal control problem. A key change is reflected in the form of the Hamilton-Pontryagin function (see Section \ref{sct:frechet}, \eqref{hpf}), which turns out to be a non-local function including weighted time integrals of the wave functions solving Schr\"odinger system and its adjoint. This is the major difference of the optimal control problem introduced in this paper from the original control problem analyzed in \cite{qst24}. Non-local structure of the Hamilton-Pontryagin function is a key factor for the regularization effect, both in terms of the enhanced regularity and simplicity of the optimal electromagnetic field input and improved stability of the algorithms to overcome the instability caused by the non-uniqueness of the optimal control. We prove a Fr\'echet differentiability of the cost functional in Hilbert space and establish a canonical form of the Fr\'echet gradient as an element of the Hilbert space in the form of the gradient of the Hamilton-Pontryagin function with respect to the control vector (Theorem~\ref{thm-frechet}, Section  \ref{sct:frechet}). Having established an explicit formula for the Fr\'echet gradient, we develop a gradient projection method (GPM) in a Hilbert space setting (Section \ref{ssct:GPM}). Next, we prove the Pontryagin Maximum Principle (PMP) and establish a bang-bang structure of the optimal control (Theorem~\ref{thm-pmp}, Section \ref{sct:pmp}). This implies that, as a solution of the filtering ODE with bang-bang input, an optimal electromagnetic field intensity is continuous and piecewise-smooth in time (Section \ref{ssct:Filtering}). Using the Pontryagin Maximum Principle, we derive an optimal control synthesis function by expressing the optimal control and the optimal electromagnetic wave through the optimal and adjoint trajectories (\eqref{Gamma-1},\eqref{Gamma-2},\eqref{optimalv}). The latter implies a closed optimality system in the form of nonlinear integro-differential equations with two-end boundary conditions for the identification of the optimal and adjoint state vector functions (\eqref{closedsystem1},\eqref{closedsystem2}). Combining with filtering ODE solution, we deduce a PMP-based explicit two-step method (EPMP) for the identification of the optimal electromagnetic field input (Section \ref{sct:pmp}). Based on the Pontryagin Maximum Principle and EPMP algorithm, we develop a novel iterative Pontryagin Maximum Principle (IPMP) method for the identification of the bang-bang optimal control. IPMP method replaces an EPMP algorithm with an iterative process, where at every iteration, a nonlinear optimality system of integro-differential equations is replaced with the solution of Schr\"odinger and its adjoint systems, followed by the identification of the bang-bang control vector via PMP condition (or optimal control synthesis equation). The IPMP method is a computationally feasible approximation of the EPMP method with enforcement of the Pontryagin Maximum Principle in the limit of the iteration process. It has some similarity with the gradient ascent method, with the difference being that instead of moving in the gradient direction in the full control space, it generates the movement in the manifold of bang-bang control vectors guided by the Pontryagin Maximum Principle.

Numerical results for the calculation of the bang-bang optimal control and continuous in time optimal electromagnetic field based on the GPM and IPMP algorithms are presented in Section~\ref{sct:num_results}. It is demonstrated that both the GPM and IPMP methods converge to the bang-bang optimal control and produce the same continuous-in-time, piecewise-smooth optimal electromagnetic field input. The IPMP algorithm significantly outperforms the GPM algorithm. The simulations show that the IPMP method converges twice as fast as the GPM method in terms of the number of iterations, particularly for higher-proton models, where the computational cost increases exponentially. Numerical simulations of up to 7-proton models with various filtering parameter ranges and with various selections of initial iterations demonstrate that 
trading off between the original non-filtered model with bang-bang optimal magnetic field and the filtered model with continuous in time and piecewise-smooth optimal magnetic field is associated with the loss of the maximum triplet-born singlet yield expressed as a maximum of the cost functional within 1\%. Moreover, despite the enhanced regularity and simplicity of the optimal electromagnetic field input, the filtered model preserves coherent oscillations of the optimal Schr\"odinger and its adjoint systems, nearly identical in the case of the non-filtered model. 
Remarkably, filtering presents a powerful regularization tool to address the non-uniqueness of the optimal control in the original no-filter model. Numerical simulations demonstrate that by choosing the filtering parameter small enough, the filtered optimal control problem has a unique bang-bang optimal control. Moreover, 
trading off between the original non-filtered model with multiple bang-bang optimal magnetic fields and the filtered model with a continuous and piecewise smooth-in-time unique optimal magnetic field is associated with the loss of the maximum singlet yield expressed as a maximum of the cost functional within 1\%.  The approach is numerically effective and significantly simpler than existing methods for finite bandwidth input approximations, e.g., such as those based on Fourier decompositions \cite{Hirose2018}.

Pontryagin Maximum Principle reveals that the key mechanism of the path from bang-bang optimal electromagnetic input to coherent oscillations of the density vector-function is hidden in the sign-changing oscillations of the Hamilton-Pontryagin function with respect to the control parameters. The latter is also a defining factor of the oscillations of the Fr\'echet gradient, which is the cause of the appearance of local extreme points and non-uniqueness. Therefore, the non-local nature of the Hamilton-Pontryagin function in the filtered optimal control problem, that is to say, its stability with respect to sign-changing oscillations, is a driving force for the regularization phenomena, both in terms of the enhanced regularity and simplicity of the optimal electromagnetic field input and improved stability of the algorithms to overcome the instability caused by the non-uniqueness of the optimal control. 

\begin{remark}
The GPM method is a first-order numerical method with very comparable performance to popular methods, such as GRAPE, Krotov, or band-limited optimal techniques in quantum optimal control problems \cite{Khaneja2005,Morzhin2019,Lewis2014}. The main methodological difference is in the paradigm between "optimize-then-discretize" (O-D) versus "discretize-then-optimize" (D-O) approaches. Our Fr\'echet gradient-based GPM method in Hilbert space is in O-D class, for it requires rigorous derivation of the Fr\'echet gradient formula before pursuing discretization of the gradient ascent algorithm in Hilbert space. On the other hand, methods like GRAPE, Krotov, or band-limited optimization belong to the class of D-O algorithms, for they require the discretization of the problem first, followed by the implementation of the gradient ascent algorithm in the finite-dimensional discrete setting. In general, each of these approaches has its own advantages and disadvantages. The GPM method in an infinite-dimensional Hilbert space provides a better physical consistency or higher accuracy, while the D-O class of methods is usually easier to implement. It is also possible to combine the benefits of both approaches; for example, through special discretization, one can introduce a discrete adjoint problem in a way that the discrete Fr\'echet gradient matches the discretization of our continuous Fr\'echet gradient in the Hilbert space. 
\end{remark}

\begin{remark}
It is essential to note that one might consider a smoothing ($C^\infty$-approximation) of the bang-bang optimal control as a candidate to resolve the major technical question formulated above in the frame of the optimal control problem analyzed in \cite{qst24}. Indeed, the solution of the Schr\"odinger system with spin Hamiltonian given by the Zeeman interaction and hyperfine coupling interaction terms, is continuously dependent on the control vector-function in Hilbert space. Precisely, for a sequence of smooth control vector functions converging in Hilbert space $L_2^3(0, T)$, the corresponding solutions of the Schr\"odinger system will converge uniformly in the time interval $[0, T]$. Since $C^\infty$ functions are dense in Hilbert space $L_2^3(0, T)$, one can establish that the $C^\infty$-approximation of the bang-bang optimal control vector-function is an approximate optimal control. However, note that such $C^\infty$-smoothing of the bang-bang electromagnetic wave will have a very sharp rise or fall around the time instances where optimal control is discontinuous, and therefore, from the practical standpoint, it will have similar inconveniences as bang-bang wave input. The main idea of this paper is to introduce a new family of regularized optimal control problems via coupling the Schr\"{o}dinger system to a control field through the first-order filtering equation for the electromagnetic field, and new control vector is taken as a linear input of the filtering ODE (see Section~\ref{sct:model}, \eqref{schrodinger},\eqref{filter}). This approach produces a continuous and piecewise-smooth-in-time optimal control with a much simpler structure. Its smoothness and simple structure are influenced by two factors: the non-local structure of the Hamilton-Pontryagin function (Section \ref{sct:frechet}, \eqref{hpf}), and Euler's formula for the solution of the filtering ODE with bang-bang input (Section \ref{ssct:Filtering}). 
\end{remark}

\section{\label{sct:model}Mathematical Model}
We consider the spin dynamics of a correlated radical pair system obtained from triplet born states governed by the \textit{filtered Schr\"odinger system}
\begin{align}
i\hbar \frac{d\psi^l}{dt} = \mathbf{H}(\mathbf{v}) \psi^l , ~ \mbox{in $(0,T]$} , 
\quad \psi^l(0) = \psi^l_\mathrm{T}, \label{schrodinger} \\
\frac{d\mathbf{v}}{dt} + \gamma \mathbf{v} = \gamma \mathbf{u}, ~ \mbox{in $(0,T]$}; \quad \mathbf{v}(0) = \mathbf{v}_0, \label{filter}
\end{align}
for $l = 1,\dots,s$, where $\hbar$ is the Plank constant, $\psi^l:[0,T]\to\mathbb{C}^n$ is the state vector, $\mathbf{H}:\mathbb{R}^3\to\mathbb{C}^{n\times n}$ is the Hamiltonian operator, $\psi^l_\mathrm{T}\in\mathbb{C}^n$ is the triplet state, $s\in\mathbb{N}$ is the number of triplet states, $\mathbf{v} \in L^3_2(0,T;\mathbb{R}^3)$ is the magnetic field vector function, $\gamma>0$ be a filtration parameter, $\mathbf{v}_0\in\mathbb{R}^3$ and $\mathbf{u} \in L^3_2(0,T;\mathbb{R}^3)$ be a control parameter. For the $p\in\mathbb{N}$ proton model with 1/2-spin, the number of triplet states is $3\cdot2^p$, and the state dimension is $n=2^{p+2}$. The triplet states can be classified as follows:

\begin{itemize}
\item[T0] states: $\psi^{j-2^p}_T = \frac{e_j+e_{j+2^p}}{\sqrt{2}}$, $j = 2^p+1,\dots,2^{p+1}$;
\item[T$+$] states: $\psi^j_T = e_{j-2^p}$, $j = 2^p+1,\dots,2^{p+1}$;
\item[T$-$] states: $\psi^j_T = e_{j+2^p}$, $j = 2^{p+1}+1,\dots,3\cdot2^p$,
\end{itemize}
where we use a notation $\{e_j\}$ for the standard orthonormal basis in $\mathbb{R}^n$. Spin Hamiltonian $\mathbf{H}$ is given as a sum of three terms 
\begin{equation}
\mathbf{H}(\mathbf{v}) = \mathbf{H}_Z(\mathbf{v}) + \mathbf{H}_{hfi} - i\mathbf{K}. \label{hamilton}
\end{equation}
where 
\begin{align}
\mathbf{H}_Z(\mathbf{v}) = \mu_B g \Big[ ( \mathbf{S}_{1x} + \mathbf{S}_{2x} ) v_x + ( \mathbf{S}_{1y} + \mathbf{S}_{2y} ) v_y \nonumber\\ 
+ ( \mathbf{S}_{1z} + \mathbf{S}_{2z} ) v_z \Big], \label{Zeeman}
\end{align}
be a Zeeman interaction term with \[ \mathbf{v}=\mathbf{v}(t)=(v_x(t),v_y(t),v_z(t))\in  C(0, T;\mathbb{R}^3), \] where the constants $\mu_B$ and $g$ stand for the Bohr's magneton and the electron's spin ratio, and $\mathbf{S}_1=(\mathbf{S}_{1i})_{i=x,y,z}$ and $\mathbf{S}_2=(\mathbf{S}_{2i})_{i=x,y,z}$ are the electron's spin operators. The static Hermitian operators $\mathbf{H}_{hfi}$ and $\mathbf{K}$ are defined as 
\begin{align}
\mathbf{H}_{hfi} = \mu_B g \sum_{j=1}^p\sum_{i=x,y,z} \mathbf{A}_{ji} \mathbf{I}_{ji} \mathbf{S}_{1i}, \label{Hyperfine} \\
\mathbf{K} = \frac{1}{2}\left( k_S \mathbf{P}_S + k_T \mathbf{P}_T \right) \label{Haberkorn}
\end{align}
where $\mathbf{A}_j=(\mathbf{A}_{ji})_{i=x,y,z} \in \mathbb{R}^3$ is the hyperfine parameters corresponding to the $j$-proton, and $\mathbf{I}_j=(\mathbf{I}_{ji})_{i=x,y,z}$ is the corresponding nucleus spin operator, for $j=1,\dots,p$. Finally, $ k_S, k_T > 0$ are diffusion rates, and $\mathbf{P}_S$ and $\mathbf{P}_T$ are the projection operators onto singlet and triplet states, respectively. For the particular case $p=1$, the electron and nucleus spin operators are given below
\begin{eqnarray}
\resizebox{0.45\textwidth}{!}{%
$\mathbf{S}_1 = \left( \frac{1}{2} \sigma_x \otimes E_2 \otimes E_2, \frac{1}{2} \sigma_y \otimes E_2 \otimes E_2 , \frac{1}{2} \sigma_z \otimes E_2 \otimes E_2 \right),$} \nonumber\\
\resizebox{0.45\textwidth}{!}{%
$\mathbf{S}_2 = \left( E_2 \otimes \frac{1}{2} \sigma_x \otimes E_2 , E_2 \otimes \frac{1}{2} \sigma_y \otimes E_2 , E_2 \otimes \frac{1}{2} \sigma_z \otimes E_2 \right),$} \nonumber\\ 
\resizebox{0.45\textwidth}{!}{%
$\mathbf{I}_1 = \left( E_2 \otimes E_2 \otimes \frac{1}{2}\sigma_x , E_2 \otimes E_2 \otimes \frac{1}{2}\sigma_y , E_2 \otimes E_2 \otimes \frac{1}{2}\sigma_z \right),$} \nonumber
\end{eqnarray}
where $E_2$ is the identity matrix of order 2, and $\sigma_i$, $i=x,y,z$ denotes the Pauli matrices. In addition to this case, the projection operators are defined as
\begin{equation}
\mathbf{P}_S = \frac{1}{4} E_8 - \mathbf{S}_1 \cdot \mathbf{S}_2, \quad 
\mathbf{P}_T = \frac{3}{4} E_8 + \mathbf{S}_1 \cdot \mathbf{S}_2, \nonumber
\end{equation}
where $E_8$ denotes the identity matrix of order 8.

\subsection{First Order Filtering} \label{ssct:Filtering}
Given $\gamma>0$, $\mathbf{v}_0\in \mathbb{R}^3$ and $\mathbf{u}\in L^3_2(0,T;\mathbb{R}^3)$, the \textit{filtering equation} \eqref{filter} provides a continuous in time electromagnetic field intensity according to the formula 
\begin{equation}
\mathbf{v}(t; \mathbf{u}) = e^{-\gamma t} \left[ \mathbf{v}_0 + \int_0^t \gamma e^{\gamma\tau} \mathbf{u}(\tau) d\tau \right], ~ t\in[0,T]. \label{magf_exp}
\end{equation}
In particular, this formula implies that if the control input $\mathbf{u}$ is piecewise-constant in time, then the optimal electromagnetic wave input $\mathbf{v}$ is continuous and piecewise-smooth in time.

\section{\label{sct:ocp}Optimal Control Problem}
Consider the optimal control problem of maximizing the amount of triplet born singlet yield over the time interval $[0, T]$
\begin{align}
&\resizebox{0.45\textwidth}{!}{%
$\mathcal{J}(\mathbf{u}) 
= \frac{k^\prime_S}{3\cdot2^{p+1}} \displaystyle\sum_{l=1}^{3\cdot2^p} \int_0^T \langle \psi^l(t;\mathbf{u}) | \mathbf{P}_S | \psi^l(t;\mathbf{u}) \rangle_{\mathbb{C}^n} dt $} \nonumber\\
&\rightarrow \max \label{cost_func}
\end{align}
on a control set 
\begin{equation}
\mathcal{V} = \left\{ 
\begin{array}{c} \displaystyle
\mathbf{u} \in L^3_2(0,T;\mathbb{R}^3) ~ :  \\ \displaystyle
\mathbf{u}(t) \in V = \prod_{i=x,y,z}[m_i,M_i] \\ \displaystyle
\mbox{ for a.e. } t \in [0,T] 
\end{array}
\right\} \label{ctrl_set}
\end{equation}
where $k^\prime=k_S / \hbar$ and $\Big ((\psi^l(t;\mathbf{u}))_{l=1}^{3\cdot 2^p}, \mathbf{v}(t; \mathbf{u})\Big )$ is the corresponding solution of the filtered Schr\"odinger system \eqref{schrodinger},\eqref{filter}.  

\section{\label{sct:frechet}Fr\'echet Differentiability in Hilbert Space}
Define the \textit{Hamilton Pontryagin function}
\begin{eqnarray}
\mathcal{H}(\psi(t),\mathbf{u}(t),\chi(t)) := \mathcal{H}(\psi,\mathbf{u},\chi)(t) \nonumber \\
=
\resizebox{0.45\textwidth}{!}{%
$\displaystyle\frac{\mu_B g / \hbar}{3\cdot2^{p-1}} \sum_{l=1}^{3\cdot2^p} \int_t^T \mathrm{Im} \langle \chi^l(\tau) | \mathbf{S}_{1} + \mathbf{S}_{2} | \psi^l(\tau)  \rangle_{\mathbb{C}^n} \gamma e^{\gamma(t-\tau)} d\tau \cdot \mathbf{u}(t) , $} \nonumber\\
\label{hpf}
\end{eqnarray}
where $\psi^l,\chi^l:[0,T] \to \mathbb{C}^n$, $l=1,\dots,3\cdot2^p$ and $u:[0,T]\to\mathbb{R}^3$.

The following is the Fr\'echet differentiability result of the cost functional $\mathcal{J}$.

\begin{theorem} \label{thm-frechet}
The functional $\mathcal{J}$ is continuously differentiable in $\mathcal{V}$, and the Fr\'echet derivative $\mathcal{J}'(\mathbf{u}) \in L^3_2(0,T;\mathbb{R}^3)$, for $\mathbf{u}\in\mathcal{V}$, is given by
\begin{align}
\mathcal{J}'(\mathbf{u}) 
= \frac{\partial\mathcal{H}}{\partial \mathbf{u}}  \nonumber\\
= \resizebox{0.45\textwidth}{!}{%
$\displaystyle\frac{\gamma \mu_B g / \hbar}{3\cdot2^{p-1}} \sum_{l=1}^{3\cdot2^p} \int_t^T \mathrm{Im} \langle \chi^l(\tau;\mathbf{u}) | \mathbf{S}_1 + \mathbf{S}_2 | \psi^l(\tau;\mathbf{u}) \rangle_{\mathbb{C}^n} e^{\gamma(t-\tau)} d\tau , $} \label{frechet-dif}
\end{align}
where $\Big ((\psi^l(t;\mathbf{u}))_{l=1}^{3\cdot 2^p}, \mathbf{v}(t; \mathbf{u})\Big )$ is the solution of \eqref{schrodinger},\eqref{filter} and $(\chi^l(\cdot;\mathbf{u}))_{l=1}^{3\cdot 2^p}$ be a solution of the corresponding adjoint Schr\"odinger system
\begin{align}
i\hbar \frac{d\chi^l}{dt} &= \mathbf{H}^\ast(\mathbf{v}) \chi^l - i\frac{k_S}{2} \mathbf{P}_S \psi^l, \quad \mbox{in $[0,T)$}; \nonumber\\ 
\chi^l(T) &= 0, \label{adjoint}
\end{align}
where 
\begin{equation}
\mathbf{H}^\ast(\mathbf{v}) = \mathbf{H}_z(\mathbf{v}) + \mathbf{H}_{hfi} + i \mathbf{K}.
\end{equation}
be an adjoint Hamiltonian.
\end{theorem}

\noindent\textit{Proof.}
Let $\mathbf{u} \in \mathcal{V}$ and consider an increment $\delta\mathbf{u}\in L^3_2(0,T;\mathbb{R}^3)$ such that $\mathbf{u} +\delta\mathbf{u}\in\mathcal{V}$. Denote $\psi^l=\psi^l(\cdot;\mathbf{u} )$ and $\overline{\psi}^l=\overline{\psi}^l(\cdot;\mathbf{u} +\delta\mathbf{u} )$ the corresponding solutions of \eqref{schrodinger} with $\psi_T=\psi^l_T$, $l=1,...,3 \cdot 2^m$. The vector-functions $\delta\psi^l=\bar{\psi}^l-\psi^l$ and $\delta\mathbf{v}$ 
is a solution of the Cauchy problem:
\begin{align}
\imath \hbar \frac{ d \delta\psi^l}{d t}  &=  \mathbf{H}(\mathbf{v}) \delta\psi^l + \mathbf{H}_{z}(\mathbf{\delta v})\bar{\psi}^l , \ 0\leq t \leq T; \nonumber\\ 
\delta\psi^l(0) &= 0 , \label{incrementedSchroedinger's Equation} \\ 
 \frac{ d \delta\mathbf{v}}{d t}+\gamma\delta\mathbf{v}  &=  \gamma\delta\mathbf{u}, \ 0\leq t \leq T; \nonumber \\ 
 \delta\mathbf{v}(0) &= 0 ,  \label{incric}
\end{align}
or, equivalently the pair $(\delta\psi^l,\delta\mathbf{v})$ solves the integral equation in $(0,T)$:
\begin{align}
\delta\psi^l(t) &= -\imath \hbar^{-1}\int_0^t\mathbf{H}(\mathbf{v}(\tau)) \delta\psi^l(\tau)\,d\tau \nonumber\\
&\quad-\imath \hbar^{-1} \int_0^t \mathbf{H}_{z}(\mathbf{\delta v}(\tau)) \bar{\psi}^l(\tau)\,d\tau, \label{est1}\\
\delta \mathbf{v}(t) &= \int_0^t \gamma e^{\gamma(\tau-t)} \delta \mathbf{u}(\tau) d\tau. \label{delta-v}
\end{align}
From \eqref{magf_exp},\eqref{est1},\eqref{delta-v} it follows that 
\begin{align}
|\mathbf{v}(t)| \leq C, \quad
|\delta\mathbf{v}(t)|\leq C\int_0^t|\delta\mathbf{u}(\tau)|\,d\tau, \label{est2}\\
|\delta\psi^l(t)|\leq C \int_0^t|\mathbf{v}(\tau)| |\delta\psi^l(\tau)|\,d\tau \nonumber\\
+ C\int_0^t|\mathbf{\delta v}(\tau)|\,d\tau, 
\end{align}
for $0\leq t\leq T$, where $C$ (and all the subsequently used constants $C_i)$ depends on parameters $\gamma, T, m_i, M_i$. 
Since $\mathbf{u}\in \mathcal{V}$, it follows that
\begin{equation}\label{est3}
|\delta\psi^l(t)|\leq C_1 \int_0^t|\delta\psi^l(\tau)|\,d\tau + C\int_0^t|\mathbf{\delta u}(\tau)|\,d\tau,
\end{equation}
for $0\leq t \leq T$. 
Using Gronwall's lemma \cite{lady1} from \eqref{est3} it follows that
\begin{equation}\label{est4}
|\delta\psi^l(t)|\leq Ce^{C_1T} \int_0^T|\mathbf{\delta u}(\tau)|\,d\tau, 
\end{equation}
for $0\leq t \leq T$.

Note that similarly, we can establish a uniform boundedness of the state and adjoined vectors.
Given $\mathbf{u}\in \mathcal{V}$, the state vector $\psi^l$, and adjoined vector $\chi^l$ satisfy the following integral equations:
\begin{eqnarray}
\psi^l(t)=-\imath \hbar^{-1}\int_0^t\mathbf{H}(\mathbf{v}(\tau)) \psi^l(\tau)\,d\tau + \psi_T^l, \nonumber\\
\chi^l(t)= \!\! \int_t^T \!\! \imath \Big [ \hbar^{-1}\mathbf{H}^*(\mathbf{v}(\tau)) \chi^l(\tau)+\frac{k_S}{2\hbar} \mathbf{P}_S \psi^l(\tau)\Big ]\,d\tau, \nonumber
\end{eqnarray}
where $\mathbf{v}$ satisfies \eqref{magf_exp}.
Applying Gronwall's lemma again, we derive the uniform bound
\begin{equation}\label{unest}
\max (|\psi^l(t)|, |\chi^l(t)|) \leq C, \ 0\leq t \leq T.
\end{equation}

Next, we transform an increment of the functional \eqref{cost_func} as follows:
\begin{align}
\mathcal{J}(\mathbf{u}+\delta \mathbf{u}) - \mathcal{J}(\mathbf{u})  \nonumber\\
= \frac{k^\prime_S}{3\cdot2^p} \sum_{l = 1}^{3\cdot2^p} \int_0^T \left[ \mathrm{Re} \langle \psi^l(t) | \mathbf{P}_S | \delta\psi^l(t) \rangle_{\mathbb{C}^n}
\right. \nonumber\\ \left. + \frac{1}{2} \langle \delta\psi^l(t) | \mathbf{P}_S | \delta\psi^l(t) \rangle_{\mathbb{C}^n} \right] dt, \label{fd-eq1}
\end{align}
where $\delta\psi^l = \overline{\psi}^l - \psi^l$. Using \eqref{adjoint},\eqref{incrementedSchroedinger's Equation} we have
\begin{align}
0 &= \int_0^T i\hbar \frac{d}{dt} \langle \chi^l | \delta\psi^l \rangle_{\mathbb{C}^n} dt \nonumber\\
&= \int_0^T \left[ \langle -i\hbar \frac{d\chi^l}{dt} | \delta\psi^l \rangle_{\mathbb{C}^n} + \langle \chi^l | i\hbar \frac{d\delta\psi^l}{dt} \rangle_{\mathbb{C}^n} \right] dt \nonumber \\
&= \int_0^T \left[ \langle - \mathbf{H}^\ast(\mathbf{v}) \chi^l | \delta\psi^l \rangle_{\mathbb{C}^n}
+ \langle \chi^l | \mathbf{H}(\mathbf{v}) \delta\psi^l \rangle_{\mathbb{C}^n} \right] \nonumber\\
& - \int_0^T i \frac{k_S}{2} \langle \psi^l | \mathbf{P}_S | \delta\psi^l \rangle_{\mathbb{C}^n} dt \nonumber \\
& + \int_0^T \langle \chi^l | \mathbf{H}_z(\delta \mathbf{v}) | \psi^l \rangle_{\mathbb{C}^n} + \langle \chi^l | \mathbf{H}_z(\delta \mathbf{v}) | \delta\psi^l \rangle_{\mathbb{C}^n} dt \nonumber
\end{align}
which implies
\begin{align}
\frac{k_S}{2} \int_0^T \mathrm{Re} \langle \psi^l | \mathbf{P}_S | \delta\psi^l \rangle_{\mathbb{C}^n} \nonumber\\
=  \int_0^T \mathrm{Im} \langle \chi^l | \mathbf{H}_z(\delta \mathbf{v}) | \psi^l \rangle_{\mathbb{C}^n} dt \nonumber\\
+  \int_0^T \mathrm{Im} \langle \chi^l | \mathbf{H}_z(\delta \mathbf{v}) | \delta\psi^l \rangle_{\mathbb{C}^n} dt . \label{fd-eq2}
\end{align}

Combining \eqref{fd-eq1} and \eqref{fd-eq2} we obtain
\begin{align}
\mathcal{J}(\mathbf{u}+\delta \mathbf{u}) - \mathcal{J}(\mathbf{u}) \nonumber\\
= \frac{1 / \hbar}{3\cdot2^{p-1}} \sum_{l=1}^{3\cdot2^p} \int_0^T \mathrm{Im} \langle \chi^l | \mathbf{H}_z(\delta \mathbf{v}) | \psi^l \rangle_{\mathbb{C}^n} dt + R, \label{fd-eq3}
\end{align}
where
\begin{align}
R = \frac{1 / \hbar}{3\cdot2^p} \sum_{l=1}^{3\cdot2^p} \int_0^T \Big[ 2 \mathrm{Im} \langle \chi^l | \mathbf{H}_z(\delta \mathbf{v}) | \delta\psi^l \rangle_{\mathbb{C}^n}  \nonumber\\ 
+ \frac{k_S}{2} \langle \delta\psi^l | \mathbf{P}_S | \delta\psi^l \rangle_{\mathbb{C}^n} \Big] dt . \label{fd-rem}
\end{align}

Substituting \eqref{delta-v} in \eqref{fd-eq3} and using \eqref{Zeeman}, and after changing the order of integration according to Fubini's theorem, we arrive at
\begin{align}
\mathcal{J}(\mathbf{u}+\delta \mathbf{u}) - \mathcal{J}(\mathbf{u}) - R \nonumber \\
= \resizebox{0.45\textwidth}{!}{%
$\displaystyle\frac{\gamma \mu_B g / \hbar}{3\cdot2^{p-1}} \sum_{l=1}^{3\cdot2^p} \sum_{i=x,y,z} \int_0^T \!\! \int_t^T \!\! \mathrm{Im} \langle \chi^l(\tau) | \mathbf{S}_{1i} + \mathbf{S}_{2i} | \psi^l(\tau) \rangle_{\mathbb{C}^n} e^{\gamma(t-\tau)} \delta u_i(t) d\tau dt $} \nonumber \\
= \resizebox{0.45\textwidth}{!}{%
$\displaystyle\left\langle \frac{\gamma \mu_B g / \hbar}{3\cdot2^{p-1}} \sum_{l=1}^{3\cdot2^p} \int_t^T \!\! \mathrm{Im} \langle \chi^l(\tau;\mathbf{u}) | \mathbf{S}_1 + \mathbf{S}_2 | \psi^l(\tau;\mathbf{u}) \rangle_{\mathbb{C}^n} e^{\gamma(t-\tau)} d\tau , \delta \mathbf{u} \right\rangle_{L^3_2(0,T;\mathbb{R}^3)} $} \label{FrGr}
\end{align}
which implies \eqref{frechet-dif}, provided that 
\begin{equation}\label{osmall}
R=o\left(\|\delta\mathbf{u}\|_{L^3_2(0,T;\mathbb{R}^3)}\right) , \ \quad\mbox{as} \ \|\delta\mathbf{u}\|_{L^3_2(0,T;\mathbb{R}^3)}\to 0.
\end{equation}

From \eqref{est2},\eqref{est4},\eqref{unest} it follows that
\begin{align}\label{Rsmall}
|R| &\leq C_3 \Big[ \|\mathbf{\delta \psi}\|^2_{C(0,T; \mathbb{C}^n)} \nonumber\\ 
&\quad + \|\mathbf{\delta v}\|_{L_1^3(0,T; \mathbb{R}^3)} \|\mathbf{\delta \psi}\|_{C(0,T; \mathbb{C}^k)} \Big] \nonumber\\
&\leq C_4\|\mathbf{\delta u}\|^2_{L_1^3(0,T;\mathbb{R}^3)}.
\end{align}

By applying Cauchy-Bunyakovsky-Schwarz (CBS) inequality, we deduce the estimation
\begin{equation}\label{cbs}
|R|\leq C_4 T \|\mathbf{\delta u}\|^2_{L_2^3(0,T; \mathbb{R}^3)}.
\end{equation}

From \eqref{FrGr}, \eqref{cbs} we deduce that
\begin{align}
\delta\mathcal{J}(\mathbf{u})= \left\langle \frac{\partial\mathcal{H}}{\partial\mathbf{u}}, \delta\mathbf{u}\right\rangle_{L_2^3(0,T;\mathbb{R}^3)} \nonumber\\
+o\left(\|\mathbf{\delta u}\|_{L_2^3(0,T;\mathbb{R}^3)}\right), \ \quad\mbox{as} \ \|\mathbf{\delta u}\|_{L_2^3}\to 0,
\end{align}
which proves the Fr\'echet differentiability and the gradient formula. 
\hfill \ensuremath{\Box}

\vskip.1in
The Fr\'echet differentiability results in Theorem \ref{thm-frechet} imply the following necessary condition for the optimality.
\begin{corollary} \label{cor-opt1}
Let $\mathbf{u}^\ast \in \mathcal{V}$ be an optimal value of $\mathcal{J}$. Then, for every $\mathbf{u} \in \mathcal{V}$ we have
\begin{align}
&\resizebox{0.45\textwidth}{!}{%
$\displaystyle\sum_{l=1}^{3\cdot2^p} 
\left\langle 
\int_t^T \mathrm{Im} \langle \chi^l(\tau;\mathbf{u}^\ast) | \mathbf{S}_1 + \mathbf{S}_2 | \psi^l(\tau;\mathbf{u}^\ast) \rangle_{\mathbb{C}^n} e^{\gamma(t-\tau)} d\tau , \mathbf{u}^\ast - \mathbf{u}
\right\rangle_{L^3_2(0,T;\mathbb{R}^3)} $} \nonumber\\ 
&\geq 0. 
\label{cor-opt1-1} 
\end{align}
\end{corollary}

\section{Pontryagin Maximum Principle}\label{sct:pmp}
The optimality condition \eqref{cor-opt1-1} is equivalent to the following integral form of the Pontryagin Maximum Principle (PMP):
\begin{corollary}[integral form of PMP] \label{cor-integralPMP}
Let $\mathbf{u}^\ast \in \mathcal{V}$ be an optimal control. Then,
\begin{align}
\max_{\mathbf{u}\in \mathcal{V}} \int_0^T\mathcal{H}( \psi(t; \mathbf{u}^\ast) , \mathbf{u} , \chi(t; \mathbf{u}^\ast) )dt \nonumber\\
 =\int_0^T\mathcal{H}( \psi(t; \mathbf{u}^\ast) , \mathbf{u}^\ast(t) , \chi(t; \mathbf{u}^\ast))dt\label{cor-integralPMP} 
\end{align}
\end{corollary}

Next, we formulate the \textit{Pontryagin Maximum Principle}:
\begin{theorem}[Pontryagin Maximum Principle] \label{thm-pmp}
Let $\mathbf{u}^\ast \in \mathcal{V}$ be an optimal control. Then, for almost every $t\in[0,T]$, we have
\begin{align}\label{PMP}
\max_{\mathbf{u}\in V} \mathcal{H}( \psi(t; \mathbf{u}^\ast) , \mathbf{u} , \chi(t; \mathbf{u}^\ast) ) \nonumber\\
= \mathcal{H}( \psi(t; \mathbf{u}^\ast) , \mathbf{u}^\ast(t) , \chi(t; \mathbf{u}^\ast)).
\end{align}
\end{theorem}

\noindent\textit{Proof.} 
Let $\mathbf{u}^\ast \in \mathcal{V}$ be an optimal control, and $\delta \mathbf{u} \in L_2^3(0,T; \mathbb{R}^3)$ be an admissible control variation, i.e. $\mathbf{u}^\ast+\delta \mathbf{u} \in \mathcal{V}$. We have 
\begin{equation}\label{varpos}
 \delta \mathcal{J}(\mathbf{u}^\ast)=\mathcal{J}(\mathbf{u}^\ast+\delta \mathbf{u})-\mathcal{J}(\mathbf{u}^\ast) \leq 0.
 \end{equation}
 Let $t\in (0, T)$ be a Lebesgue point for the optimal control $\mathbf{u}^\ast$. For any fixed vector $\mathbf{v}\in V$, and for all sufficiently small $\epsilon>0$, consider a special control variation
\begin{equation}\label{incrctrl}
\delta \mathbf{u}(t)=
\left\{\begin{array}{l}
\mathbf{v}-\mathbf{u}^\ast(\tau), \ t-\epsilon \leq \tau \leq t+\epsilon,\\
0, \ \tau \in [0,T]\setminus [t-\epsilon, t+\epsilon]. 
\end{array}\right.
\end{equation}
From \eqref{FrGr} it follows that
\begin{align}
\delta\mathcal{J}(\mathbf{u}^\ast)=\int\limits_{t-\epsilon}^{t+\epsilon}\Big [ \mathcal{H}(\psi(\tau; \mathbf{u}^\ast), \mathbf{v}, \chi(\tau; \mathbf{u}^\ast)) \nonumber\\ 
- \mathcal{H}(\psi(\tau; \mathbf{u}^\ast), \mathbf{u}^\ast(\tau), \chi(\tau; \mathbf{u}^\ast))\Big ]\,d\tau + R. \label{est10}
\end{align}
Since vector functions $\psi(\cdot; \mathbf{u}^\ast)$ and $\chi(\cdot; \mathbf{u}^\ast)$ are continuous on $[0,T]$, $t$ is a Lebesgue point of the integrand in \eqref{est10}.
Therefore, we have 
\begin{eqnarray}
\resizebox{0.45\textwidth}{!}{%
$\displaystyle\lim\limits_{\epsilon \to 0} \frac{1}{2\epsilon}\int\limits_{t-\epsilon}^{t+\epsilon}\Big [ \mathcal{H}(\psi(\tau; \mathbf{u}^\ast), \mathbf{v}, \chi(\tau; \mathbf{u}^\ast)) - \mathcal{H}(\psi(\tau; \mathbf{u}^\ast), \mathbf{u}^\ast(\tau), \chi(\tau; \mathbf{u}^\ast))\Big ]\,d\tau $} \nonumber\\
\resizebox{0.45\textwidth}{!}{%
$= \mathcal{H}(\psi(t; \mathbf{u}^\ast), \mathbf{v}, \chi(t; \mathbf{u}^\ast)) 
-\mathcal{H}(\psi(t; \mathbf{u}^\ast), \mathbf{u}^\ast(t), \chi(t; \mathbf{u}^\ast)). $} \nonumber\\ \label{est11}
\end{eqnarray}
By applying CBS inequality, from \eqref{Rsmall} we deduce that
\begin{equation}\label{est12}
\frac{1}{2\epsilon}R\leq C_4\|\mathbf{\delta u}\|^2_{L_2^3(t-\epsilon,t+\epsilon;\mathbb{R}^3)}\to 0, \ \quad\mbox{as} \ \epsilon\to 0. 
\end{equation}
Dividing \eqref{est10} by $2\epsilon$, passing to the limit as $\epsilon\to 0$, and using \eqref{varpos}, \eqref{est11} and \eqref{est12} it follows that
\begin{align}\label{est14}
\resizebox{0.45\textwidth}{!}{%
$\mathcal{H}(\psi(t; \mathbf{u}^\ast), \mathbf{v}, \chi(t; \mathbf{u}^\ast)) 
\leq \mathcal{H}(\psi(t; \mathbf{u}^\ast), \mathbf{u}^\ast(t), \chi(t; \mathbf{u}^\ast)). $}
\end{align}
Since $\mathbf{v}\in V$ is arbitrary, and Lebesgue points of $\mathbf{u}^\ast$ are dense in $[0,T]$, {\it Pontryagin maximum principle} \eqref{PMP} follows.
\hfill \ensuremath{\Box}

\vskip.1in
From PMP \eqref{PMP} it follows that the optimal control is of \textit{bang-bang} type, meaning that each component switches between lower and upper bounds of prism $V$. Precisely, it follows from Theorem \ref{thm-pmp} that if $\mathbf{u}^\ast \in \mathcal{V}$ is an optimal control, then 
\begin{equation}
\resizebox{0.45\textwidth}{!}{%
$ u^\ast_i 
= \left\{ \begin{array}{ll}\label{bangbangformula}
M_i , \!\! & \mbox{if} ~ \textrm{sign} \displaystyle\sum_{l=1}^{3\cdot2^p} \int_t^T \!\!\mathrm{Im} \langle \chi^l(\tau) | \mathbf{S}_{1i} + \mathbf{S}_{2i} | \psi^l(\tau) \rangle_{\mathbb{C}^n} e^{\gamma(t-\tau)} d\tau > 0 ; \\
m_i , \!\! & \mbox{if} ~ \textrm{sign} \displaystyle\sum_{l=1}^{3\cdot2^p} \int_t^T \!\!\mathrm{Im} \langle \chi^l(\tau) | \mathbf{S}_{1i} + \mathbf{S}_{2i} | \psi^l(\tau) \rangle_{\mathbb{C}^n} e^{\gamma(t-\tau)} d\tau < 0 , 
\end{array} \right. $}
\end{equation}
for $i=x,y,z$, where $\psi = \psi(\cdot;\mathbf{u}^\ast)$ and $\chi = \chi(\cdot;\mathbf{u}^\ast)$ are the corresponding optimal states. Hence, the formula \eqref{bangbangformula} expresses optimal control $\mathbf{u_*}$ in terms of optimal states $\psi = \psi(\cdot;\mathbf{u}^\ast)$ and $\chi = \chi(\cdot;\mathbf{u}^\ast)$: 
\begin{equation}
\mathbf{u}^\ast = \Gamma(\psi,\chi), \label{Gamma-1} 
\end{equation}
where $\Gamma=(\Gamma_i, i=x,y,z)$ be a vector-function with components
\begin{align}
\Gamma_i(\psi,\chi) 
= \frac{M_i+m_i}{2} \nonumber\\ + \resizebox{0.45\textwidth}{!}{%
$\displaystyle\frac{M_i - m_i}{2} \mathrm{sign} \sum_{l=1}^{3\cdot2^p} \int_t^T \hspace{-.2cm}\mathrm{Im} \langle \chi^l(\tau) | \mathbf{S}_{1i} + \mathbf{S}_{2i} | \psi^l(\tau) \rangle_{\mathbb{C}^n} e^{\gamma(t-\tau)}dt $} , \label{Gamma-2}
\end{align}
for $i=x,y,z$.

$\Gamma$ is an \textit{optimal control synthesizing function}. Using expression \eqref{magf_exp}, one expresses the corresponding optimal magnetic field in terms of the optimal states $\psi=\psi(\cdot;\mathbf{u}^\ast)$ and $\chi=\chi(\cdot;\mathbf{u}^\ast)$:
\begin{equation}\label{optimalv}
\mathbf{v}(\cdot;\mathbf{u}^\ast) = \mathbf{v}(\cdot;\Gamma(\psi,\chi)). 
\end{equation}
$\Gamma$ is an \textit{optimal control synthesizing function}. Using expression \eqref{magf_exp}, one expresses the corresponding optimal magnetic field in terms of the optimal states $\psi=\psi(\cdot;\mathbf{u}^\ast)$ and $\chi=\chi(\cdot;\mathbf{u}^\ast)$:
\begin{equation}\label{optimalv}
\mathbf{v}(\cdot;\mathbf{u}^\ast) = \mathbf{v}(\cdot;\Gamma(\psi,\chi)). 
\end{equation}
By substituting \eqref{optimalv} in \eqref{schrodinger} and \eqref{adjoint} we deduce that the optimal states $\psi$ and $\chi$ must satisfy the following nonlinear closed-loop system of $3\cdot 2^{2(p+1)}$ integro-differential equations with two-end boundary values:
\begin{align}
i\hbar \frac{d\psi^l}{dt} = \mathbf{H}(\mathbf{v}(t;\Gamma(\psi,\chi))), ~ \mbox{in $(0,T]$} ; \nonumber\\ 
\psi^l(0) = \psi^l_T \in \mathbb{C}^n ; \label{closedsystem1} \\
i\hbar \frac{d\chi^l}{dt} = \mathbf{H}^\ast(\mathbf{v}(t;\Gamma(\psi,\chi))) - i \frac{k_S}{2} \mathbf{P}_S \psi^l, ~ \mbox{in $(0,T]$} ; \nonumber\\
\chi^l(T) = 0 \in \mathbb{C}^n, \label{closedsystem2}
\end{align}
for $l = 1,\dots,3\cdot2^p$.

Hence, \textit{Pontryagin Maximum Principle} implies the following explicit two-step method (EPMP) for the identification of the optimal electromagnetic field input:
\begin{description}
\item[{ Step 1.}] Solve the nonlinear closed system of integro-differential equations \eqref{closedsystem1},\eqref{closedsystem2} to find optimal state and adjoined vector-functions 
\[ (\psi,\chi): [0,T]\to \mathbb{C}^{3\cdot2^{2(p+1)}}\times \mathbb{C}^{3\cdot2^{2(p+1)}}; \]
\item[{ Step 2.}] Use \textit{optimal control synthesizing function} $\Gamma$ to find a bang-bang optimal control $\mathbf{u}^\ast$ from \eqref{Gamma-1}, and continuous in-time optimal electromagnetic field input $\mathbf{v}^\ast$ from \eqref{optimalv}.
\end{description}
The EPMP method is computationally expensive. In the next section, we suggest an iterative method based on the PMP.

\begin{remark} It is essential to note that our model assumes that the radical pair system is a pure system described by the Schr\"{o}dinger
system, and doesn't take into account many factors, including the interaction with the environment \cite{KOC2022}, decoherence, and spin relaxation
phenomena \cite{Kom2009,Tie2012,Luo2024,Poo2015,Wor2016,Fay2021}. Therefore, the natural question is what is the limitation of the characterization of the {\it quantum coherence} through {\it Pontryagin Maximum Principle} in terms of system complexity? Remarkably, the applicability of the Pontryagin Maximum Principle and characterization of the quantum coherence via bang-bang optimal control is relevant for a broad class of quantum optimal control frameworks. Mathematically, it is hidden in the bilinear structure of the Hamilton-Pontryagin function with respect to state and control variables. In particular, replacing the phenomenological reaction term $\mathbf{K}$ in \eqref{Haberkorn} via the fundamental quantum dynamical evolution of radical pair 
recombination reaction term derived from the quantum measurement theory and incorporating quantum Zeno effect \cite{Kom2009} would preserve relevance of the Pontryagin Maximum Principle for the characterization of the quantum coherence and bang-bang structure of the optimal control. 
\end{remark}

\section{Algorithms} \label{sct:alg}

\subsection{Gradient Projective Method (GPM) in Hilbert Space} \label{ssct:GPM}
Fr\'echet differentiability result Theorem \ref{thm-frechet} implies the following gradient method in the Hilbert space $L^3_2(0,T;\mathbb{R}^3)$.

\vskip.1in \noindent\textit{Step 1.}
Set $N=0$ and define initial control $\mathbf{u}^0 \in \mathcal{V}$.

\vskip.1in \noindent\textit{Step 2.}
Compute magnetic field $\mathbf{v}^N=\mathbf{v}^N(\cdot;\mathbf{u}^N)$ using filter expression \eqref{magf_exp} for given $\mathbf{v}_0\in\mathbb{R}^3$ and filter parameter $\gamma>0$.

\vskip.1in \noindent\textit{Step 3.}
Solve Schr\"odinger equation \eqref{schrodinger} using $\mathbf{v}^N$ to find $\psi^N=\psi(\cdot;\mathbf{u}^N)$ and compute corresponding cost $\mathcal{J}(\mathbf{u}^N)$.

\vskip.1in \noindent\textit{Step 4.}
If $N=0$, move to the next step. Otherwise, check the following criteria
\begin{align}
\left| \frac{\mathcal{J}(\mathbf{u}^N) - \mathcal{J}(\mathbf{u}^{N-1})}{\mathcal{J}(\mathbf{u}^N)} \right| < \epsilon_1 \quad \mbox{and}  \nonumber\\
\frac{ \| \mathbf{u}^N - \mathbf{u}^{N-1} \|_{L^3_2(0,T;\mathbb{R}^3)} }{\| \mathbf{u}^N \|_{L^3_2(0,T;\mathbb{R}^3)}}  < \epsilon_2 , \label{tol}
\end{align}
where $\epsilon_1,\epsilon_2>0$ are required tolerance errors. If \eqref{tol} is verified, then terminate the iteration process. Otherwise, move to the next step.

\vskip.1in \noindent\textit{Step 5.}
Use $\psi^N$ and $v^N$ to solve the adjoint problem \eqref{adjoint} in order to find $\chi^N = \chi^N(\cdot;u^N)$ and compute gradient $\mathcal{J}'(u^N)$ using expression \eqref{frechet-dif}.

\vskip.1in \noindent\textit{Step 6.}
Choose step-size parameter $\lambda_N>0$ and compute new control $\mathbf{u}^{N+1}$ as follows
\begin{equation}
\mathbf{u}^{N+1} = \mathbf{u}^N + \lambda_N \cdot \mathcal{J}'(\mathbf{u}^N) .
\end{equation}

\vskip.1in \noindent\textit{Step 7.}
Replace $\mathbf{u}^{N+1}=\mathrm{proj}_V(\mathbf{u}^{N+1})$, where
\begin{equation}
\mathrm{proj}_V(\mathbf{u}^{N+1}) := \left\{ \begin{array}{ll}
m_i, & \mbox{if} ~ \mathbf{u}^{N+1}_i \leq m_i ; \\
u^{N+1}_i , & \mbox{if} ~ m_i < \mathbf{u}^{N+1}_i < M_i ; \\
M_i , & \mbox{if} ~ u^{N+1}_i \geq M_i ;
\end{array} \right.
\end{equation}
for $i = x,y,z$. Then, replace $N$ with $N+1$ and move to Step 2.

\vskip.1in
\noindent\textit{Remark.} The learning rate $\lambda_N$ in Step 6 is calculated for each iteration and based on the Barzilai-Borwein method. Indeed, we compute $\lambda_N$ proportional to $\lambda_{BB}^N$, where
\begin{align}
\resizebox{0.45\textwidth}{!}{%
$\lambda_{BB}^N  = \left| \frac{ \langle \mathbf{u}^{N}-\mathbf{u}^{N-1} , \mathcal{J}'(\mathbf{u}^N)-\mathcal{J}'(\mathbf{u}^{N-1}) \rangle_{L^3_2(0,T;\mathbb{R}^3)} }{ \| \mathcal{J}'(\mathbf{u}^N)-\mathcal{J}'(\mathbf{u}^{N-1}) \|_{L^3_2(0,T;\mathbb{R}^3)} } \right| . $}
\end{align}

\subsection{Iterative Pontryagin Maximum Principle (IPMP) Method} \label{ssct:IPMP}

\textit{Pontryagin Maximum Principle} and the two-step EPMP method suggest the following iterative algorithm to solve the optimal control problem.

\vskip.1in \noindent\textit{Step 1.}
Set $N=0$ and define initial control $\mathbf{u}^0 \in \mathcal{V}$.

\vskip.1in \noindent\textit{Step 2.}
Compute magnetic field $\mathbf{v}^N=\mathbf{v}^N(\cdot;\mathbf{u}^N)$ using filter expression \eqref{magf_exp} for given $\mathbf{v}_0\in\mathbb{R}^3$ and filter parameter $\gamma>0$.

\vskip.1in \noindent\textit{Step 3.}
Solve Schr\"odigner system \eqref{schrodinger} and its adjoint \eqref{adjoint} using $\mathbf{v}^N$ to find $\psi^N=\psi(\cdot;\mathbf{u}^N)$ and $\chi^N=\chi^N(\cdot;\mathbf{u}^N)$, respectively. 

\vskip.1in \noindent\textit{Step 4.}
Compute a new control vector by using \eqref{Gamma-1}
\begin{equation}\label{itersynthes}
\mathbf{u}^{N+1} = \Gamma(\psi^N,\chi^N), 
\end{equation}
i.e.
\begin{equation}
u^{N+1}_i 
= \left\{ \begin{array}{ll}\label{iterbangbangformula}
M_i , & \mbox{if} ~ \Gamma_i(\psi(\cdot; \mathbf{u}^N), \chi(\cdot; \mathbf{u}^N) ) > 0 ; \\
m_i , & \mbox{if} ~ \Gamma_i(\psi(\cdot; \mathbf{u}^N), \chi(\cdot; \mathbf{u}^N) ) > 0 , 
\end{array} \right.
\end{equation}
for $i=x,y,z$.

\vskip.1in \noindent\textit{Step 5.}
If $\mathbf{u}^{N+1}=\mathbf{u}^N$, terminate the iteration process. Otherwise, replace $N$ with $N+1$ and move to Step 2.

Note that the IPMP algorithm at every iteration solves the linear Schr\"odinger system \eqref{schrodinger} and its adjoint \eqref{adjoint}, and produces a new bang-bang control vector by \eqref{itersynthes}. Whenever $\mathbf{u}^{N+1}=\mathbf{u}^N$ is fulfilled, then $\mathbf{u}^N$ satisfies the PMP relation \eqref{Gamma-1}, \eqref{Gamma-2}, and therefore is a candidate to be a bang-bang optimal control, and accordingly $\mathbf{v}^N=\mathbf{v}(\cdot; \mathbf{u}^N)$ is a candidate to be a continuous in-time optimal electromagnetic field input.

\section{Numerical Results}\label{sct:num_results}

This section presents numerical results obtained by implementing the algorithms described in Section \ref{sct:alg}. Here, we considered the dynamics for the model as described in Section \ref{sct:model} for $p=1,\dots,7$ proton cases with spin-1/2 and time length $T=0.5~\mathrm{\mu s}$. Constant rates for the decay are chosen as $k_S = k_T = 0.5~\hbar~\mathrm{\mu s}^{-1}$. Constant-in-time hyperfine $\mathbf{A}=(\mathbf{A}_j)_{j=1,\dots,p}$ is set as follows:
\begin{eqnarray*}
\mathbf{A}_1 = ( -0.234 , -0.234 , 0.117 ) ; \\
\mathbf{A}_2 = ( -0.030 , -0.022 , 0.688 ) ; \\
\mathbf{A}_3 = ( 0.238 , 0.357 , 0.117 ) ; \\
\mathbf{A}_j = ( -0.218 , -0.202 , -0.054 ) , 
\end{eqnarray*}
for $j = 4,\dots,p$, all in millitesla $(\mathrm{mT})$. 

A 200 points discretization of the time interval $[0, T]$ was considered, and solutions of the Schr\"odinger system \eqref{schrodinger} and its adjoint \eqref{adjoint} were obtained using an adaptation of the 4th order Runge-Kutta method.

\vskip.1in
\noindent\textit{Remark.} No-filter model was considered in \cite{qst24}, where adapted algorithms were derived and implemented. Yield values for the approximated optimal controls for the no-filter case mentioned in this section were calculated based on the results of \cite{qst24}.

\subsection{Prism Case 1.}
In this subsection, we choose a prism $V_1$ with positive bounds $m_i=3~\mathrm{\mu T}$ and $M_i=6~\mathrm{\mu T}$ and consider a control set
\begin{equation*}
\mathcal{V}_1 := \left\{ 
\begin{array}{c}
\mathbf{u} \in L^3_2(0,T;\mathbb{R}^3) ~ : \\
\mathbf{u}(t) \in V_1 = \prod_{i=x,y,z} [3,6] , \\
\mbox{a.e.} ~ t \in [0,T] 
\end{array}
\right\} .
\end{equation*}

Next, we present numerical results for the GPM and IPMP algorithms on $p$-proton model cases with $p \leq 7$. Our numerical analysis focuses on the convergence of both methods to a unique optimal solution, robustness of the methods with respect to the selection of the initial iteration, comparison with the no-filter model, effect of the free filtering parameters $\gamma > 0$ (in $\mathrm{MHz}$) and $\mathbf{v}_0\in \mathbb{R}^3$ (in $\mathrm{mT}$).  

\begin{figure}[h!]
\centering
\includegraphics[width=.45\textwidth]{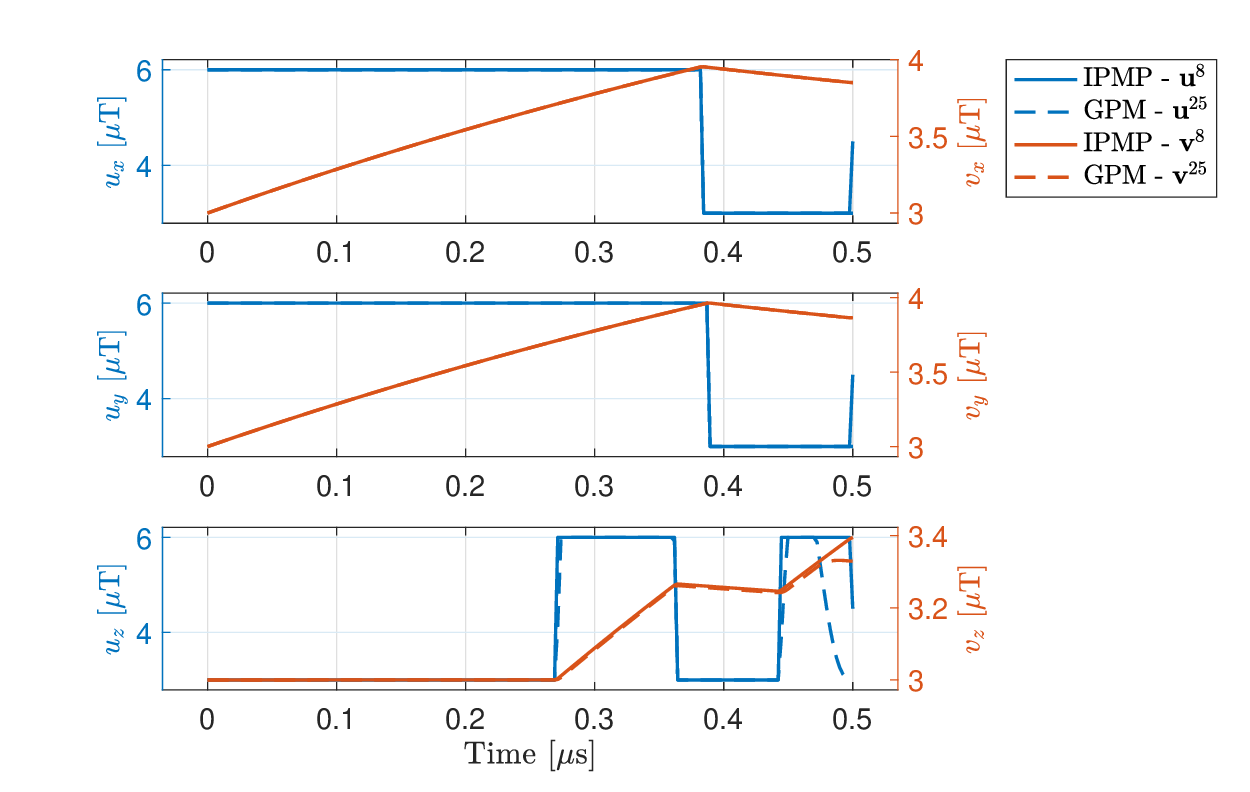}
\vskip-.1in
\caption{Approximated control (blue) and magnetic field (red) coordinates obtained for the IPMP (solid) and GPM (dashed) methods at the last iteration for the 1-proton case, and initial control $\mathbf{u}^0(t)=(3,3,3)~\mathrm{\mu T}$. Filtering parameters are $\mathbf{v}_0 = (3,3,3)~\mathrm{\mu T}$ and $\gamma = 1~\mathrm{MHz}$. The number of iterations was 25 for the GPM algorithm and 8 for the IPMP algorithm.}
\label{fig:1P_01}
\end{figure}

\vskip.1in\noindent\textbf{1-Proton Case.}
We start by comparing the approximate solutions of IPMP and GPM. Filtering parameters are set to $\mathbf{v}_0=(3,3,3)~\mathrm{\mu T}$ and $\gamma=1~\mathrm{MHz}$. Figure \ref{fig:1P_01} shows the calculated optimal control (left) and magnetic field (right) for both IPMP and GPM algorithms with an initial constant-in-time control vector being selected as $\mathbf{u}^0(t)=(3,3,3)~\mathrm{\mu T}$. The GPM solution was obtained after 25 iterations, whereas the IPMP method converged only in 8 iterations.

\begin{figure}[h]
\centering
\includegraphics[width=.4\textwidth]{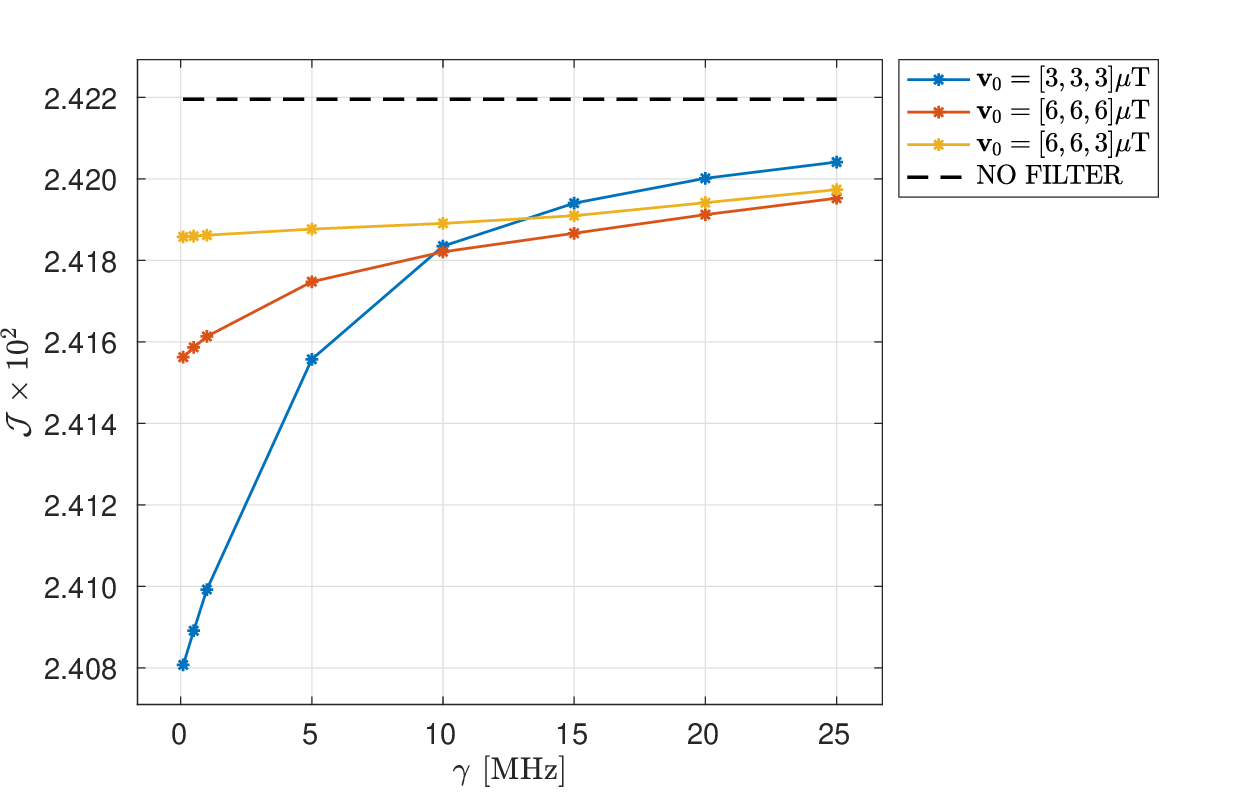}
\caption{1-proton case. Asymptotic of $\mathcal{J}$ for a sample of increasing values of $\gamma$ in $\mathrm{MHz}$. The dashed line corresponds to the cost value of the approximated optimal control for the no-filter model. Minimum and maximum yield loss (in $\%$) for each case of $\mathbf{v}_0$ is shown in Table \ref{tb1}.}
\label{fig:1P_02}
\end{figure}

Next, we compare the results with the no-filter case. Figure \ref{fig:1P_02} demonstrates the dependence of the scaled maximum value of the cost functional $\mathcal{J}$ on a sample of increasing filter parameter values $\gamma$;  here, we set $\mathbf{u}^0(t)=(3,3,3)\mathrm{\mu T}$ and use the IPMP algorithm only. The dashed line in Figure \ref{fig:1P_02} corresponds to the scaled maximum value of the cost functional in the no-filter model, i.e., the scaled maximum value of the quantum singlet yield.
Colored graphs correspond to various filtered cases with different choices of the free filtering parameter $\mathbf{v}_0$. 

\begin{figure*}[t]
\centering
\begin{subfigure}[b]{.495\textwidth}
\centering
\includegraphics[width=.7\linewidth]{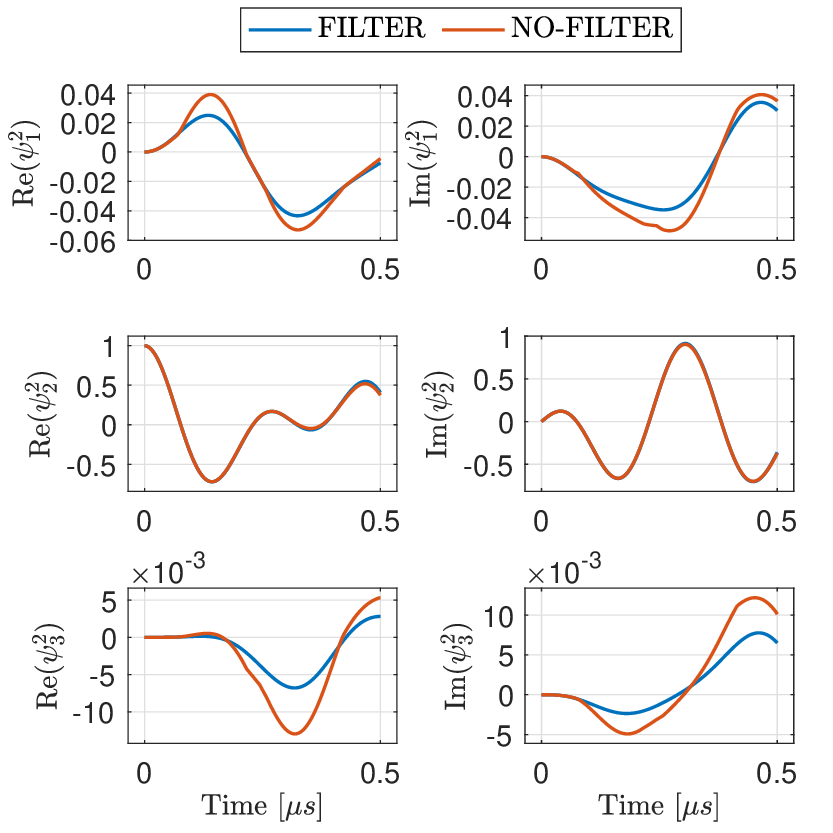}
\caption{} \label{fig:AP_V1_1P_a}
\end{subfigure}
\begin{subfigure}[b]{.495\textwidth}
\centering
\includegraphics[width=.7\linewidth]{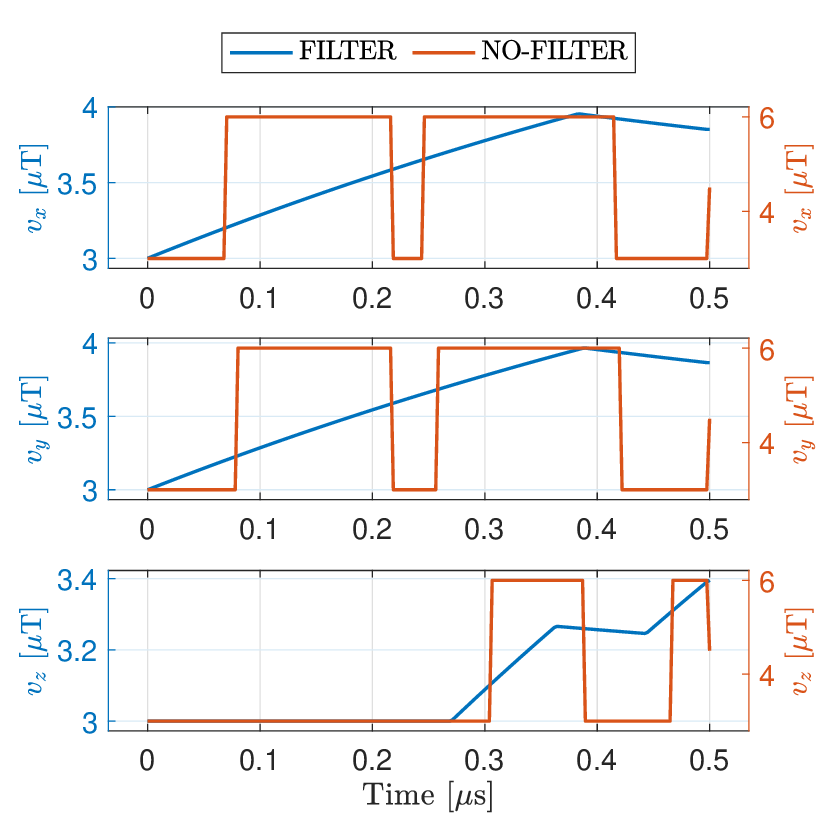}
\caption{} \label{fig:AP_V1_1P_b}
\end{subfigure}
\caption{1-Proton Case. (a) Real (left) and imaginary (right) parts of the coordinates 1, 2, and 3 of the wave function (state) $\psi^2$ obtained for the FILTER (blue) and NO-FILTER (red) cases. Filter parameters are $\gamma=1~\mathrm{MHz}$, $\mathbf{v}_0=(3,3,3)~\mathrm{\mu T}$ and $\mathbf{u}_0(t)=(3,3,3)~\mathrm{\mu T}$. (b) Corresponding optimized magnetic field for the FILTER (blue) and NO-FILTER (red) cases.}
\label{fig:AP_V1_1P}
\end{figure*}

\begin{table*}[t]
\caption{\label{tb1} Minimum and Maximum yield loss (in $\%$), concerning $\gamma$, relative to the yield of the corresponding approximated optimal control for the no-filter model, for different initial controls $\mathbf{v}_0$ and each case of number of protons.}
\centering
\begin{tabular}{lcccccc}
\hline
$\mathbf{u}_0$ & \multicolumn{2}{c}{$[3,3,3]\mbox{$\mu$T}$} & \multicolumn{2}{c}{$[6,6,6]\mbox{$\mu$T}$} & \multicolumn{2}{c}{$[6,6,3]\mbox{$\mu$T}$} \\
Proton Case & min & max & min & max & min & max \\ \hline 
1 & 0.0636 & 0.5732 & 0.1003 & 0.2614 & 0.0915 & 0.1394 \\
2 & 0.3344 & 2.4134 & 0.0197 & 0.0312 & 0.0283 & 0.0519 \\
3 & 0.1745 & 1.1543 & 0.0061 & 0.1383 & -0.0009 & 0.0000 \\
4 & 0.2096 & 1.3905 & 0.0153 & 0.0267 & 0.0195 & 0.0266 \\
5 & 0.1762 & 1.1903 & 0.0039 & 0.0455 & 0.0045 & 0.0054 \\
6 & 0.1840 & 1.2097 & 0.0118 & 0.0245 & 0.0154 & 0.0197 \\
7 & 0.1646 & 1.0945 & 0.0050 & 0.0261 & 0.0067 & 0.0087 \\ 
\hline
\end{tabular}
\end{table*}

First, we observe that as the filtering parameter $\gamma$ increases, the maximum of the cost functional asymptotically approaches the maximum in the no-filter case. To assess the effect of filtering, it is essential to note that all the graphs for the filtered cases are below the dashed line, meaning that the trade-off between the original non-filtered model with bang-bang optimal magnetic field and the filtered model with continuous in time optimal magnetic field is associated with some loss of the maximum singlet yield expressed as a maximum of the cost functional.

In Table \ref{tb1}, we collected the minimum and maximum yield loss by applying filtering with all selected values of parameters $\gamma$ and $\mathbf{v}_0$ shown in Figure \ref{fig:1P_02}, and with various selections of initial iteration $\mathbf{u}_0$. Remarkably, in all cases, loss of the quantum singlet yield is less than 1\%. This is a very strong argument on behalf of the filtered model, with the gain of regularity of the optimal electromagnetic field input.  

Numerical results in Figure \ref{fig:1P_02} also demonstrate the role and importance of the selection of the filtering parameter $\mathbf{v}_0$: minimal yield loss is achieved if the initial value $\mathbf{v}_0$ of the magnetic field in all filtered cases is selected as a value of the bang-bang optimal magnetic field at initial moment $t=0$ in non-filtered model. 

Figure \ref{fig:AP_V1_1P_a} shows the real and imaginary parts of the waves $\psi^l_k$, for the state $l=2$ (out of 6) and coordinates $k=1,2,3$ (out of 8) obtained using the IPMP algorithm for the filter (blue) and no-filter (red) models. Corresponding optimized magnetic field coordinates are displayed in Figure \ref{fig:AP_V1_1P_b}. It is demonstrated here that, despite the enhanced regularity and simplicity of the optimal magnetic field in the filter model, coherent oscillations of the waves determined by the Schr\"odinger system in both models are preserved and nearly identical. \\

\begin{figure*}[t]
\centering
\begin{subfigure}[b]{.45\textwidth}
\centering
\includegraphics[width=\textwidth]{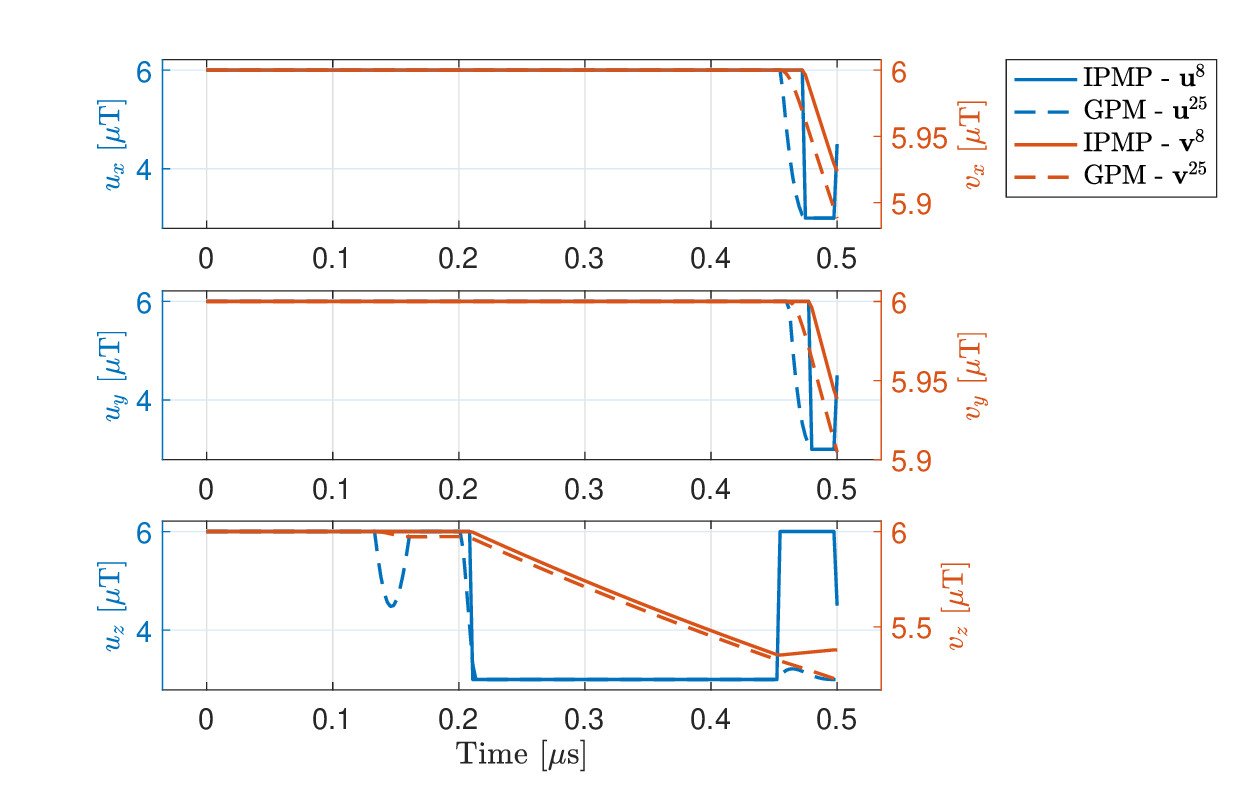}
\caption{} \label{fig:2P_F01_a}
\end{subfigure}
\hfill
\begin{subfigure}[b]{.45\textwidth}
\centering
\includegraphics[width=\textwidth]{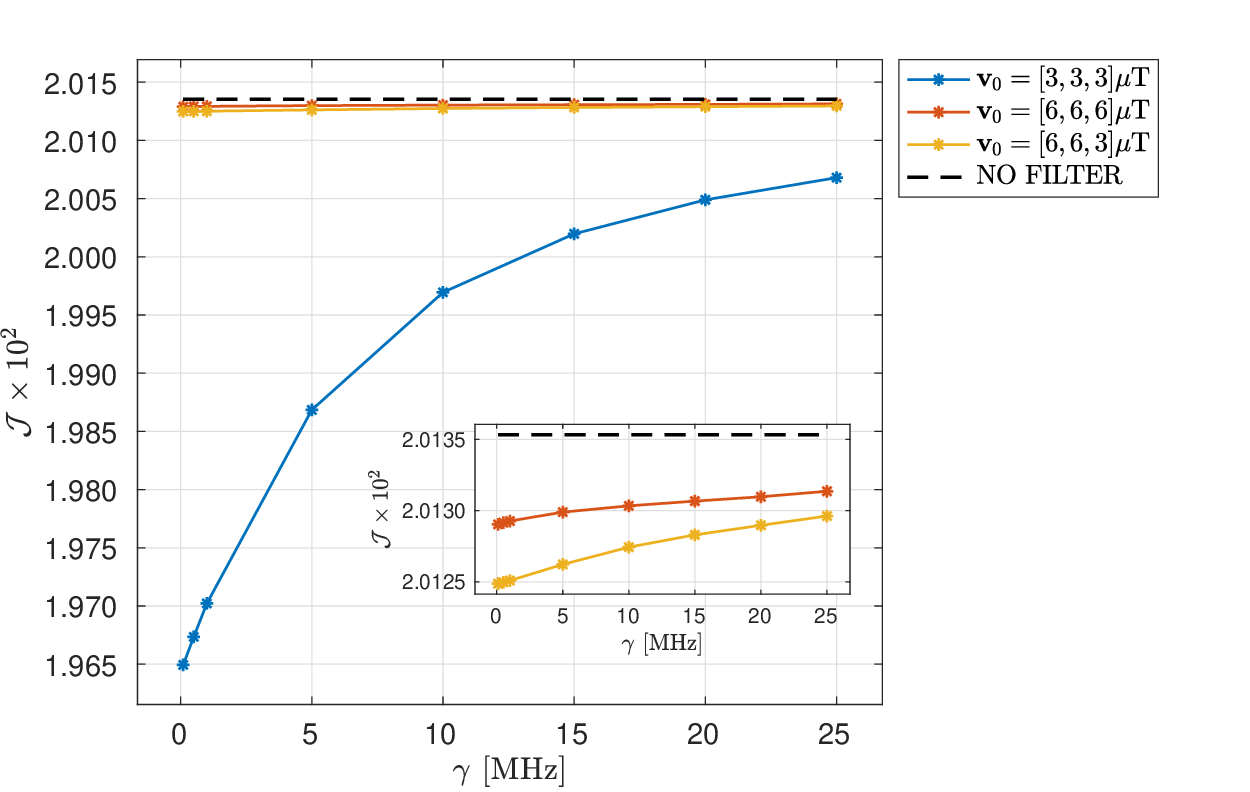}
\caption{} \label{fig:2P_F01_b}
\end{subfigure}
\caption{2-Proton Case. (a) Control (blue) and magnetic field (red) coordinates obtained for the IPMP (solid) and GPM (dashed) algorithms at the last iteration. Parameters are $\mathbf{u}^0=(3,3,3)~\mathrm{\mu T}$, $\mathbf{v}_0=(6,6,6)~\mathrm{\mu T}$ and $\gamma=1~\mathrm{MHz}$. (b) Asymptotics of $\mathcal{J}$ for a sample of increasing values of $\gamma$ in $\mathrm{MHz}$. The dashed black line corresponds to the optimized cost value of the approximated optimal control for the non-filtered model.}
\label{fig:2P_F01}
\end{figure*}

\begin{figure*}[t]
\centering
\begin{subfigure}[b]{.45\textwidth}
\centering
\includegraphics[width=.7\linewidth]{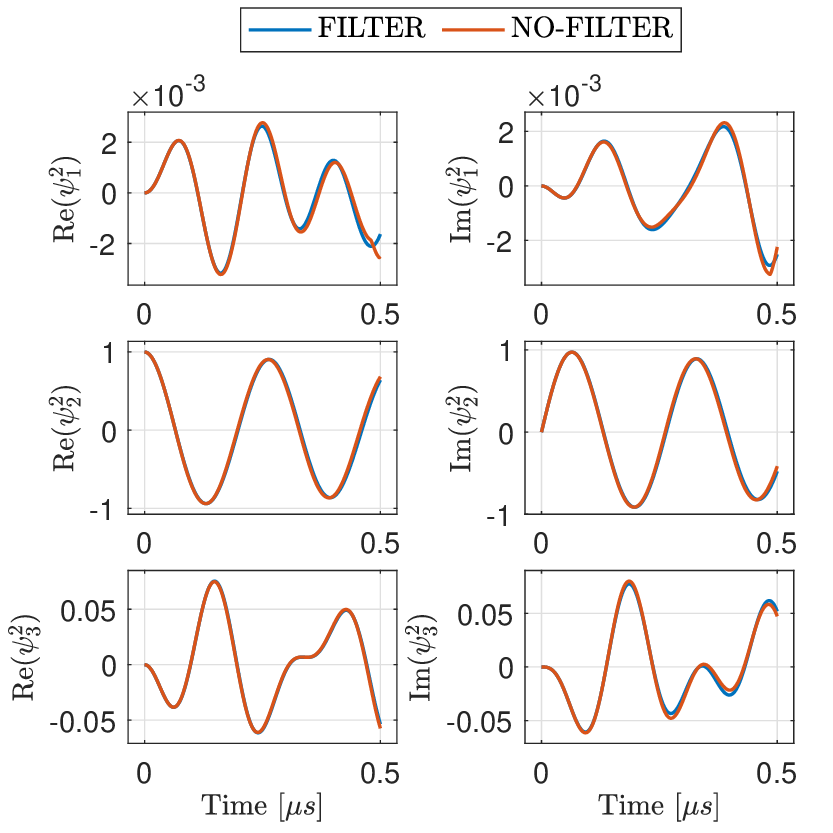}
\caption{} \label{fig:2P_F02_a}
\end{subfigure}
\begin{subfigure}[b]{.45\textwidth}
\centering
\includegraphics[width=.7\linewidth]{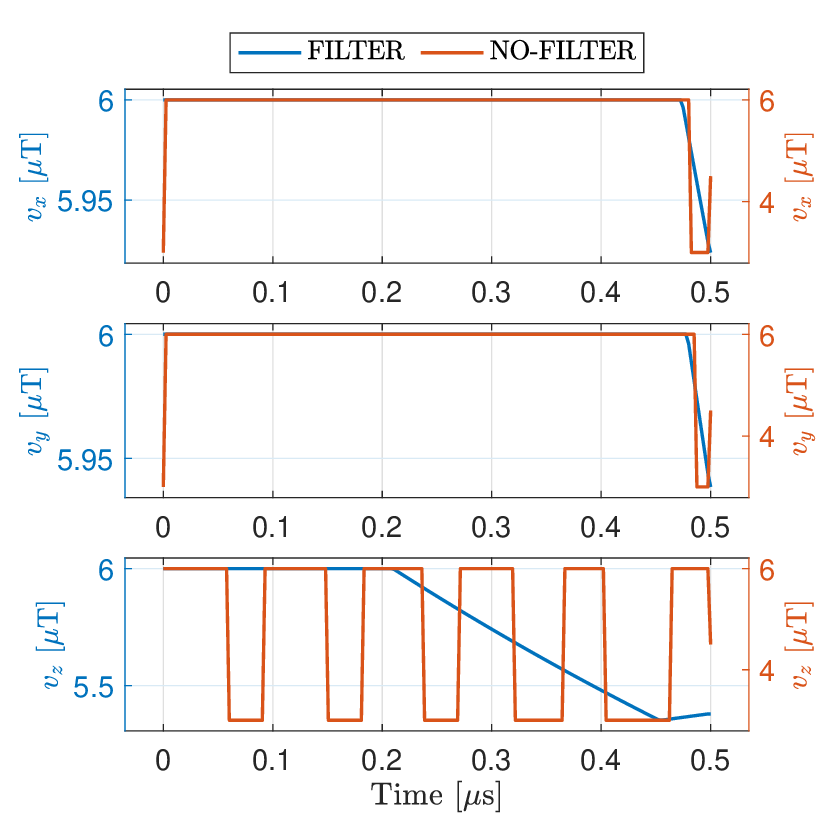}
\caption{} \label{fig:2P_F02_b}
\end{subfigure}
\caption{2-Proton Case. (a) Real (left) and imaginary (right) parts of the coordinates 1, 2, and 3 of the wave function (state) $\psi^2$ obtained for the filtered (blue) and non-filtered (red) cases. Filter parameters are $\gamma=1~\mathrm{MHz}$, $\mathbf{v}_0=(6,6,6)~\mathrm{\mu T}$ and $\mathbf{u}_0(t)=(3,3,3)~\mathrm{\mu T}$. (b) Corresponding optimized magnetic field for the filtered (blue) and non-filtered (red) cases.}
\label{fig:2P_F02}
\end{figure*}

\begin{figure*}[t]
\centering
\begin{subfigure}[b]{.45\textwidth}
\centering
\includegraphics[width=\textwidth]{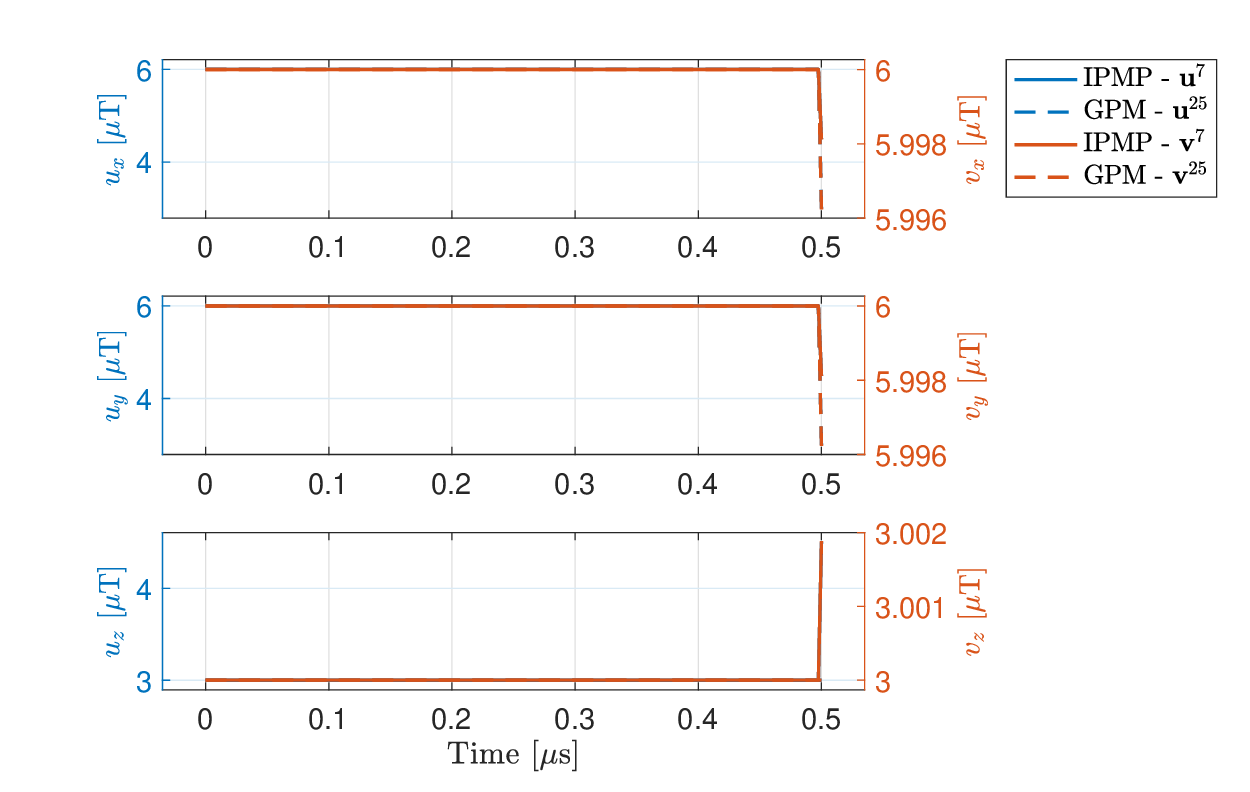}
\caption{} \label{fig:3P_F01_a}
\end{subfigure}
\hfill
\begin{subfigure}[b]{.45\textwidth}
\centering
\includegraphics[width=\textwidth]{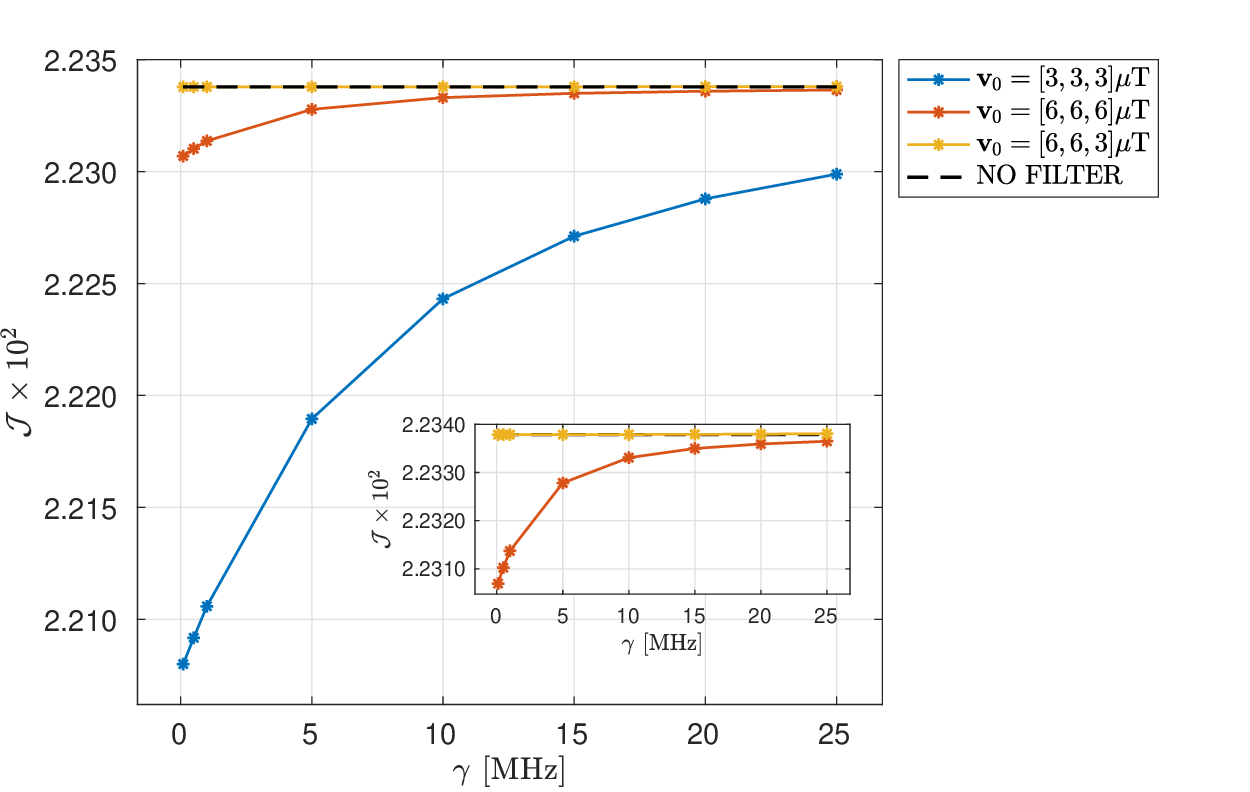}
\caption{} \label{fig:3P_F01_b}
\end{subfigure}
\caption{3-Proton Case. (a) Control (blue) and magnetic field (red) coordinates obtained for the IPMP (solid) and GPM (dashed) algorithms at the last iteration. Parameters are $\mathbf{u}^0=(3,3,3)~\mathrm{\mu T}$, $\mathbf{v}_0=(6,6,3)~\mathrm{\mu T}$ and $\gamma=1~\mathrm{MHz}$. (b) Asymptotics of $\mathcal{J}$ for a sample of increasing values of $\gamma$ in $\mathrm{MHz}$. The dashed black line corresponds to the optimized cost value of the approximated optimal control for the non-filtered model.}
\label{fig:3P_F01}
\end{figure*}

\begin{figure*}[t]
\centering
\begin{subfigure}[b]{.45\textwidth}
\centering
\includegraphics[width=.7\linewidth]{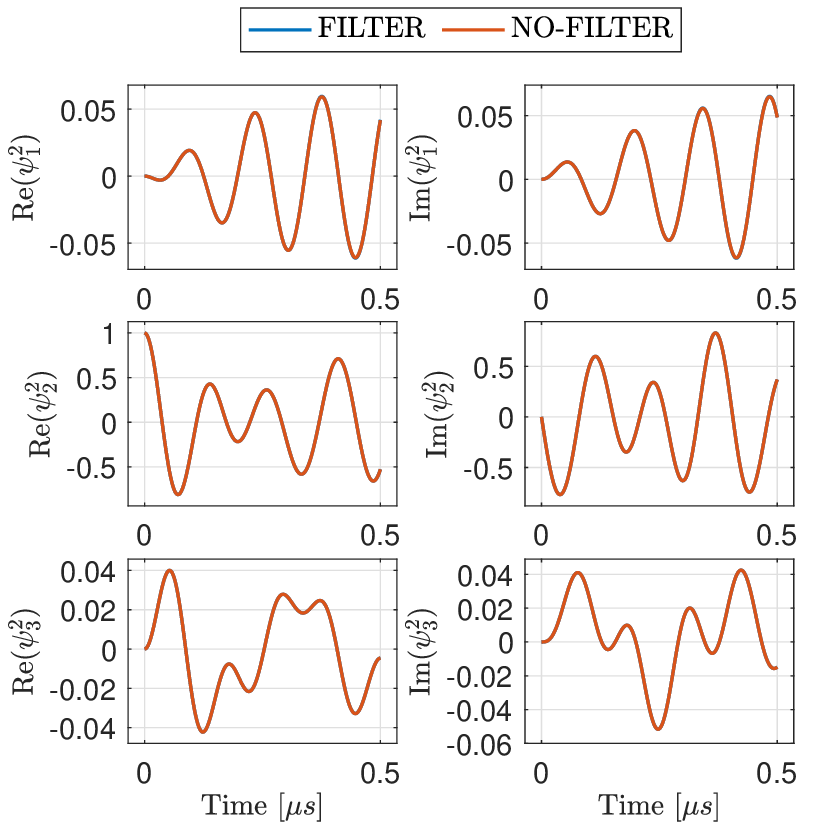}
\caption{} \label{fig:3P_F02_a}
\end{subfigure}
\begin{subfigure}[b]{.45\textwidth}
\centering
\includegraphics[width=.7\linewidth]{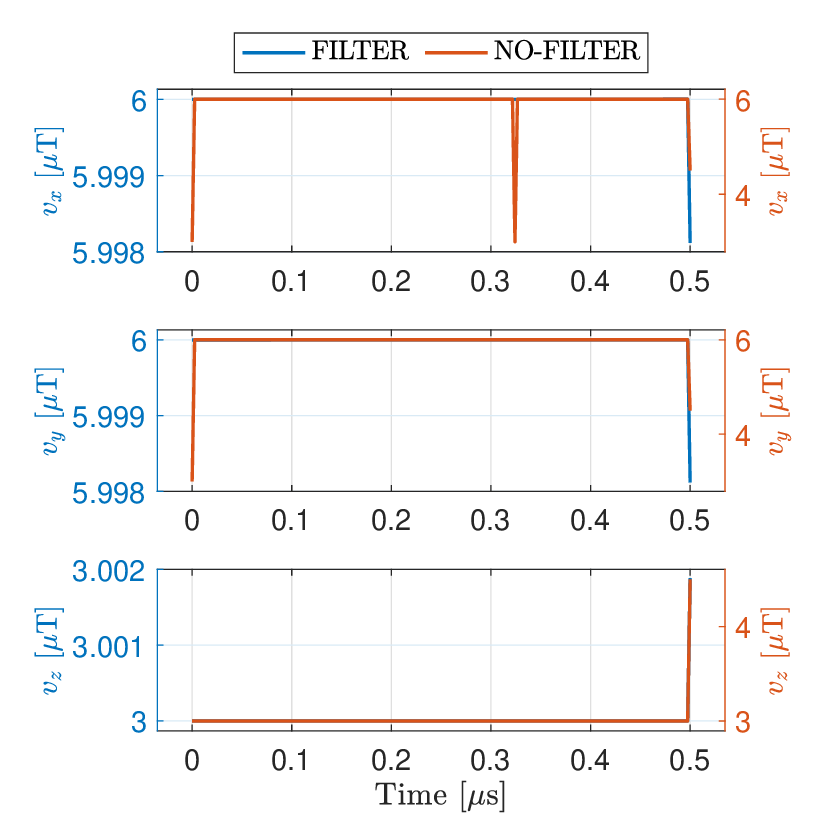}
\caption{} \label{fig:3P_F02_b}
\end{subfigure}
\caption{3-Proton Case. (a) Real (left) and imaginary (right) parts of the coordinates 1, 2, and 3 of the wave function (state) $\psi^2$ obtained for the FILTER (blue) and NO-FILTER (red) cases. Filter parameters are $\gamma=1~\mathrm{MHz}$, $\mathbf{v}_0=(6,6,3)~\mathrm{\mu T}$ and $\mathbf{u}_0(t)=(3,3,3)~\mathrm{\mu T}$. (b) Corresponding optimized magnetic field for the filtered (blue) and non-filtered (red) cases.}
\label{fig:3P_F02}
\end{figure*}

\begin{figure*}[t]
\centering
\begin{subfigure}[b]{.45\textwidth}
\centering
\includegraphics[width=\textwidth]{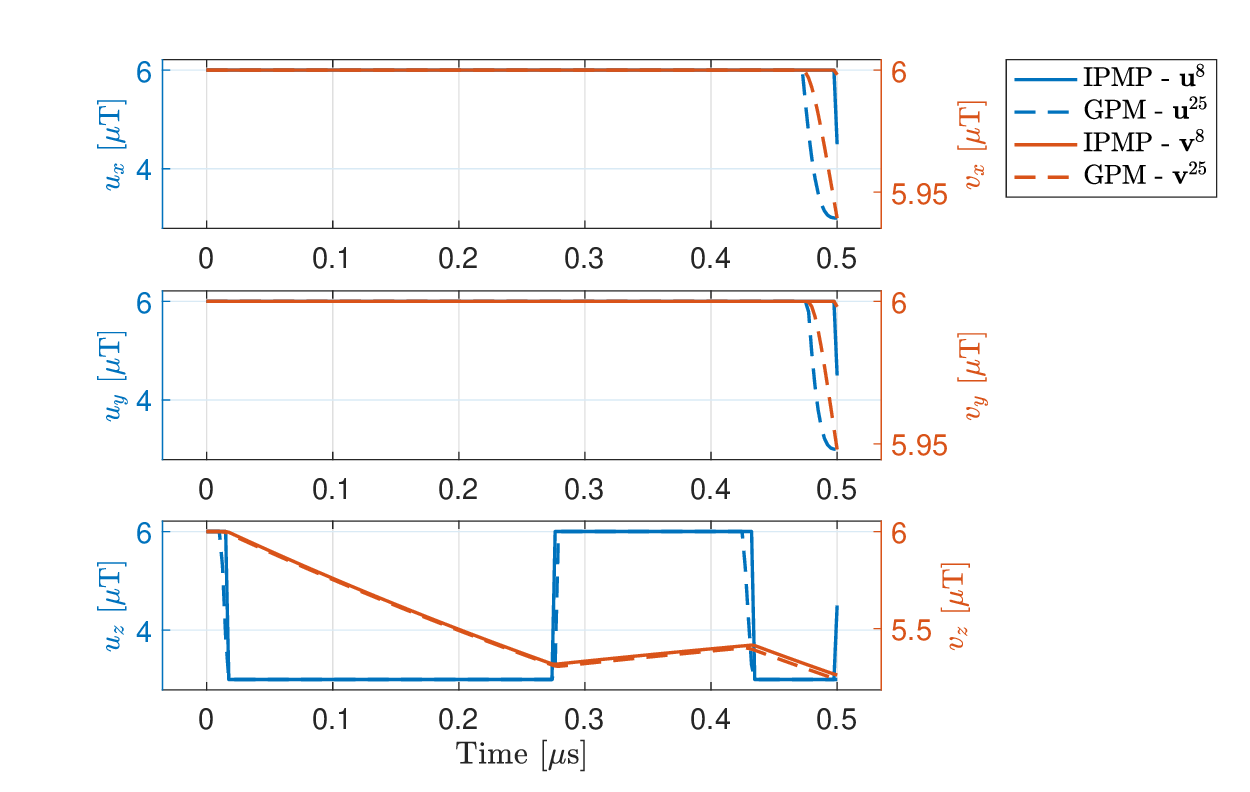}
\caption{} \label{fig:4P_F01_a}
\end{subfigure}
\hfill
\begin{subfigure}[b]{.45\textwidth}
\centering 
\includegraphics[width=\textwidth]{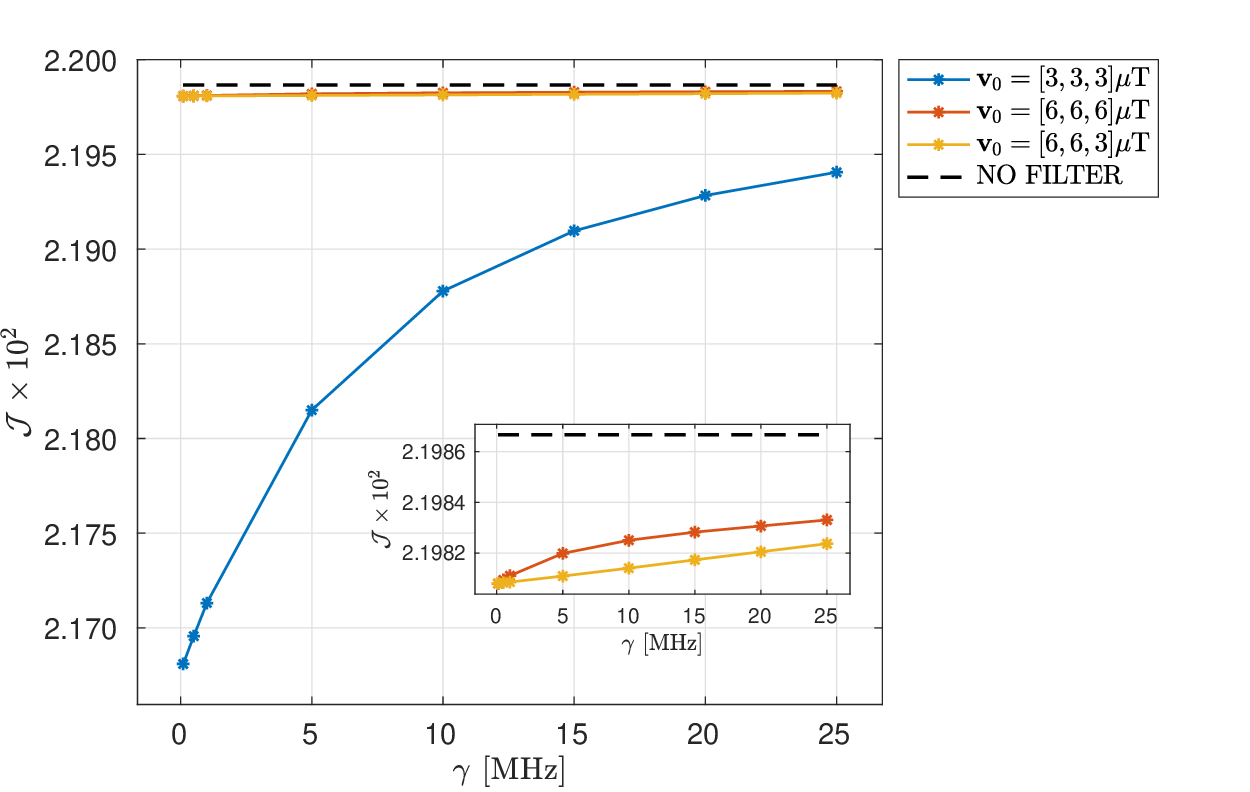}
\caption{} \label{fig:4P_F01_b}
\end{subfigure}
\caption{4-Proton Case. (a) Control (blue) and magnetic field (red) coordinates obtained for the IPMP (solid) and GPM (dashed) algorithms at the last iteration. Parameters are $\mathbf{u}^0=(3,3,3)~\mathrm{\mu T}$, $\mathbf{v}_0=(6,6,6)~\mathrm{\mu T}$ and $\gamma=1~\mathrm{MHz}$. (b) Asymptotics of $\mathcal{J}$ for a sample of increasing values of $\gamma$ in $\mathrm{MHz}$. The dashed black line corresponds to the optimized cost value of the approximated optimal control for the non-filtered model.}
\label{fig:4P_F01}
\end{figure*}

\begin{figure*}[t]
\centering
\begin{subfigure}[b]{.45\textwidth}
\centering
\includegraphics[width=0.7\linewidth, height=5cm]{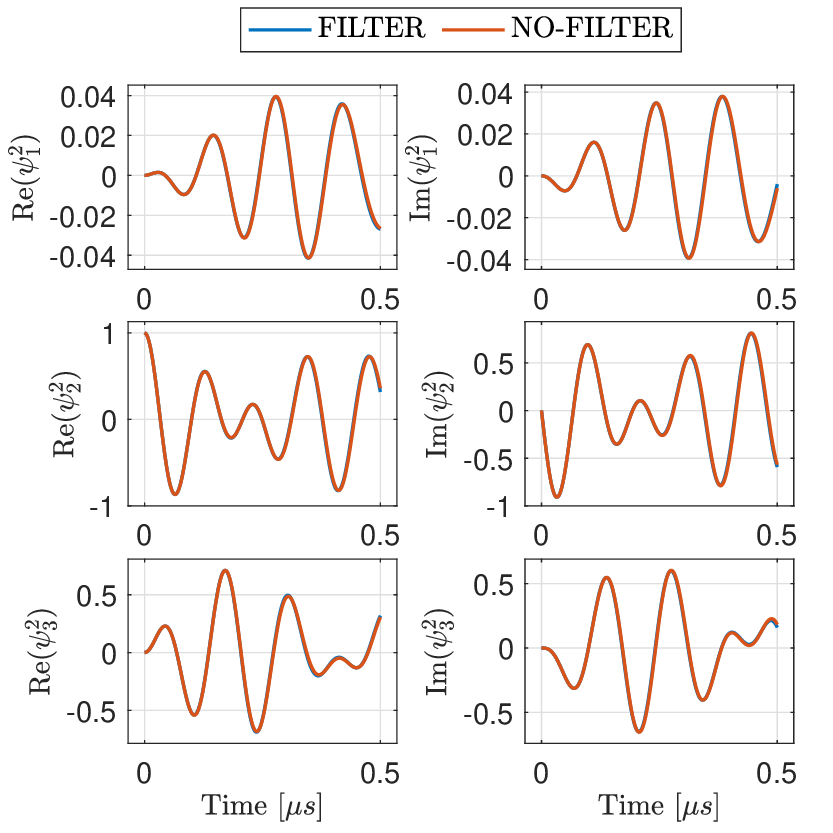}
\caption{} \label{fig:4P_F02_a}
\end{subfigure}
\begin{subfigure}[b]{.45\textwidth}
\centering
\includegraphics[width=0.7\linewidth, height=5cm]{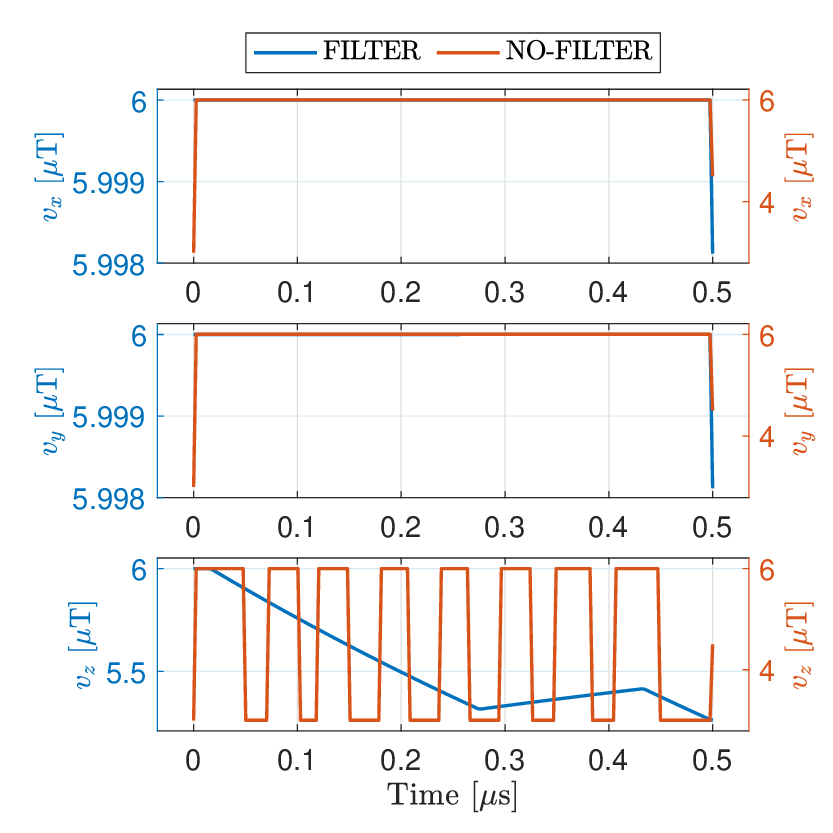}
\caption{} \label{fig:4P_F02_b}
\end{subfigure}
\caption{4-Proton Case. (a) Real (left) and imaginary (right) parts of the coordinates 1, 2, and 3 of the wave function (state) $\psi^2$ obtained for the filtered (blue) and non-filtered (red) cases. Filter parameters are $\gamma=1~\mathrm{MHz}$, $\mathbf{v}_0=(6,6,6)~\mathrm{\mu T}$ and $\mathbf{u}_0(t)=(3,3,3)~\mathrm{\mu T}$. (b) Corresponding optimized magnetic field for the filtered (blue) and non-filtered (red) cases.}
\label{fig:4P_F02}
\end{figure*}

\begin{figure*}[t]
\centering
\begin{subfigure}[b]{.45\textwidth}
\centering
\includegraphics[width=\textwidth]{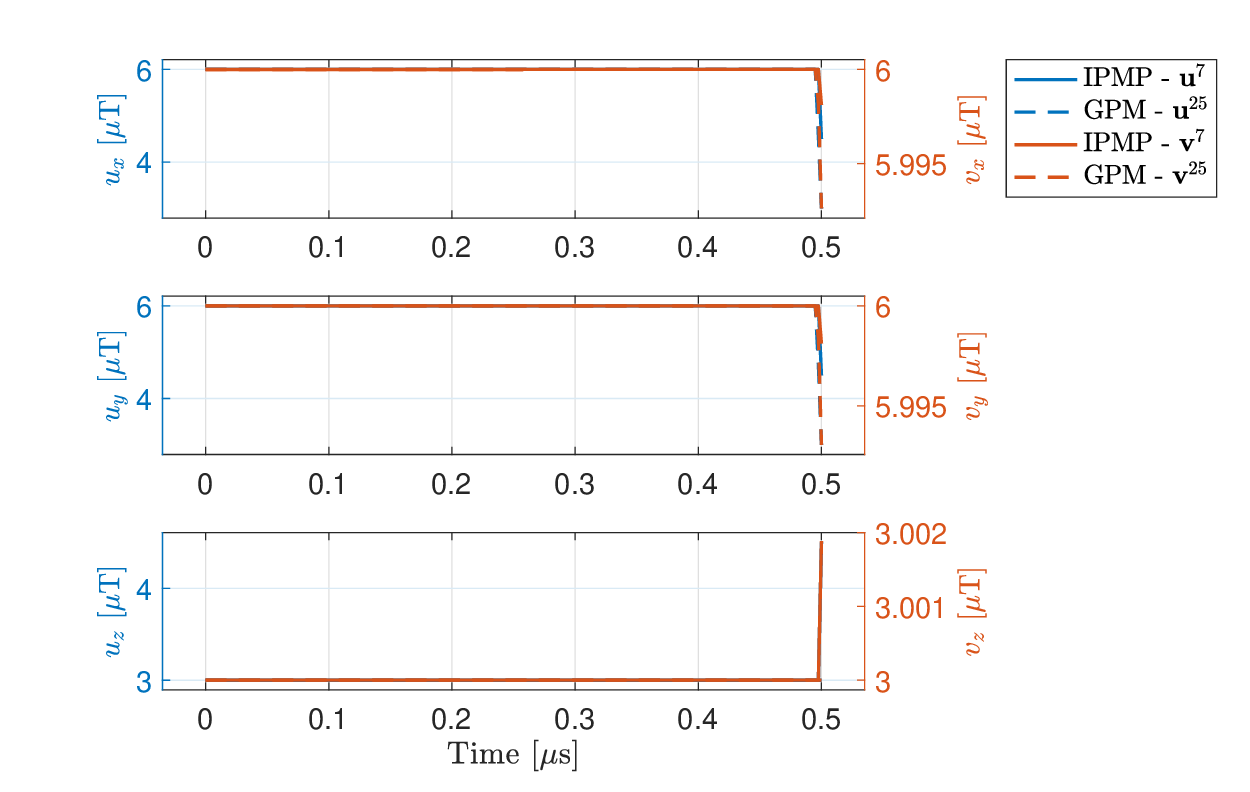}
\caption{} \label{fig:5P_F01_a}
\end{subfigure}
\hfill
\begin{subfigure}[b]{.45\textwidth}
\centering
\includegraphics[width=\textwidth]{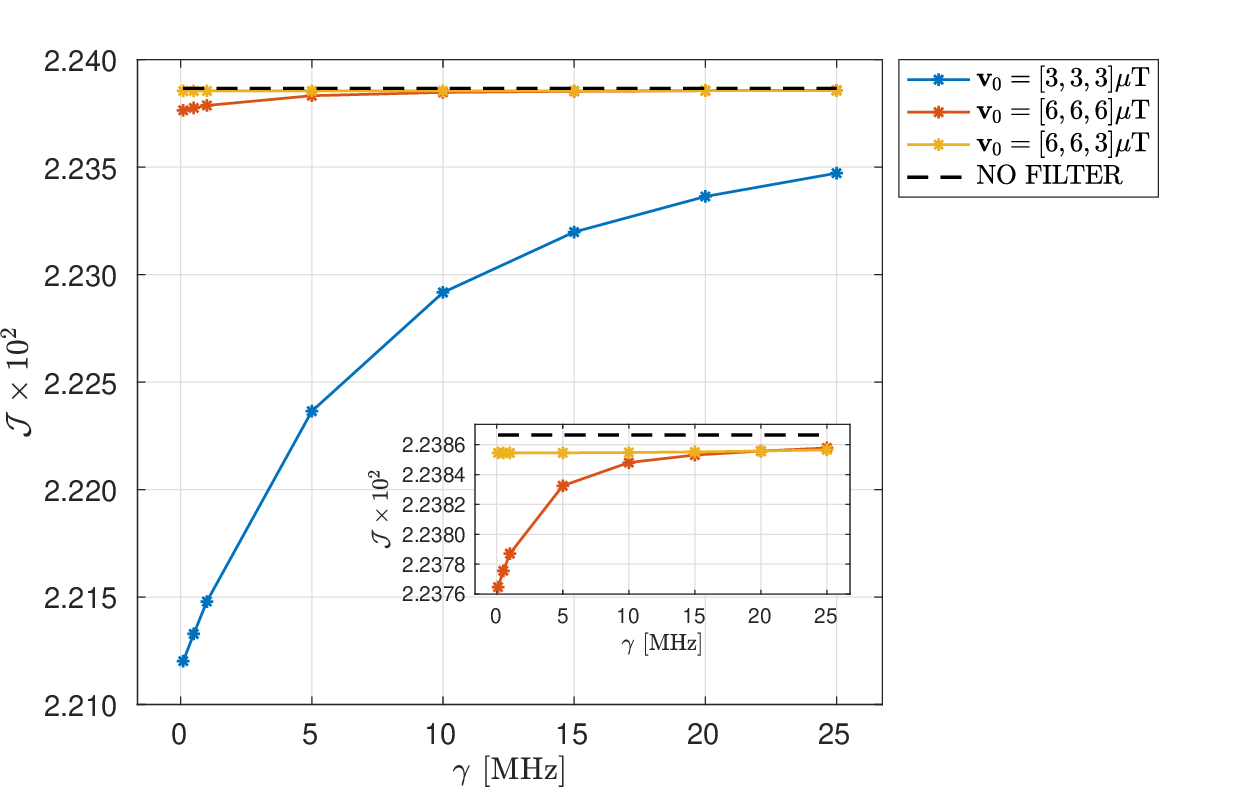}
\caption{} \label{fig:5P_F01_b}
\end{subfigure}
\caption{5-Proton Case. (a) Control (blue) and magnetic field (red) coordinates obtained for the IPMP (solid) and GPM (dashed) algorithms at the last iteration. Parameters are $\mathbf{u}^0=(3,3,3)~\mathrm{\mu T}$, $\mathbf{v}_0=(6,6,3)~\mathrm{\mu T}$ and $\gamma=1~\mathrm{MHz}$. (b) Asymptotics of $\mathcal{J}$ for a sample of increasing values of $\gamma$ in $\mathrm{MHz}$. The dashed black line corresponds to the optimized cost value of the approximated optimal control for the non-filtered model.}
\label{fig:5P_F01}
\end{figure*}

\begin{figure*}[t]
\centering
\begin{subfigure}[b]{.45\textwidth}
\centering
\includegraphics[width=0.7\linewidth, height=5cm]{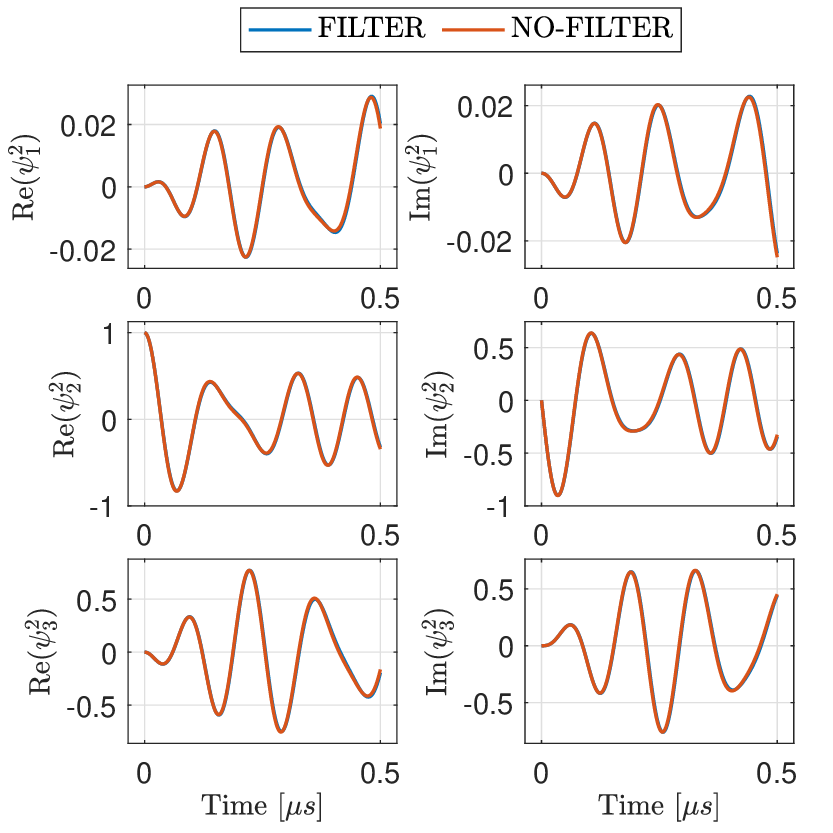}
\caption{} \label{fig:5P_F02_a}
\end{subfigure}
\begin{subfigure}[b]{.45\textwidth}
\centering
\includegraphics[width=0.7\linewidth, height=5cm]{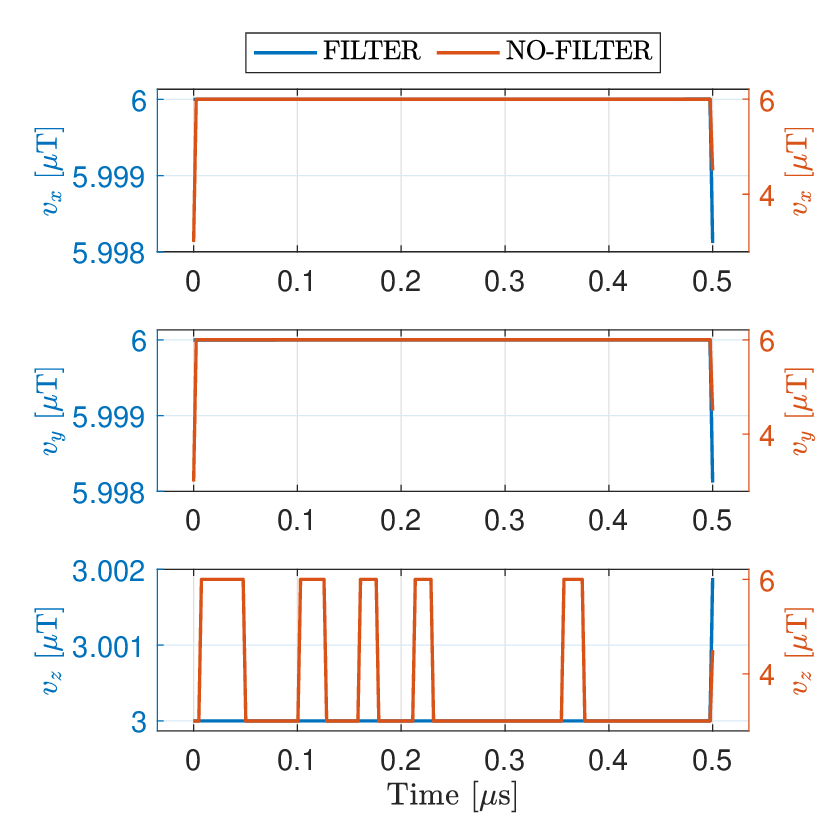}
\caption{} \label{fig:5P_F02_b}
\end{subfigure}
\caption{5-Proton Case. (a) Real (left) and imaginary (right) parts of the coordinates 1, 2, and 3 of the wave function (state) $\psi^2$ obtained for the filtered (blue) and non-filtered (red) cases. Filter parameters are $\gamma=1~\mathrm{MHz}$, $\mathbf{v}_0=(6,6,3)~\mathrm{\mu T}$ and $\mathbf{u}_0(t)=(3,3,3)~\mathrm{\mu T}$. (b) Corresponding optimized magnetic field for the filtered (blue) and non-filtered (red) cases.}
\label{fig:5P_F02}
\end{figure*}

\begin{figure*}[t]
\centering
\begin{subfigure}[b]{.45\textwidth}
\centering
\includegraphics[width=\textwidth]{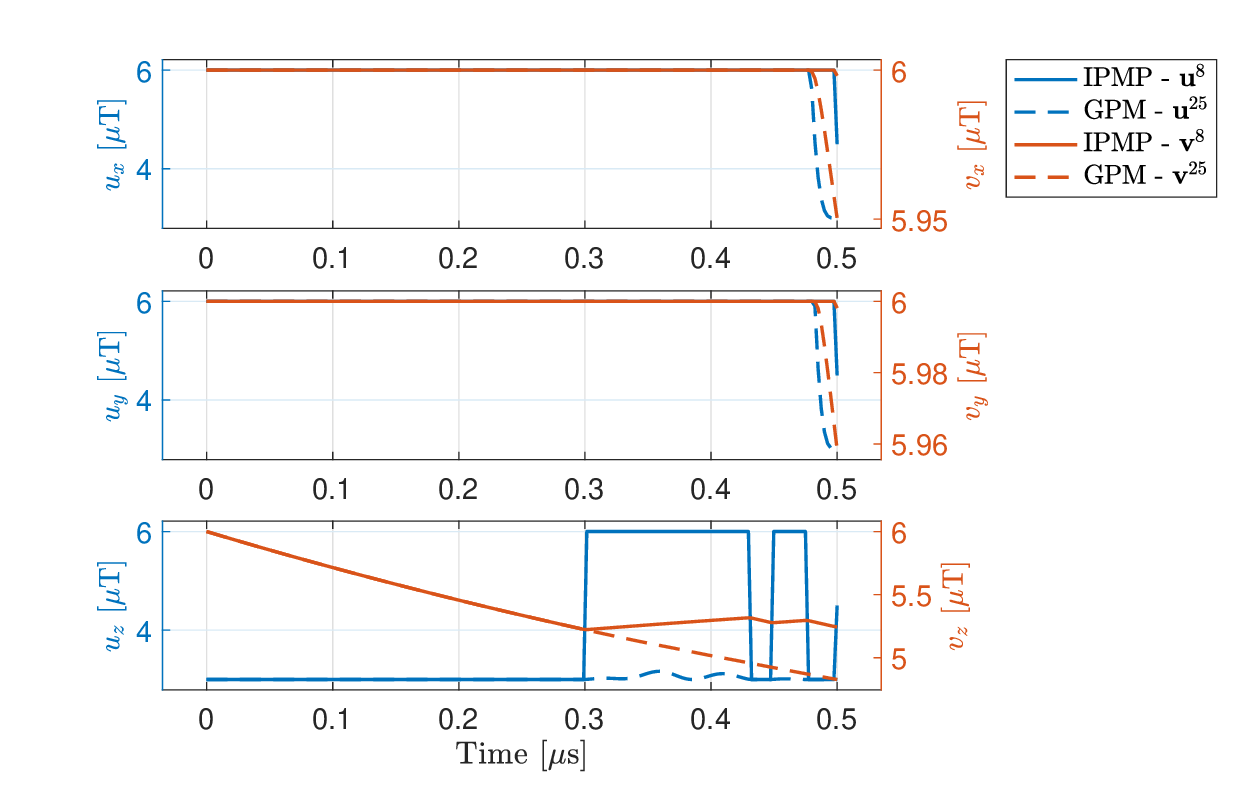}
\caption{} \label{fig:6P_F01_a}
\end{subfigure}
\hfill
\begin{subfigure}[b]{.45\textwidth}
\centering
\includegraphics[width=\textwidth]{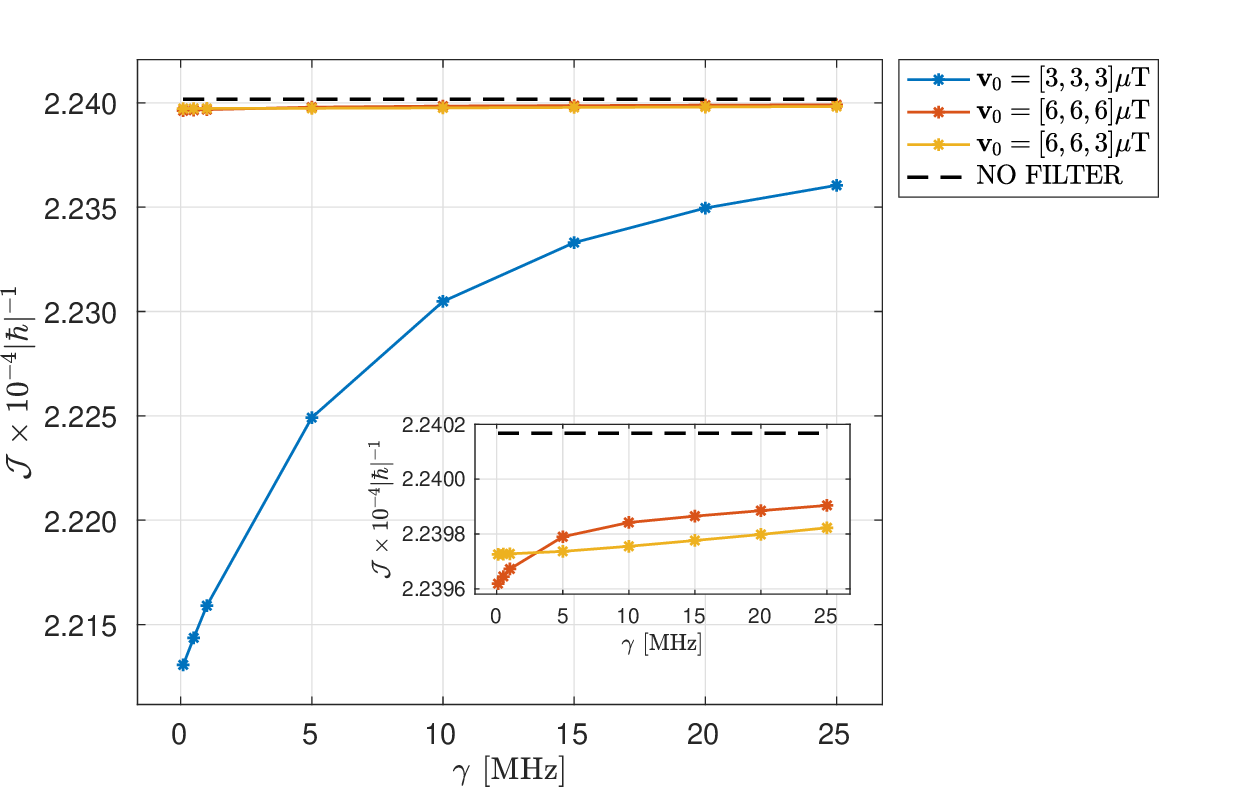}
\caption{} \label{fig:6P_F01_b}
\end{subfigure}
\caption{6-Proton Case. (a) Control (blue) and magnetic field (red) coordinates obtained for the IPMP (solid) and GPM (dashed) algorithms at the last iteration. Parameters are $\mathbf{u}^0=(3,3,3)~\mathrm{\mu T}$, $\mathbf{v}_0=(6,6,6)~\mathrm{\mu T}$ and $\gamma=1~\mathrm{MHz}$. (b) Asymptotics of $\mathcal{J}$ for a sample of increasing values of $\gamma$ in $\mathrm{MHz}$. The dashed black line corresponds to the optimized cost value of the approximated optimal control for the non-filtered model.}
\label{fig:6P_F01}
\end{figure*}

\begin{figure*}[t]
\centering
\begin{subfigure}[b]{.45\textwidth}
\centering
\includegraphics[width=0.7\linewidth, height=5cm]{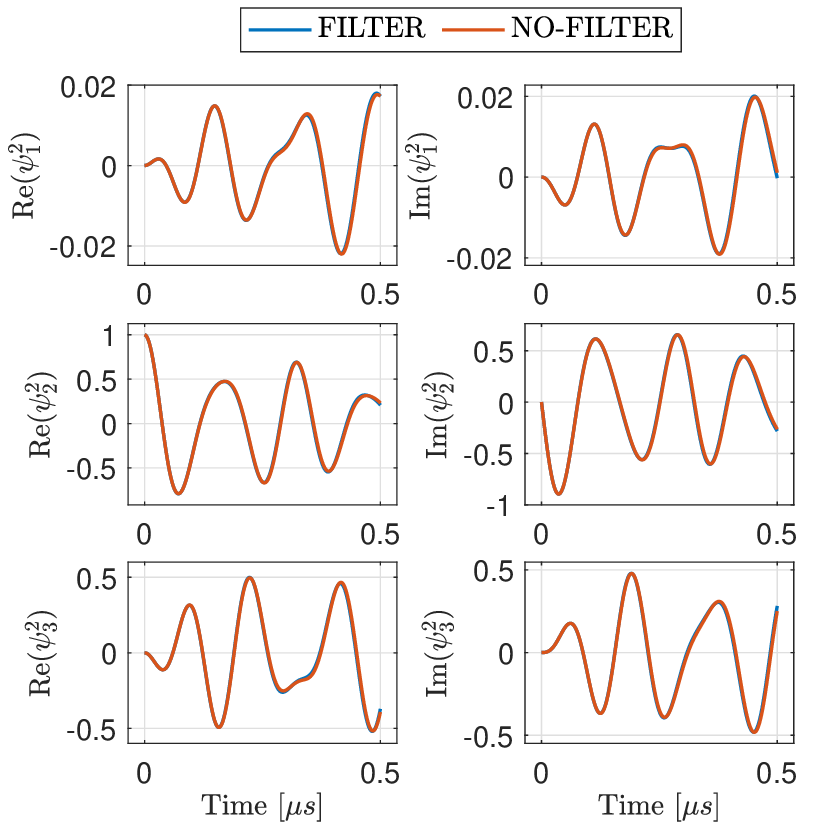}
\caption{} \label{fig:6P_F02_a}
\end{subfigure}
\begin{subfigure}[b]{.45\textwidth}
\centering
\includegraphics[width=0.7\linewidth, height=5cm]{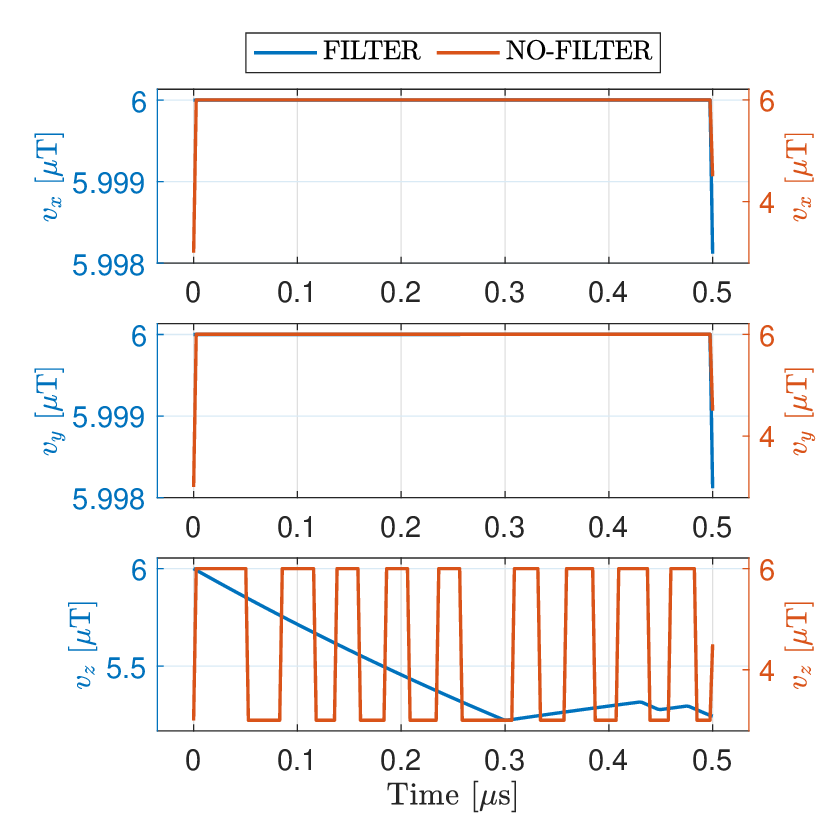}
\caption{} \label{fig:6P_F02_b}
\end{subfigure}
\caption{6-Proton Case. (a) Real (left) and imaginary (right) parts of the coordinates 1, 2, and 3 of the wave function (state) $\psi^2$ obtained for the FILTER (blue) and NO-FILTER (red) cases. Filter parameters are $\gamma=1~\mathrm{MHz}$, $\mathbf{v}_0=(6,6,6)~\mathrm{\mu T}$ and $\mathbf{u}_0(t)=(3,3,3)~\mathrm{\mu T}$. (b) Corresponding optimized magnetic field for the filtered (blue) and non-filtered (red) cases.}
\label{fig:6P_F02}
\end{figure*}

\begin{figure*}[t]
\centering
\begin{subfigure}[b]{.45\textwidth}
\centering
\includegraphics[width=\textwidth]{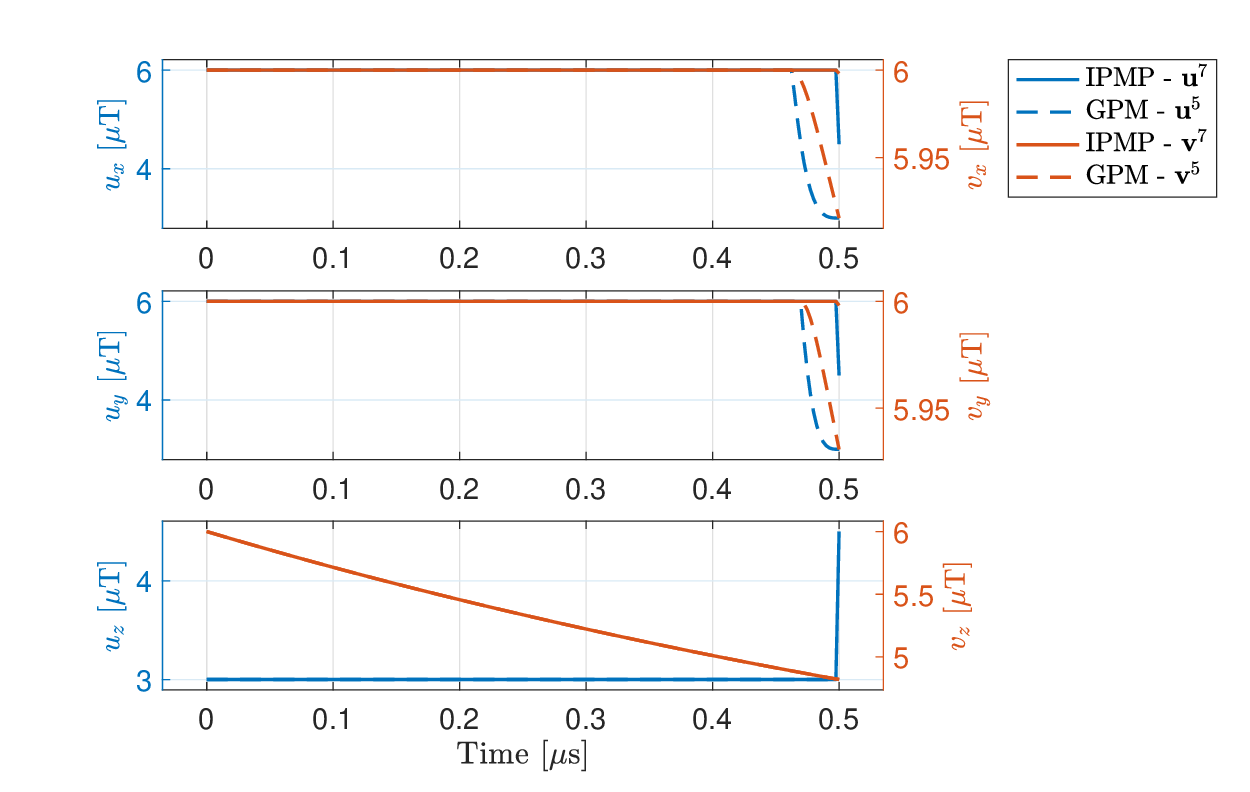}
\caption{} \label{fig:7P_F01_a}
\end{subfigure}
\hfill
\begin{subfigure}[b]{.45\textwidth}
\centering
\includegraphics[width=\textwidth]{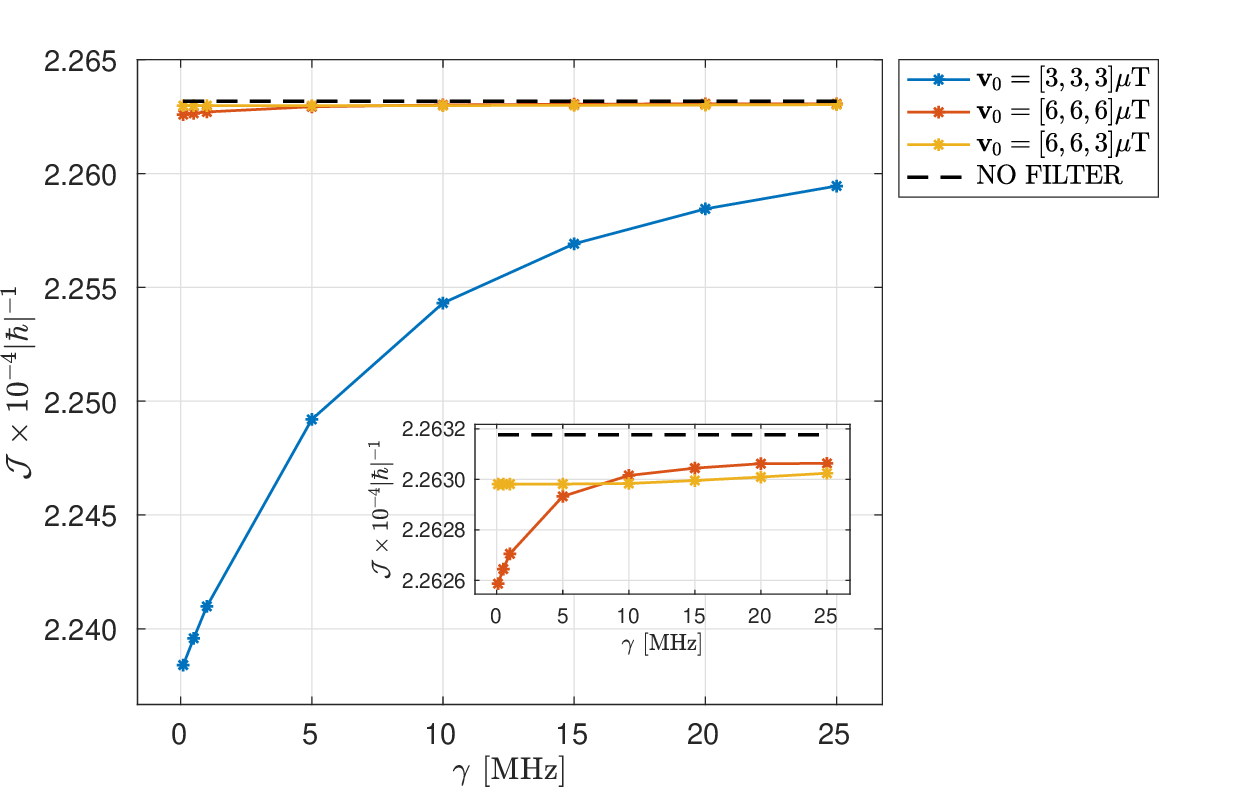}
\caption{} \label{fig:7P_F01_b}
\end{subfigure}
\caption{7-Proton Case. (a) Control (blue) and magnetic field (red) coordinates obtained for the IPMP (solid) and GPM (dashed) algorithms at the last iteration. Parameters are $\mathbf{u}^0=(3,3,3)~\mathrm{\mu T}$, $\mathbf{v}_0=(6,6,6)~\mathrm{\mu T}$ and $\gamma=1~\mathrm{MHz}$. (b) Asymptotics of $\mathcal{J}$ for a sample of increasing values of $\gamma$ in $\mathrm{MHz}$. The dashed black line corresponds to the optimized cost value of the approximated optimal control for the non-filtered model.}
\label{fig:7P_F01}
\end{figure*}

\begin{figure*}[t]
\centering
\begin{subfigure}[b]{.45\textwidth}
\centering
\includegraphics[width=0.7\linewidth, height=5cm]{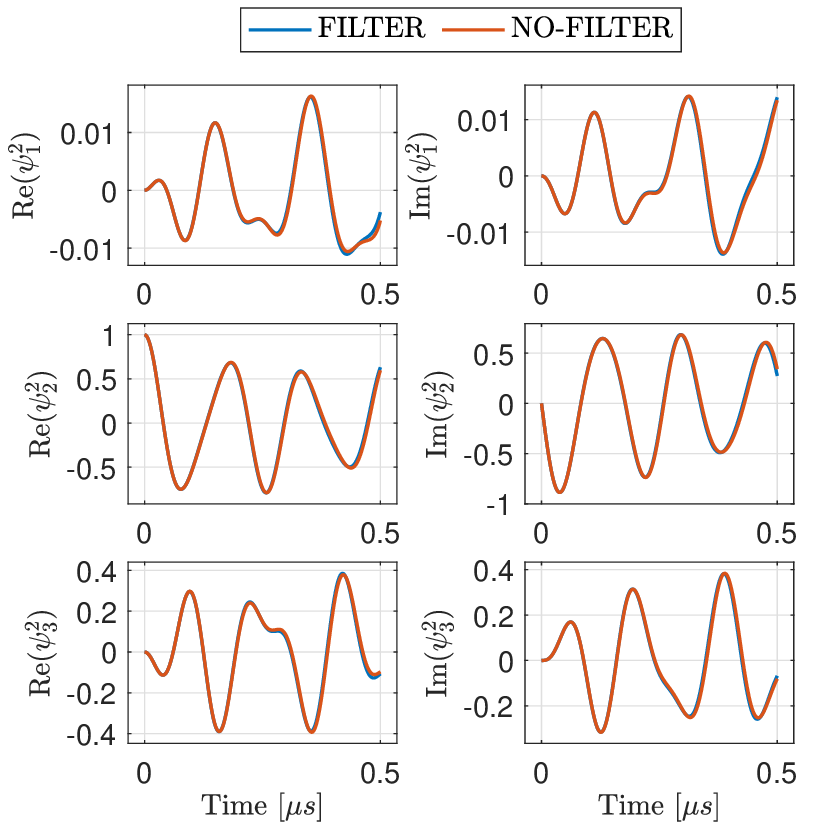}
\caption{} \label{fig:7P_F02_a}
\end{subfigure}
\begin{subfigure}[b]{.45\textwidth}
\centering
\includegraphics[width=0.7\linewidth, height=5cm]{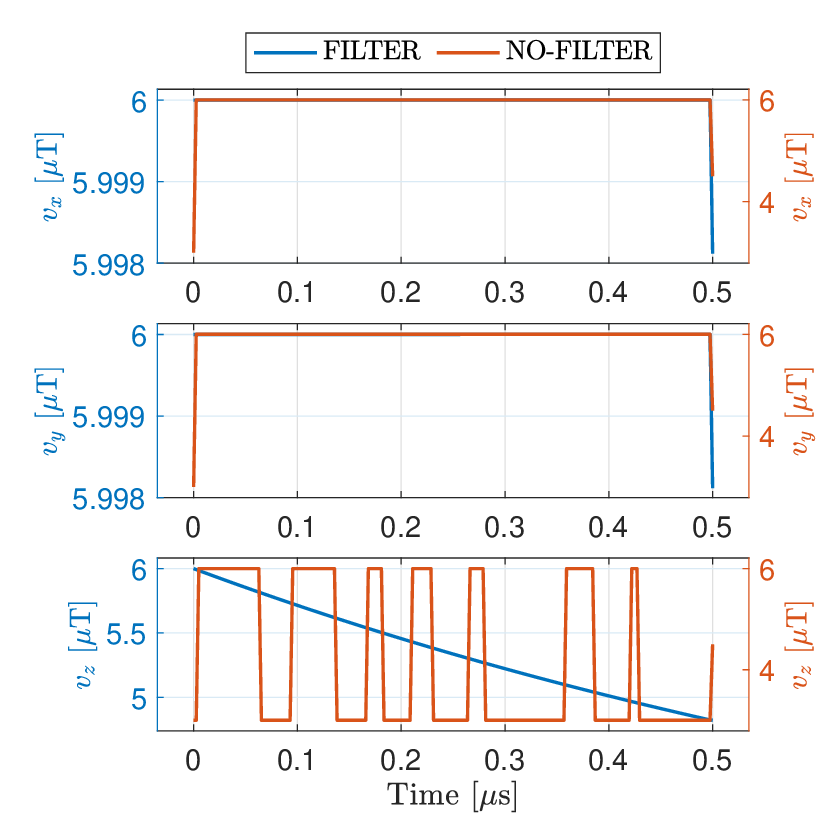}
\caption{} \label{fig:7P_F02_b}
\end{subfigure}
\caption{7-Proton Case. (a) Real (left) and imaginary (right) parts of the coordinates 1, 2, and 3 of the wave function (state) $\psi^2$ obtained for the filtered (blue) and non-filtered (red) cases. Filter parameters are $\gamma=1~\mathrm{MHz}$, $\mathbf{v}_0=(6,6,6)~\mathrm{\mu T}$ and $\mathbf{u}_0(t)=(3,3,3)~\mathrm{\mu T}$. (b) Corresponding optimized magnetic field for the filtered (blue) and non-filtered (red) cases.}
\label{fig:7P_F02}
\end{figure*}

\vskip.1in
\noindent\textbf{2-7 Proton Cases.}
All the main conclusions outlined in 1 Proton Case are confirmed in 2-7 Proton Cases. The results are presented in Figures~\ref{fig:2P_F01}-\ref{fig:7P_F02} and Table~\ref{tb1}. 

Figures \ref{fig:2P_F01_a},\ref{fig:3P_F01_a},\ref{fig:4P_F01_a},\ref{fig:5P_F01_a},\ref{fig:6P_F01_a},\ref{fig:7P_F01_a}
show a comparison of the optimal control (blue) and corresponding magnetic field (red) coordinates obtained using both the IPMP and GPM algorithms as described in Section \ref{sct:alg}, for the 2-7 proton cases, respectively.
Additionally, Figures \ref{fig:2P_F01_b},\ref{fig:3P_F01_b},\ref{fig:4P_F01_b},\ref{fig:5P_F01_b},\ref{fig:6P_F01_b},\ref{fig:7P_F01_b} show the asymptotic behavior of the cost functional values $\mathcal{J}$ with respect to the filter parameter $\gamma$, at the last iteration using IPMP.
The trade-off between the filter and no-filter models mentioned earlier is illustrated by Figures \ref{fig:2P_F02_a},\ref{fig:3P_F02_a},\ref{fig:4P_F02_a},\ref{fig:5P_F02_a},\ref{fig:6P_F02_a},\ref{fig:7P_F02_a}, showing the coordinates $k=1,2,3$ (out of $2^{p+2}$) of the optimal wave/state $\psi^l$ for the particular case $l=2$ (out of $3\cdot2^p$), and for each proton case $p=2,\dots,7$. All the other components demonstrate similar overlap between filtered and non-filtered cases. 
Finally, the corresponding optimized magnetic field coordinates are displayed in Figures \ref{fig:2P_F02_b},\ref{fig:3P_F02_b},\ref{fig:4P_F02_b},\ref{fig:5P_F02_b},\ref{fig:6P_F02_b},\ref{fig:7P_F02_b} for the filter (blue) and no-filter (red) models.

Remarkably, significantly simplified and smoothed electromagnetic field input in the filtered model implies coherent oscillations of the wave functions almost identical to those of the non-filtered model.  Moreover, Table~\ref{tb1} demonstrates that the loss of the maximum of the triplet-born singlet yield is within 1\%.  

\clearpage

The following are the main outcomes of numerical simulations of up to 7-proton models with a positive range of the control parameter:
\begin{itemize}
\item[-] Both IPMP and GPM methods converge to the unique bang-bang optimal control and produce the same unique continuous in time optimal electromagnetic field input. The simulations show that IPMP provides an approximate optimal control within 5-10 iterations, whereas GPM requires at least 2 times as many iterations. This advantage of IPMP over the GPM method is even more apparent in higher-proton models, where the computational cost increases exponentially. 
\item[-] Numerical simulations of up to 7-proton models with filtering parameter range $0.1\leq \gamma \leq 25$, and with various selection of initial iterations demonstrate that trading off between the original non-filtered model with bang-bang optimal magnetic field and filtered model with continuous and piecewise-smooth in time optimal magnetic field is associated with the loss of the maximum singlet yield expressed as a maximum of the cost functional within 1\%. 
Besides providing the enhanced regularity and simplicity of the optimal electromagnetic field input, the filtered model preserves coherent oscillations of the optimal Schr\"odinger and its adjoint systems, nearly identical in the case of the non-filtered model.  Hence, filtering presents a powerful regularization tool to replace optimal bang-bang electromagnetic field input with a continuous and piecewise-smooth in time electromagnetic field wave, which produces almost the exact quantum singlet yield, and preserves the coherent oscillations of the radical pair system. 
\item[-] Numerical simulations demonstrate that the best selection of filtering parameters $\gamma$ and $\mathbf{v}_0$ in terms of minimal loss of quantum singlet yield is achieved by choosing $\gamma$ large, and by selecting initial value of the magnetic field $\mathbf{v}_0$ to match the initial value of the optimal bang-bang magnetic field in no-filter model.
\end{itemize}

\subsection{Prism Case 2.}
To assess the effect of the control parameter with a sign-changing range, in this subsection, we choose a prism $V_2$ with bounds $m_i=3~\mathrm{\mu T}$, $M_i,=6~\mathrm{\mu T}$, for $i=x,y$, and $m_z=-1~\mathrm{\mu T}$ and $M_z=2~\mathrm{\mu T}$ and consider a control set
\begin{equation*}
\mathcal{V}_2 = \left\{ 
\begin{array}{c}
\mathbf{u} \in L^3_2(0,T;\mathbb{R}^3) ~ : \\ 
\mathbf{u}(t) \in V_2 = [3,6]^2\times[-1,2] , \\
 \mbox{a.e.} ~ t \in [0,T] 
\end{array}
\right\} .
\end{equation*}

In \cite{qst24} (section 7.3), it is demonstrated that in the case of a control parameter range chosen as $V_2$, there is a non-uniqueness of the optimal control, and the numerical methods based on the PMP algorithm present only local convergence and stability. Here, our goal is to analyze the effect of filtering on the phenomenon of non-uniqueness of the optimal control.  

Let us consider the following two grids with 27 points in each built around the vertices $(6,6,-1)$ and $(6,6,2)$ respectively.  
\begin{eqnarray}
\mathbf{u}^1_{ijk} = (6,6,-1) + 0.5 \cdot ( 1-i , 1-j , 1 - k ) ; \nonumber \\
\mathbf{u}^2_{ijk} = (6,6,2) + 0.5 \cdot ( 1-i , 1-j , 1 - k ) ;  \nonumber
\end{eqnarray}
for $i,j,k=0,1,2$. In \cite{qst24}, it is demonstrated that choosing initial control parameters from two different grids implies the convergence to two different bang-bang optimal controls in the 1-proton model. Next, we present numerical results for the IPMP algorithm and analyze the effects of the free filtering parameters $\gamma>0$ and $\mathbf{v}\in\mathbb{R}^3$ on the non-uniqueness of the optimal control. 

\begin{figure*}[t]
\centering
\begin{subfigure}[b]{.45\textwidth}
\centering
\includegraphics[width=.7\linewidth]{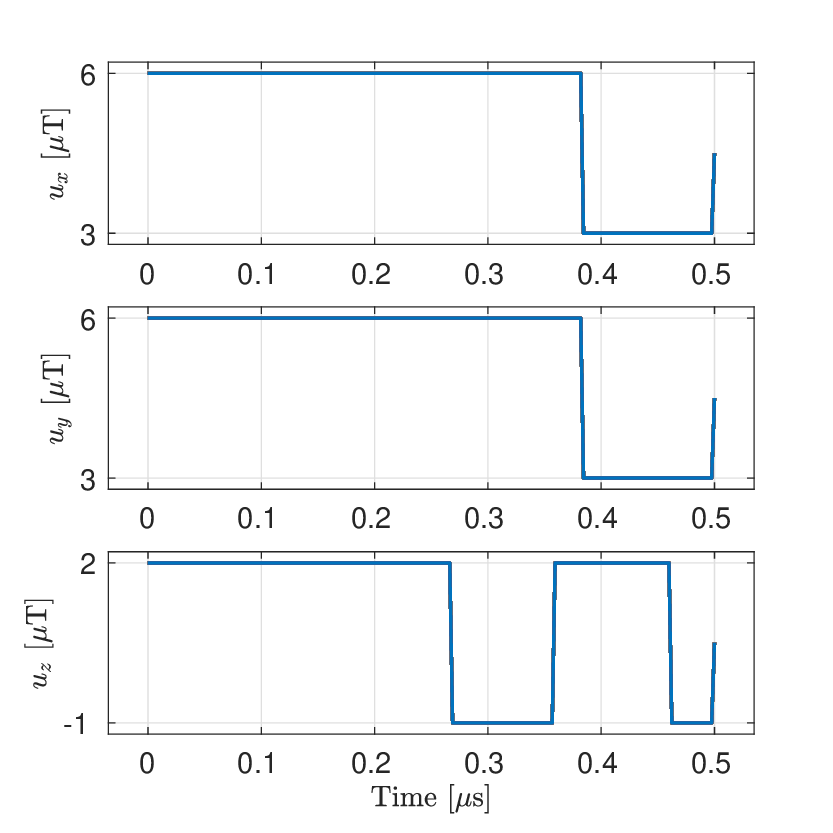}
\caption{Case $\mathbf{v}_0=(6,6,-1)~\mathrm{\mu T}$.}
\label{fig:Vm_1P_01_a}
\end{subfigure}
\hfill
\begin{subfigure}[b]{.45\textwidth}
\centering
\includegraphics[width=.7\linewidth]{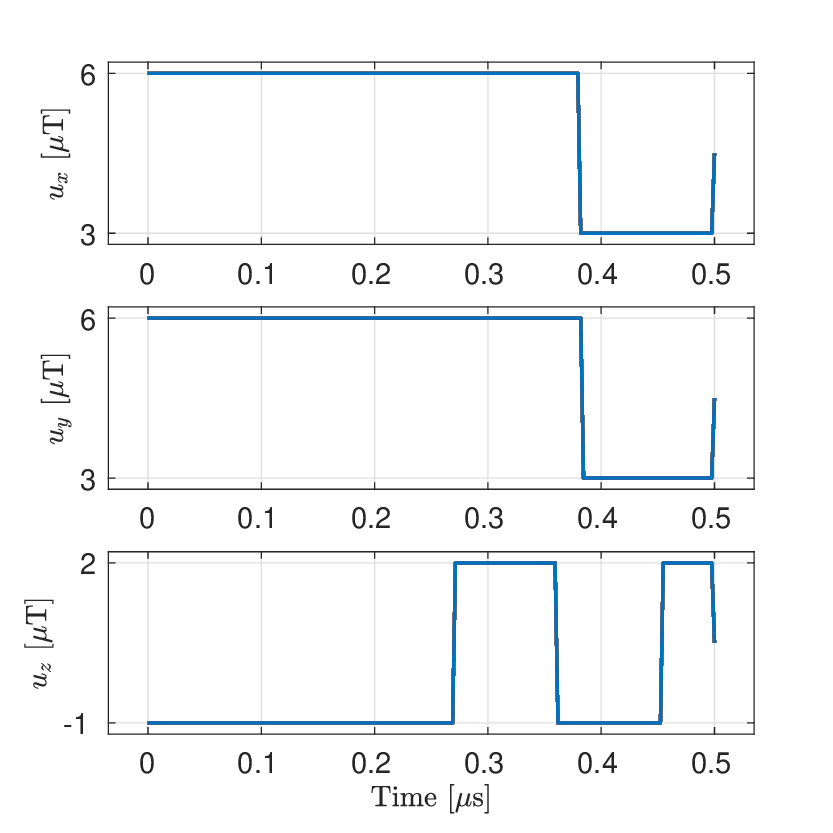}
\caption{Case $\mathbf{v}_0=(6,6,2)~\mathrm{\mu T}$.}
\label{fig:Vm_1P_01_b}
\end{subfigure}
\caption{1-proton. Approximated optimal controls obtained for initial control in the grid $\mathbf{u}^1_{ijk}$ and for each case of the initial magnetic field $\mathbf{v}_0$.}
\label{fig:Vm_1P_01}
\end{figure*}

\begin{figure*}[]
\centering
\begin{subfigure}[b]{.45\textwidth}
\centering
\includegraphics[width=0.7\linewidth]{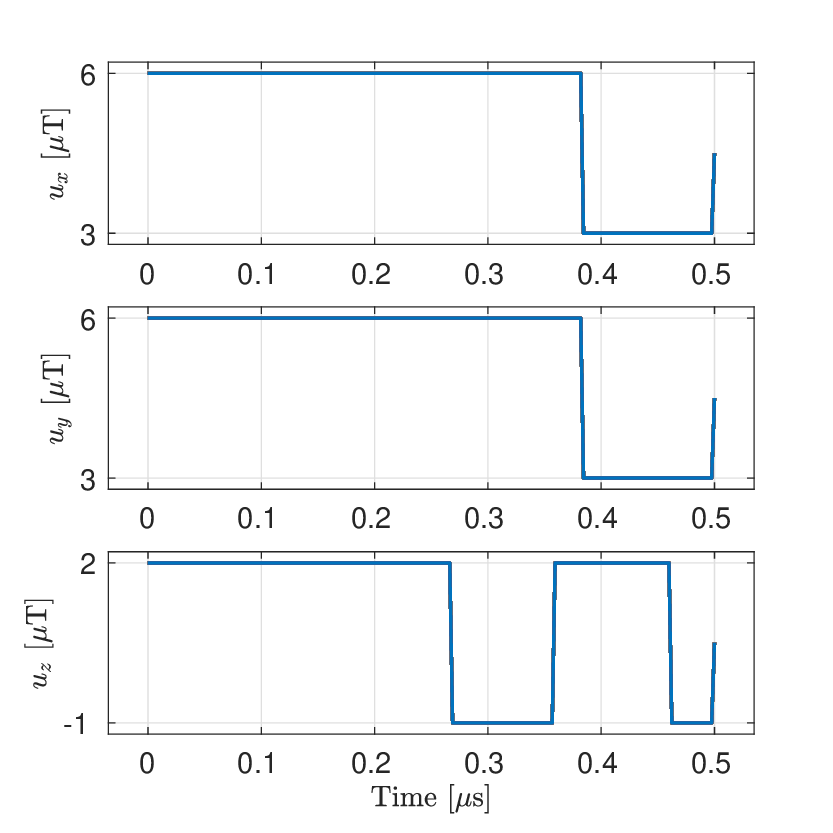}
\caption{Case $\mathbf{v}_0=(6,6,-1)~\mathrm{\mu T}$.}
\label{fig:Vm_1P_01_2_a}
\end{subfigure}
\hfill
\begin{subfigure}[b]{.45\textwidth}
\centering
\includegraphics[width=0.7\linewidth]{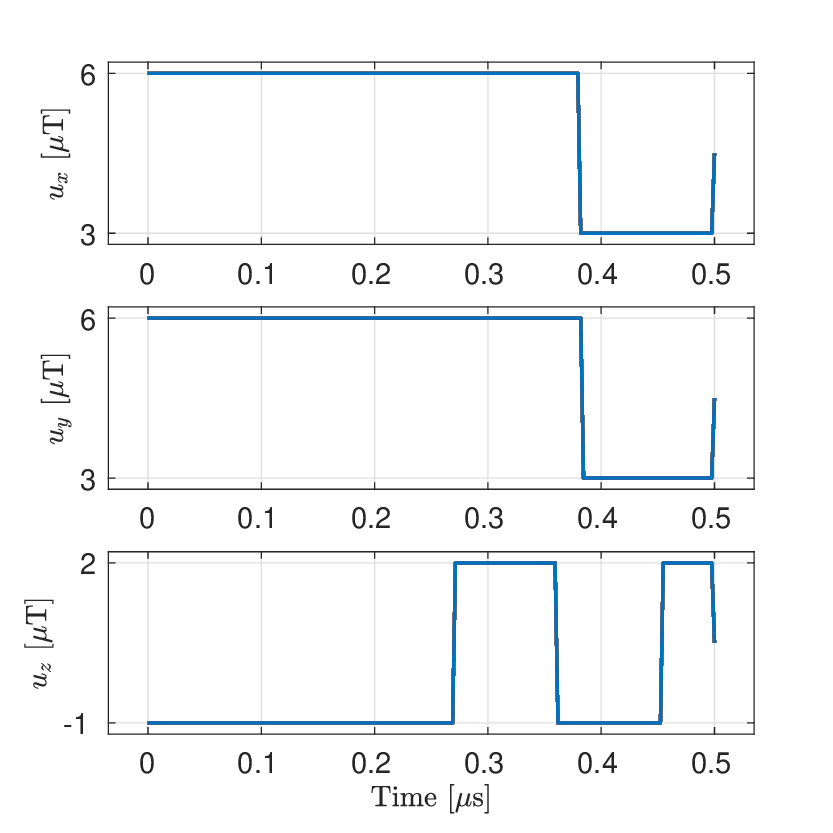}
\caption{Case $\mathbf{v}_0=(6,6,2)~\mathrm{\mu T}$.}
\label{fig:Vm_1P_01_2_b}
\end{subfigure}
\caption{1-proton. Approximated optimal controls obtained for initial control in the grid $\mathbf{u}^2_{ijk}$ and for each case of the initial magnetic field $\mathbf{v}_0$.}
\label{fig:Vm_1P_01_2}
\end{figure*}

\vskip.1in
\noindent\textbf{1-Proton Case.}
We choose a filter parameter $\gamma=1~\mathrm{MHz}$. Applying the IPMP algorithm, we calculate optimal bang-bang control $\mathbf{u}_{ijk}$ (respectively $\mathbf{\tilde{u}}_{ijk}$) with initial iteration chosen as a constant in-time control vector equal to grid points $\mathbf{u}^1_{ijk}$ (respectively $\mathbf{u}^2_{ijk}$). Figure \ref{fig:Vm_1P_01} (respectively Figure \ref{fig:Vm_1P_01_2}) shows the coordinates of optimal controls $\mathbf{u}_{ijk}$ (respectively $\mathbf{\tilde{u}}_{ijk}$) for each case of the initial magnetic field $\mathbf{v}_0=(6,6,-1)~\mathrm{\mu T}$ and $\mathbf{v}_0=(6,6,2)~\mathrm{\mu T}$. 

\vskip.1in
Then, with computer accuracy, we have
\begin{align}
\max_{\mathbf{u}^1\in\{\mathbf{u}_{ijk}\}, \mathbf{u}^2\in\{\mathbf{\tilde{u}}_{ijk}\}} \frac{ \| \mathbf{u}^{1} - \mathbf{u}^{2} \|_{L^3_2(0,T;\mathbb{R}^3)} }{ \| \mathbf{u}^1 \|_{L^3_2(0,T;\mathbb{R}^3)}}=0,  ~ \mbox{and} \nonumber\\
  \max_{\mathbf{u}^1\in\{\mathbf{u}_{ijk}\}, \mathbf{u}^2\in\{\mathbf{\tilde{u}}_{ijk}\}} \left| \frac{ \mathcal{J}(\mathbf{u}^1) - \mathcal{J}(\mathbf{u}^2) }{ \mathcal{J}(\mathbf{u}^1) } \right| = 0, \nonumber
\end{align} 
for both cases $\mathbf{v}_0=(6,6,-1)~\mathrm{\mu T}$ and $\mathbf{v}_0=(6,6,2)~\mathrm{\mu T}$.
\textit{Hence, by fixing filtering parameters as $\gamma=1$, and $\mathbf{v}_0$ as one of the vertices $(6,6,-1)$ or $(6,6,2)$ we observe a convergence to unique bang-bang optimal control when initial iteration is selected from different grids.} 

\vskip.1in
Next, we verify the amount of quantum-singlet yield loss associated with filtering relative to the non-filtering model. 
Let $\mathbf{\hat{u}}_{ijk}$ (respectively $\mathbf{\check{u}}_{ijk}$) be a bang-bang control obtained using the IPMP algorithm for the points in the grid $\mathbf{u}^1_{ijk}$ (respectively  $\mathbf{u}^2_{ijk}$) selected as initial iteration in the no-filter model. 
\begin{equation*}
\resizebox{0.45\textwidth}{!}{%
$\displaystyle\max_{\mathbf{u}^1\in\{\mathbf{u}_{ijk}\},\mathbf{u}^2\in\{\mathbf{\hat{u}}_{ijk}\}} \left| \frac{ \mathcal{J}(\mathbf{u}^1) - \mathcal{J}(\mathbf{u}^2) }{ \mathcal{J}(\mathbf{u}^1) } \right| \leq 1.0805\times10^{-3} , $}
\end{equation*}
\begin{equation*}
\resizebox{0.45\textwidth}{!}{%
$\displaystyle\max_{\mathbf{u}^1\in\{\mathbf{\tilde{u}}_{ijk}\},\mathbf{u}^2\in\{\mathbf{\check{u}}_{ijk}\}} \left| \frac{ \mathcal{J}(\mathbf{u}^1) - \mathcal{J}(\mathbf{u}^2) }{ \mathcal{J}(\mathbf{u}^1) } \right| \leq 1.2351\times10^{-3} , $}
\end{equation*}
both in the case $\mathbf{v}_0=(6,6,-1)~\mathrm{\mu T}$.
\textit{Hence, the gained uniqueness of the optimal control in the filtering model is associated with the loss of quantum singlet yield less than 1$\%$.}

As demonstrated in \cite{qst24}, the optimal control is not unique in the original model without filtering. To demonstrate that numerically, let $\mathbf{u}^1:=\mathbf{\hat{u}}_{111}$ (respectively $\mathbf{u}^2:=\mathbf{\check{u}}_{111}$) is the bang-bang optimal control obtained by IPMP algorithm with initial iteration being a central point of the grid system $\{\mathbf{u}^1_{ijk}\}$ (respectively $\{\mathbf{u}^2_{ijk}\}$) in a no-filter model. 
The following discrepancy between the bang-bang controls $\mathbf{u}^1$ and $\mathbf{u}^2$, and their associated cost values, demonstrates that the IPMP algorithm converges to two different bang-bang optimal controls with approximately the same maximum value of the quantum singlet yield: 
\begin{align}
\left |\frac{ \mathcal{J}(\mathbf{u}^1) - \mathcal{J}(\mathbf{u}^2) }{ \mathcal{J}(\mathbf{u}^1) }\right |
= 1.5435\times10^{-4} , ~ \mbox{and} \nonumber\\
\frac{ \| \mathbf{u}^1 - \mathbf{u}^2 \|_{L^3_2(0,T;\mathbb{R}^3)} }{ \| \mathbf{u}^1 \|_{L^3_2(0,T;\mathbb{R}^3)} }
= 0.3511. \nonumber
\end{align}

\begin{figure*}[t]
\centering
\begin{subfigure}[b]{.3\textwidth}
\centering
\includegraphics[width=\textwidth]{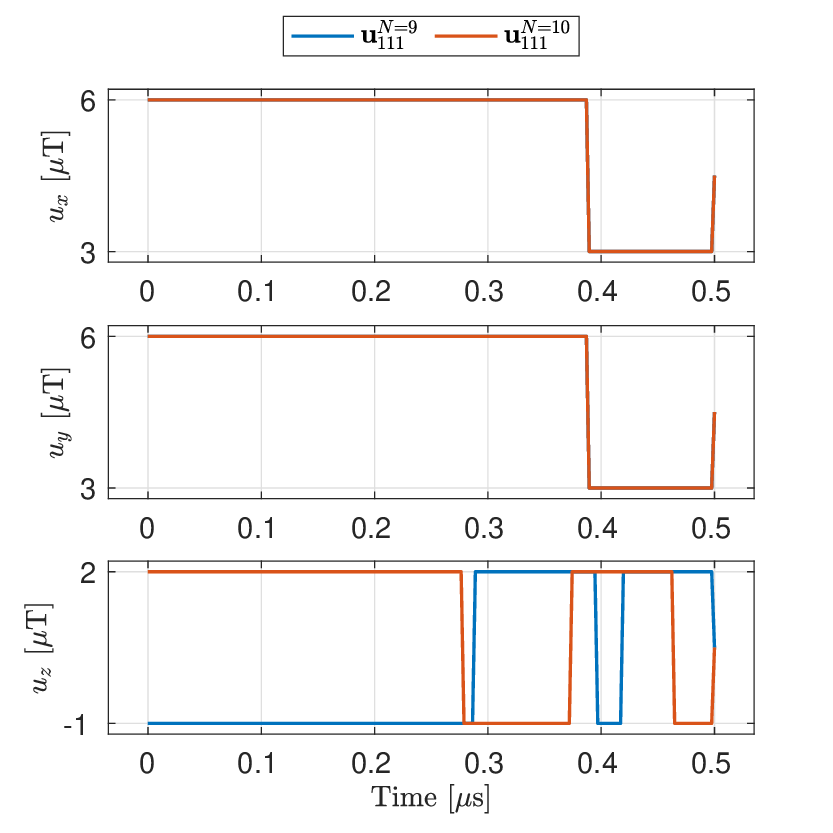}
\caption{Case $\mathbf{v}_0=(6,6,-1)~\mathrm{\mu T}$.}
\label{fig:Vm_1P_02_a}
\end{subfigure}
\hfill
\begin{subfigure}[b]{.3\textwidth}
\centering
\includegraphics[width=\textwidth]{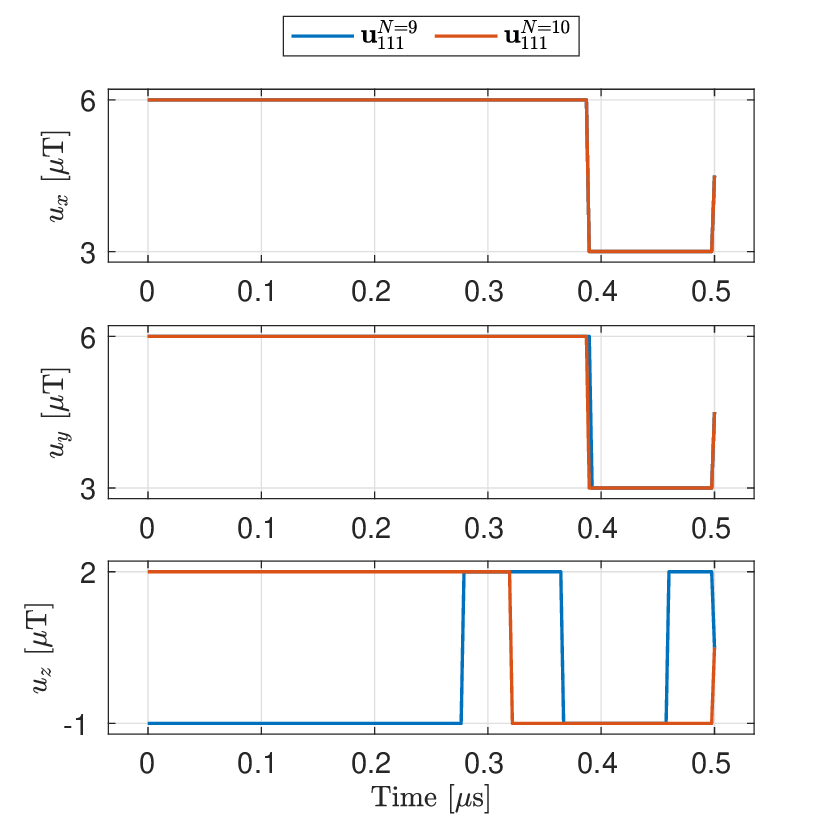}
\caption{Case $\mathbf{v}_0=(6,6,2)~\mathrm{\mu T}$.}
\label{fig:Vm_1P_02_b}
\end{subfigure}
\hfill
\begin{subfigure}[b]{.3\textwidth}
\centering
\includegraphics[width=\textwidth]{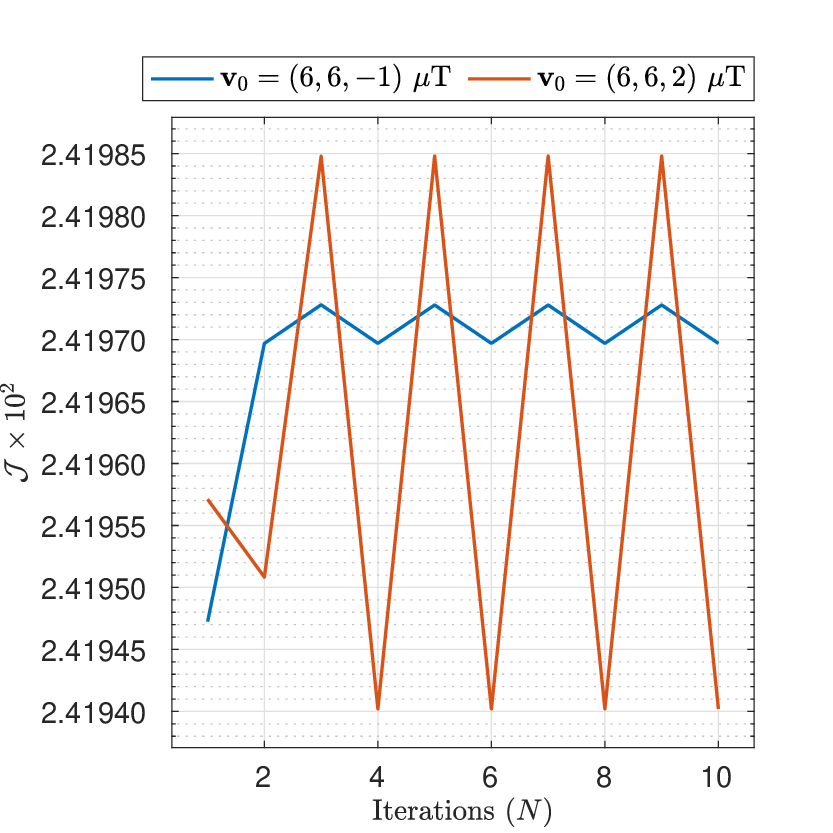}
\caption{Cost functional $\mathcal{J}$.}
\label{fig:Vm_1P_02_c}
\end{subfigure}
\caption{1-proton. Approximated bang-bang optimal control $\mathbf{u}_{111}$ at iterations $N=9$ and $N=10$ for each case of $\mathbf{v}_0$ (left and center). Similar results were obtained for the remaining points of the grid $\mathbf{u}^1_{ijk}$. Cost functional at each iteration and for each $\mathbf{v}_0$ case.}
\label{fig:Vm_1P_02}
\end{figure*}

\begin{figure*}[t]
\centering
\begin{subfigure}[b]{.3\textwidth}
\centering
\includegraphics[width=\textwidth]{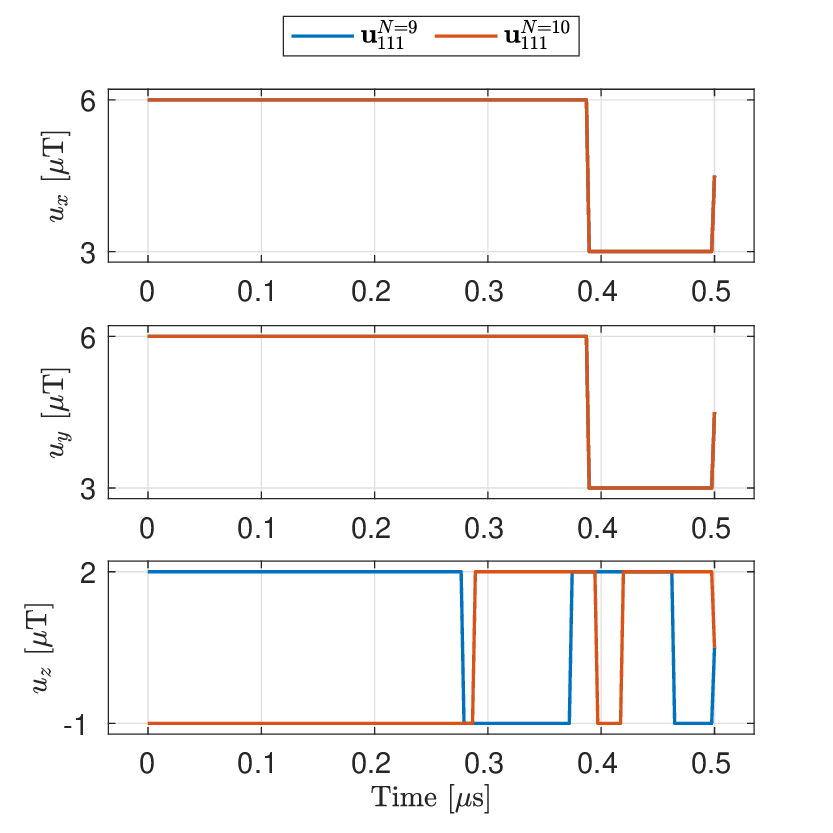}
\caption{Case $\mathbf{v}_0=(6,6,-1)~\mathrm{\mu T}$.}
\label{fig:Vm_1P_02_2_a}
\end{subfigure}
\hfill
\begin{subfigure}[b]{.3\textwidth}
\centering
\includegraphics[width=\textwidth]{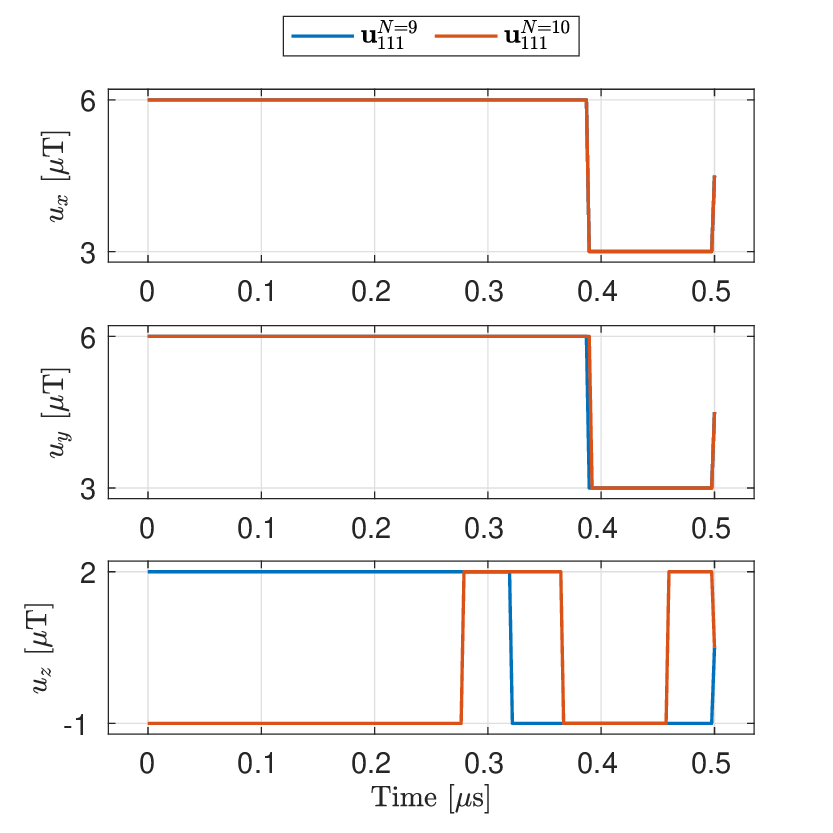}
\caption{Case $\mathbf{v}_0=(6,6,2)~\mathrm{\mu T}$.}
\label{fig:Vm_1P_02_2_b}
\end{subfigure}
\hfill
\begin{subfigure}[b]{.3\textwidth}
\centering
\includegraphics[width=\textwidth]{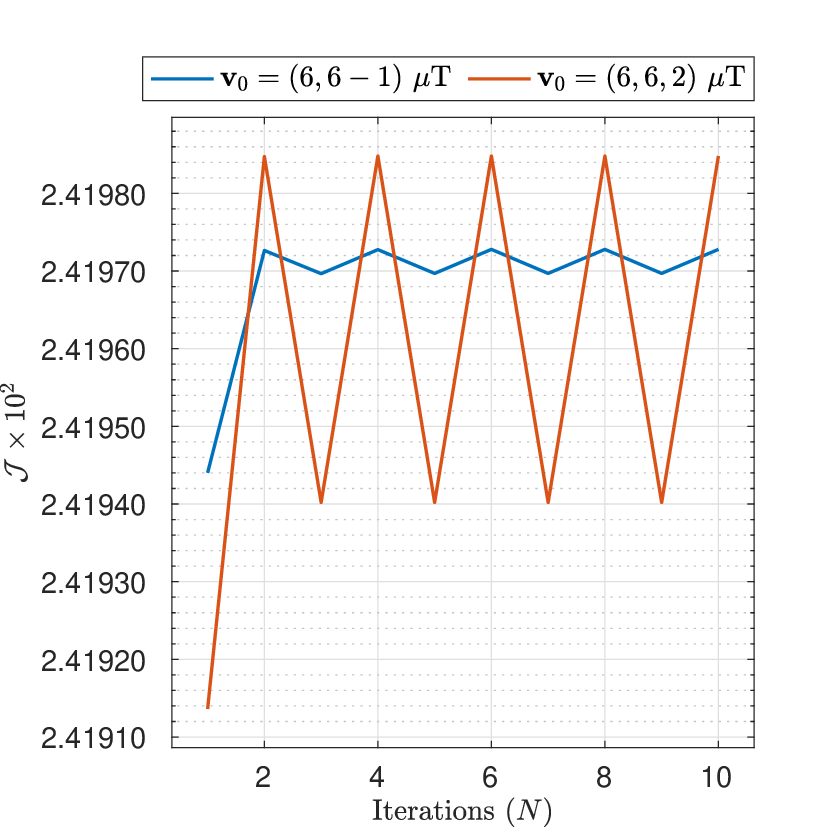}
\caption{Cost functional $\mathcal{J}$.}
\label{fig:Vm_1P_02_2_c}
\end{subfigure}
\caption{1-proton. Approximated bang-bang optimal control $\mathbf{\tilde{u}}_{111}$ at iterations $N=9$ and $N=10$ for each case of $\mathbf{v}_0$ (left and center). Similar results were obtained for the remaining points of the grid $\mathbf{u}^2_{ijk}$. Cost functional at each iteration and for each $\mathbf{v}_0$ case.}
\label{fig:Vm_1P_02_2}
\end{figure*}

Next, we consider the filtered model with $\gamma=10~\mathrm{\mu T}$ and apply the IPMP algorithm to calculate optimal bang-bang controls $\mathbf{u}_{ijk}$ (respectively $\mathbf{\tilde{u}}_{ijk}$) for each initial data in the grid $\mathbf{u}^1_{ijk}$ (respectively $\mathbf{u}^2_{ijk}$).
Numerical results demonstrate that in all cases, there is no convergence to optimal control, but there are oscillations between two optimal controls.
Figure \ref{fig:Vm_1P_02} (respectively Figure \ref{fig:Vm_1P_02_2})  shows iterations $N=9$ and $N=10$ of the calculated bang-bang control $\mathbf{u}_{111}$ (respectively $\mathbf{\tilde{u}}_{111}$) and the correspondig cost functional, for both cases $\mathbf{v}_0=(6,6,-1)~\mathrm{\mu T}$ and $\mathbf{v}_0=(6,6,2)~\mathrm{\mu T}$.

Similar behavior holds for all grid points, and the calculation of the relative cost functional discrepancy versus control parameter discrepancy for the two collections of grid points reveals the following estimates:
\begin{align*}
1.278177\times10^{-5} 
\leq 
\left| \frac{ \mathcal{J}(\mathbf{u}^{N=20}_{ijk}) - \mathcal{J}(\mathbf{u}^{N=19}_{ijk}) }{ \mathcal{J}(\mathbf{u}^{N=20}_{ijk}) } \right| \\
\leq 
1.843221\times10^{-4} ,
\end{align*}
\begin{align*}
3.234418\times10^{-1}
\leq
\frac{ \| \mathbf{u}^{N=20}_{ijk} - \mathbf{u}^{N=19}_{ijk} \|_{L^3_2(0,T;\mathbb{R}^3)} }{ \| \mathbf{u}^{N=20}_{ijk} \|_{L^3_2(0,T;\mathbb{R}^3)} } \\
\leq 
3.462466\times10^{-1} , 
\end{align*}
\begin{align*}
1.278161\times10^{-5} 
\leq
\left| \frac{ \mathcal{J}(\mathbf{\tilde{u}}^{N=20}_{ijk}) - \mathcal{J}(\mathbf{\tilde{u}}^{N=19}_{ijk}) }{ \mathcal{J}(\mathbf{\tilde{u}}^{N=20}_{ijk}) } \right| \\
\leq 
1.842881\times10^{-4} ,
\end{align*}
\begin{align*}
3.260902\times10^{-1} 
\leq
\frac{ \| \mathbf{\tilde{u}}^{N=20}_{ijk} - \mathbf{\tilde{u}}^{N=19}_{ijk} \|_{L^3_2(0,T;\mathbb{R}^3)} }{ \| \mathbf{\tilde{u}}^{N=20}_{ijk} \|_{L^3_2(0,T;\mathbb{R}^3)} } \\
\leq 
3.492381\times10^{-1} .
\end{align*}

This demonstrates that the uniqueness of the optimal control fails in a filtered model with $\gamma=10~\mathrm{MHz}$, and that, for every initial iteration chosen from the two different collections of grid points, the IPMP algorithm converges to two different approximate optimal control parameters. 

Similar behavior is observed in $p$-proton cases with $p\geq 2$. Hence, filtering presents a powerful regularization tool to address the non-uniqueness of the optimal control. 

The following are the main outcomes of numerical simulations of the $m$-proton model with a sign-changing range of the control parameter:
\begin{itemize}
\item[-] In general, there is a non-uniqueness of the optimal control, and both IPMP and GPM methods demonstrate the local convergence and stability, due to the non-uniqueness of the optimal control. IPMP has an advantage over GPM in terms of convergence rate and efficiency. 
\item[-] Filtering presents a powerful regularization tool to address the non-uniqueness of the optimal control in the original no-filter model. Numerical simulations demonstrate that, by choosing the filtering parameter $\gamma$ sufficiently small, the filtered optimal control problem has a unique bang-bang optimal control. Moreover, trading off between the original, no-filtered model with multiple bang-bang optimal magnetic fields and the filtered model with a continuous-in-time, unique optimal magnetic field is associated with a loss of the maximum singlet yield, defined as the maximum of the cost functional, within 1\%. If the filtering parameter is chosen to be large, then the filtered model is close to the no-filter model and inherits the non-uniqueness of the optimal control.
\end{itemize}

\section{Conclusions} \label{sct:conc}

We consider the quantum optimal control problem to devise the shape of the external electromagnetic field that drives the spin dynamics of radical pairs to a quantum coherent state through maximization of the triplet-singlet yield in biochemical reactions. The mathematical model is a Schr\"{o}dinger system with a spin Hamiltonian given by the sum of Zeeman interaction and hyperfine coupling interaction terms. In a recent paper \cite{qst24}, we proved the Pontryagin Maximum Principle and established a bang-bang structure of the optimal electromagnetic field intensity. 
Despite its great importance and potential for experimental applications in magnetoreception, the result of \cite{qst24} raised a major question about the practicality of generating a bang-bang electromagnetic wave on a short time scale, and the numerical stability of the algorithms due to the non-uniqueness of the bang-bang optimal control. 
 The main goal of this paper is to introduce a novel regularization of the quantum optimal control problem for the identification of the external electromagnetic field that drives the spin dynamics of radical pairs to a coherent state. We introduce a one-parameter family of optimal control problems by coupling the Schr\"{o}dinger system to a control field through filtering equations for the electromagnetic field. New regularized methods with enhanced regularity of the optimal electromagnetic field input, and improved numerical stability of the algorithms based on Fr\'echet differentiability and the Pontryagin Maximum Principle in Hilbert space are developed. The following are the main results:
\begin{itemize}
\item Fr\'echet differentiability in Hilbert spaces is proved, and the formula for the first-order Fr\'echet derivative is derived, and the gradient projection method (GPM) in Hilbert space is developed. 
Pontryagin Maximum Principle (PMP) in Hilbert space is proved, and the bang-bang structure of the optimal control is established. As an output of the filtering equation, the optimal electromagnetic field input produced by PMP is continuous and piecewise smooth in time. 
\item A closed optimality system of nonlinear integro-differential equations for the identification of the bang-bang optimal control is revealed. A new two-step algorithm for the calculation of the bang-bang optimal control, called \textbf{explicit Pontryagin maximum principle} (EPMP) method, is developed.
\item Approximation of the EPMP method, so called \textbf{iterative Pontryagin maximum principle} (IPMP) method is developed. It consists of an iterative process of solving initial value problems for the Schr\"odinger system, and it is adjoined with the subsequent derivation of the next iteration of the bang-bang optimal control via Pontryagin maximum principle.  
\item Software is developed, and numerical simulations of up to $7$-proton models are pursued, implementing IPMP and GPM algorithms. It is demonstrated that if the electromagnetic field range has a fixed sign, the IPMP and GPM methods converge to the unique bang-bang optimal control and produce the same unique continuous-in-time optimal electromagnetic field input. The simulations show that the IPMP method converges twice as fast as the GPM method in terms of the number of iterations, which is especially apparent for higher proton models when the computational cost increases exponentially. 
\item Numerical simulations of up to 7-proton models with filtering parameter range $0.1\leq \gamma \leq 25$, and with various selection of initial iterations demonstrate that 
\textbf{trading off between the original non-filtered model with bang-bang optimal magnetic field and filtered model with continuous in time optimal magnetic field is associated with the loss of the maximum singlet yield expressed as a maximum of the cost functional within 1\%. Moreover, despite the enhanced regularity and simplicity of the optimal electromagnetic field input, the non-filtered model preserves coherent oscillations of the optimal Schr\"odinger and its adjoint systems, nearly identical in the case of the non-filtered model.} Hence, filtering presents a powerful regularization tool to replace optimal bang-bang electromagnetic field input with a continuous and piecewise-smooth in time electromagnetic field wave, which produces almost the exact quantum singlet yield, and preserves the coherent oscillations of the radical pair system. 
\item Numerical simulations demonstrate that the best selection of filtering parameters $\gamma$ and $\mathbf{v}_0$ in terms of minimal loss of quantum singlet yield is achieved by choosing $\gamma$ large, and by selecting the initial value of the magnetic field $\mathbf{v}_0$ to match the initial value of the optimal bang-bang magnetic field in no-filter model.
\item In general, if the electromagnetic field range has a changing sign, there is a non-uniqueness of the optimal control, and both IPMP and GPM methods demonstrate the local convergence and stability, due to the non-uniqueness of the optimal control. IPMP has an advantage over GPM in terms of local convergence rate and efficiency. 
\item Filtering provides a powerful regularization tool to address the non-uniqueness of the optimal control in the original, no-filter model. Numerical simulations demonstrate that by choosing the filtering parameter $\gamma$ small enough, the filtered optimal control problem has a unique bang-bang optimal control. Moreover, 
\textbf{trading off between the original non-filtered model with multiple bang-bang optimal magnetic fields and the filtered model with a continuous-in-time unique optimal magnetic field is associated with the loss of the maximum singlet yield expressed as a maximum of the cost functional within 1\%.} 
\item The results open a venue for a potential experimental work on magnetoreception as a manifestation of quantum biological phenomena. 
\end{itemize}

\vskip.1in
\noindent\textbf{Acknowledgments.}
We are grateful for the help and support provided by the Scientific Computing
and Data Analysis section of Core Facilities at OIST.

\vskip.1in
\noindent\textbf{Data availability}
Data were generated by the authors and included in the article. The datasets generated and/or analyzed during the current study are publicly available in the following repository: \url{https://gitlab.com/oist_proj/pmp_filtering.git.}

\clearpage

\bibliographystyle{quantum}
\bibliography{myref}

\onecolumn\newpage
\appendix

\section*{Appendix}
To demonstrate the robustness of the developed methods, here we consider the performance of the GPM and IPMP algorithms described in Section \ref{sct:alg} for the identification of the bang-bang optimal control and, consequently, the optimal magnetic field, with a magnitude similar to Earth's constant-in-time magnetic field, given by the following vector $\mathbf{v}_E=(30,10,40)~\mathrm{\mu T}$. 

Let us consider the prism $V_E^1 = [ \mathbf{v}_E - 2 , \mathbf{v}_E + 2]$ in $\mathrm{\mu T}$ and define the control set
\begin{equation*}
\mathcal{V}_E^1 := \left\{ \mathbf{u} \in L_2^3(0,T;\mathbf{R}^3) : \mathbf{u}(t) \in V_E^1, ~ \mbox{a.e.} ~ t \in [0,T] \right\} .
\end{equation*}
We fix the initial constant-in-time control $\mathbf{u}_0(t) = (28,08,38)~\mathrm{\mu T}$, for $t\in[0,T]$, with $T=0.5~\mathrm{\mu s}$.

\begin{figure}[h!]
\centering
\begin{subfigure}[b]{.45\textwidth}
\centering
\includegraphics[width=\linewidth, height=5.5cm]{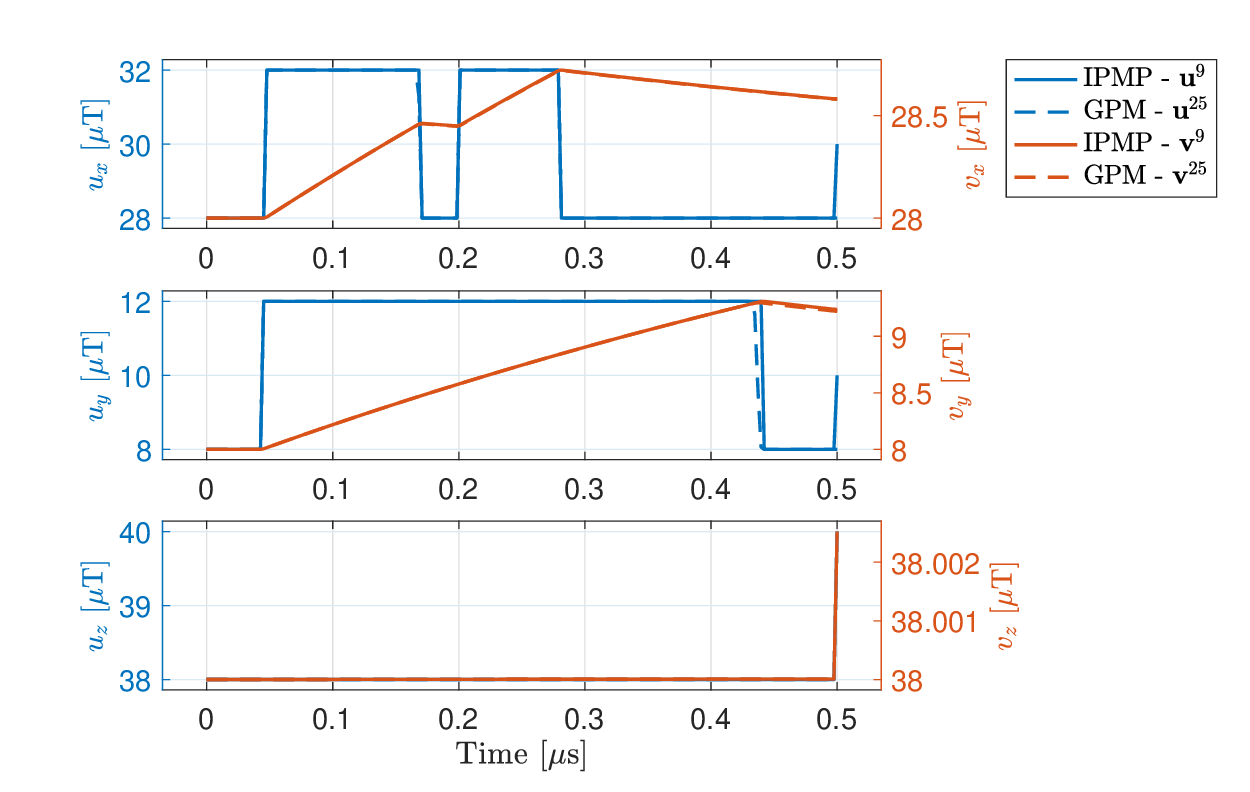}
\caption{Case $\mathbf{v}_0=(28,08,38)~\mathrm{\mu T}$.}
\label{fig:VE_1P_F01_01_a}
\end{subfigure}
\begin{subfigure}[b]{.45\textwidth}
\centering
\includegraphics[width=\linewidth, height=5.5cm]{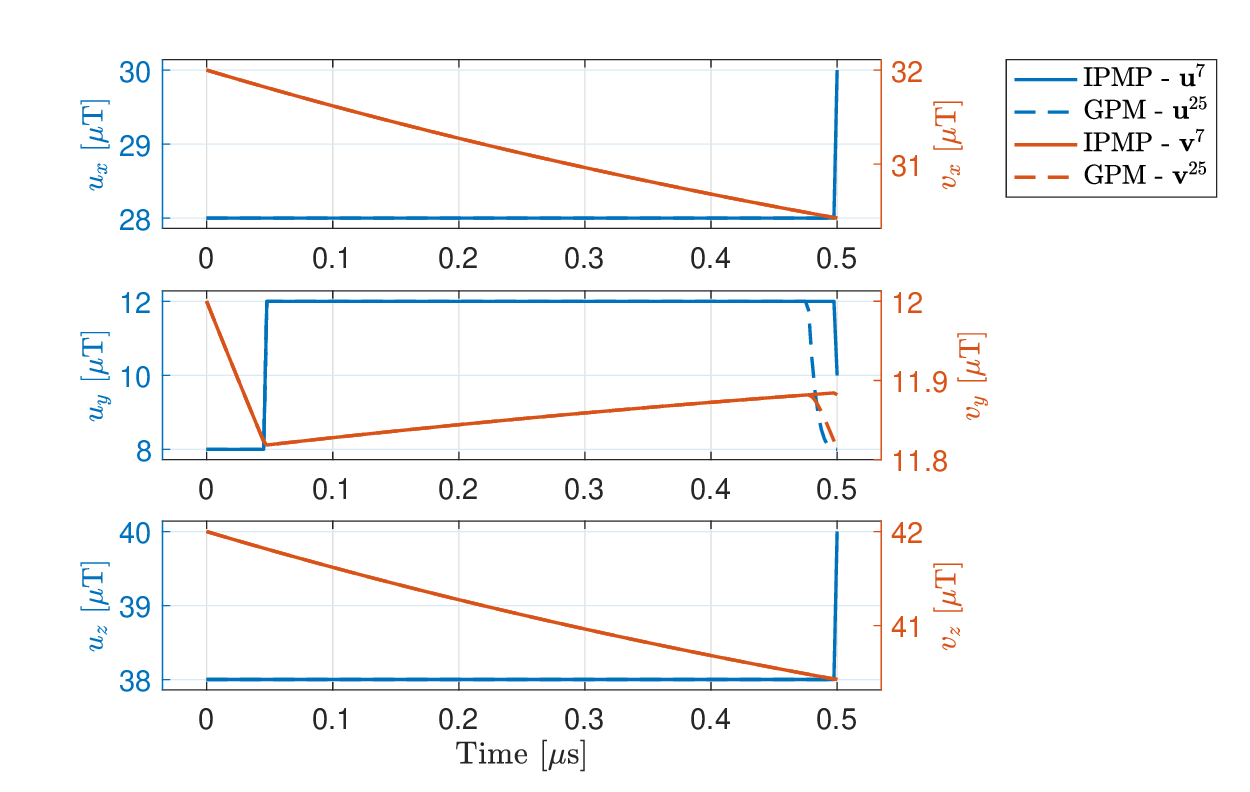}
\caption{Case $\mathbf{v}_0=(32,12,42)~\mathrm{\mu T}$.}
\label{fig:VE_1P_F01_01_b}
\end{subfigure}
\caption{1-Proton. Control coordinates (blue) and corresponding magnetic field (red) obtained using the IPMP (solid) and GPM (dashed), for different values of initial magnetic field $\mathbf{v}_0$. Filter parameter here is $\gamma=1~\mathrm{MHz}$.}
\label{fig:VE_1P_F01}
\end{figure}

\begin{figure}[h!]
\centering
\begin{subfigure}[b]{.45\textwidth}
\centering
\includegraphics[width=.7\linewidth]{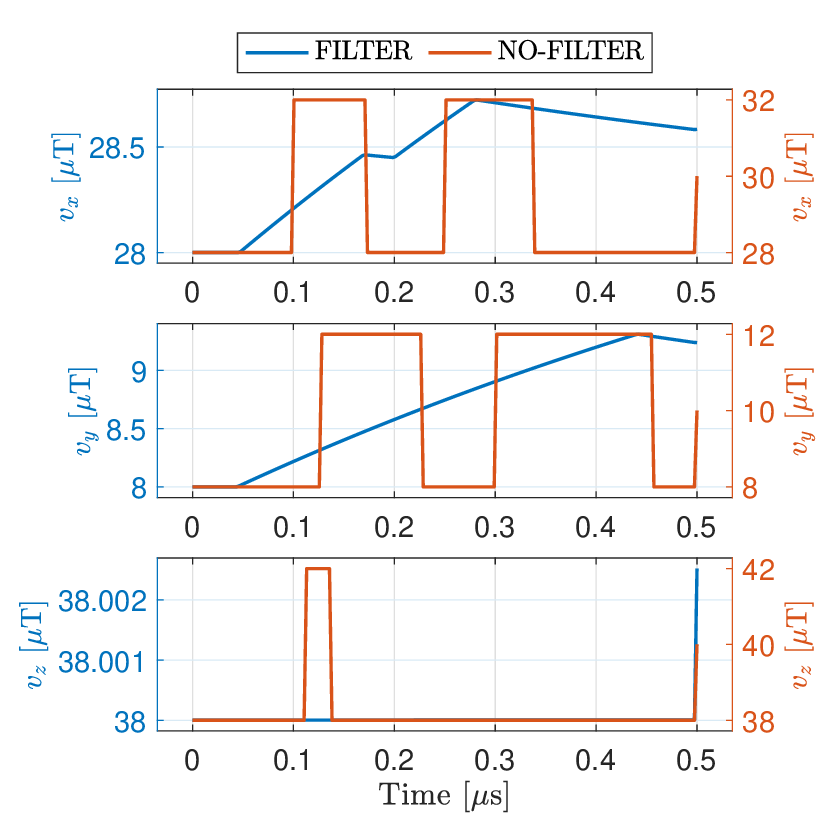}
\caption{Magnetic field coordinates}
\label{fig:VE_1P_F02_01_a}
\end{subfigure}
\hfill
\begin{subfigure}[b]{.45\textwidth}
\centering
\includegraphics[width=.7\linewidth]{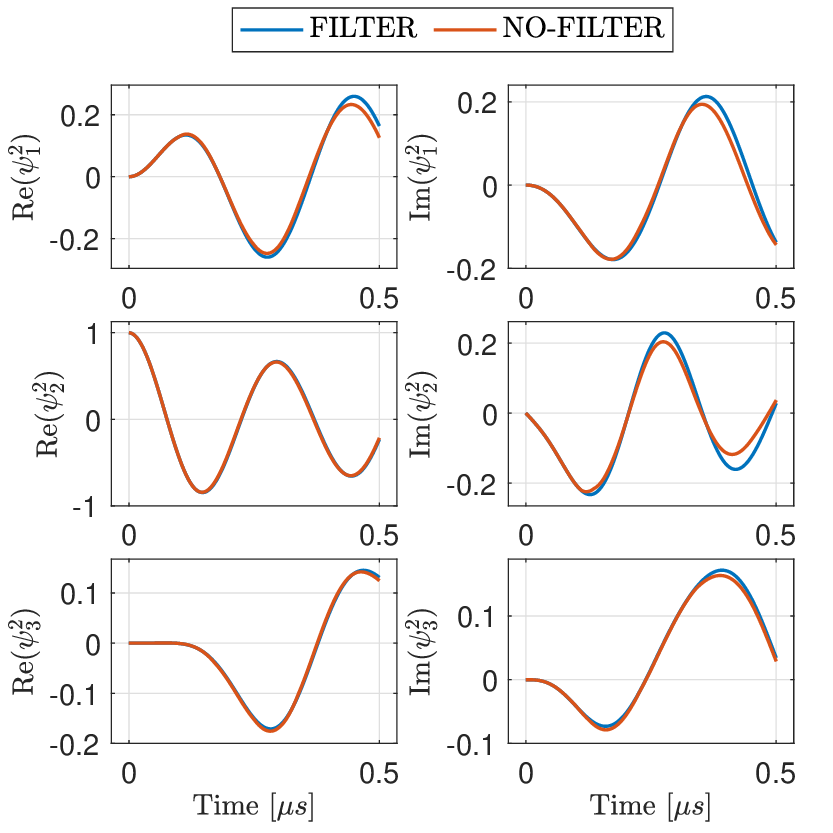}
\caption{Wave coordinates}
\label{fig:VE_1P_F02_01_b}
\end{subfigure}
\caption{1-Proton. (a) Optimized magnetic field coordinates for the filtered (blue) and non-filtered (red) models. The filter parameters are $\gamma=1~\mathrm{MHz}$ and $\mathbf{v}_0=(28,08,38)~\mathrm{\mu T}$, and initial constant-in-time control is $\mathbf{u}_0(t)=(28,08,38)~\mathrm{\mu T}$ for both cases. (b) Corresponding wave coordinates (real and imaginary parts) $\psi^l_k$, with $l=2$ and $k=1,2,3$.}
\label{fig:VE_1P_F02}
\end{figure}

\vskip.1in
We start by considering the filter parameter $\gamma=1~\mathrm{MHz}$. Figure \ref{fig:VE_1P_F01} shows the optimized control coordinates and corresponding magnetic fields obtained using IPMP and GPM for different values of initial magnetic field $\mathbf{v}_0$ (lower and upper bound). In both cases, we observed the convergence of both methods to the same control.
A comparison between the filtered and non-filtered models is displayed in Figure \ref{fig:VE_1P_F02}. In this scenario, the loss in quantum yield is (approximately) 0.7540\%.

\begin{figure}[h!]
\centering
\begin{subfigure}[b]{.5\textwidth}
\centering
\includegraphics[width=1.2\linewidth]{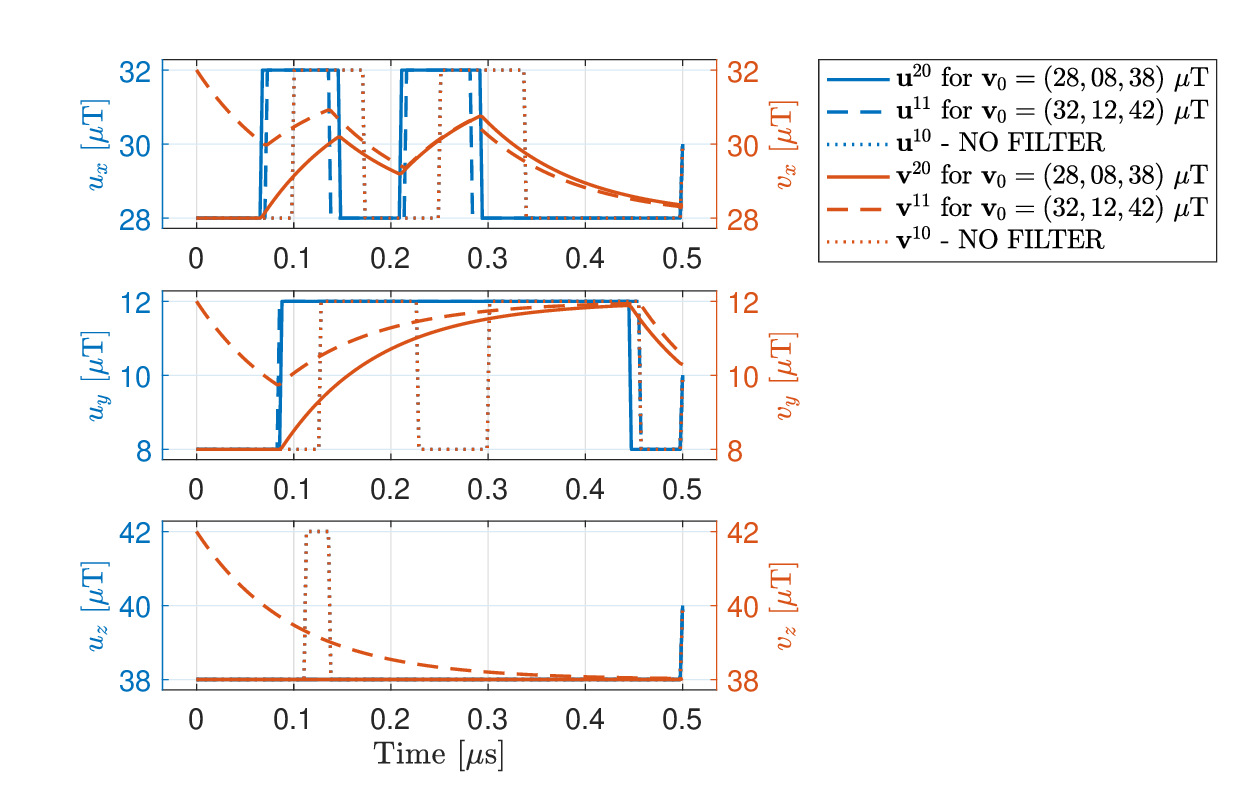}
\caption{} \label{fig:VE_1P_F03_a}
\end{subfigure}
\hfill
\begin{subfigure}[b]{.4\textwidth}
\centering
\includegraphics[width=\linewidth]{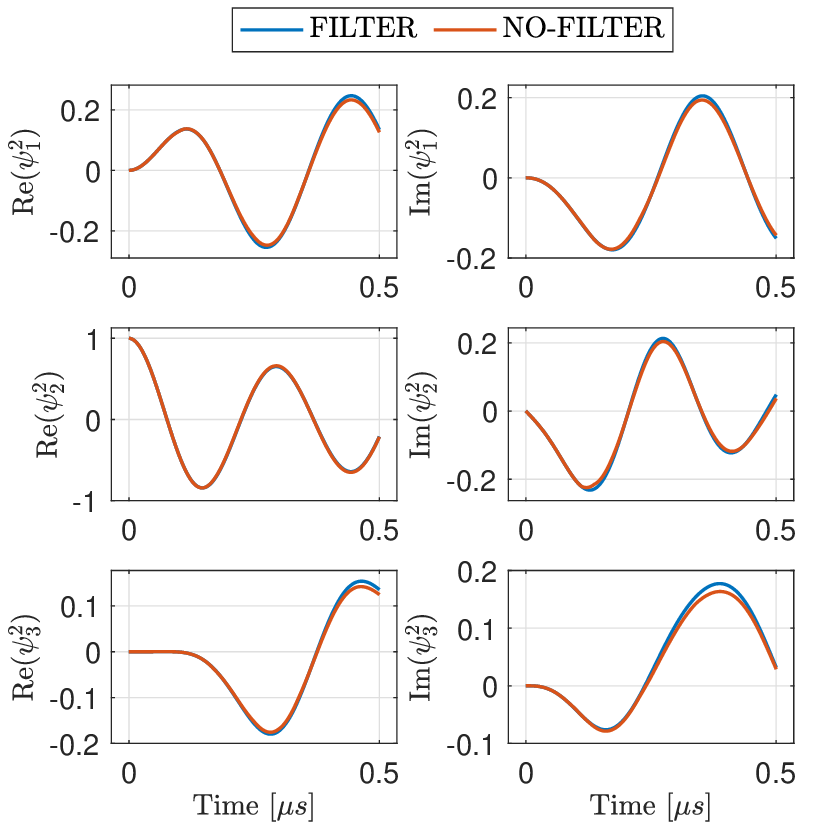}
\caption{} \label{fig:VE_1P_F03_b}
\end{subfigure}
\caption{1-Proton. (a) Control coordinates (blue) and corresponding magnetic field (red) obtained using the IPMP in the case of filter parameter $\gamma=10~\mathrm{MHz}$ (solid and dashed lines), and for both cases of initial magnetic field value $\mathbf{v}_0$ (lower and upper bounds). Dot lines represent the optimized control/magnetic field for the non-filtered case. (b) Corresponding wave coordinates (real and imaginary parts) $\psi^l_k$ for $l=2$ and $k=1,2,3$. Initial magnetic field value for the filtered model is $\mathbf{v}_0=(28,08,38)~\mathrm{\mu T}$.}
\label{fig:VE_1P_F03}
\end{figure}

Next, we consider the dynamics for the filter parameter $\gamma=10~\mathrm{MHz}$. In this case, similar results were obtained using both the IPMP and GPM algorithms. Figure \ref{fig:VE_1P_F03_a} shows the optimized control coordinates (blue) and corresponding magnetic field (red) for each case of initial magnetic field value $\mathbf{v}_0$ (lower and upper bounds), in comparison with the optimized control/magnetic field obtained using the IPMP for the non-filtered model. In this case, the maximum loss of quantum yield is (approximately) 0.3983\%.

\begin{figure}[h!]
\centering
\begin{subfigure}[b]{.45\textwidth}
\centering
\includegraphics[width=\linewidth, height=6cm]{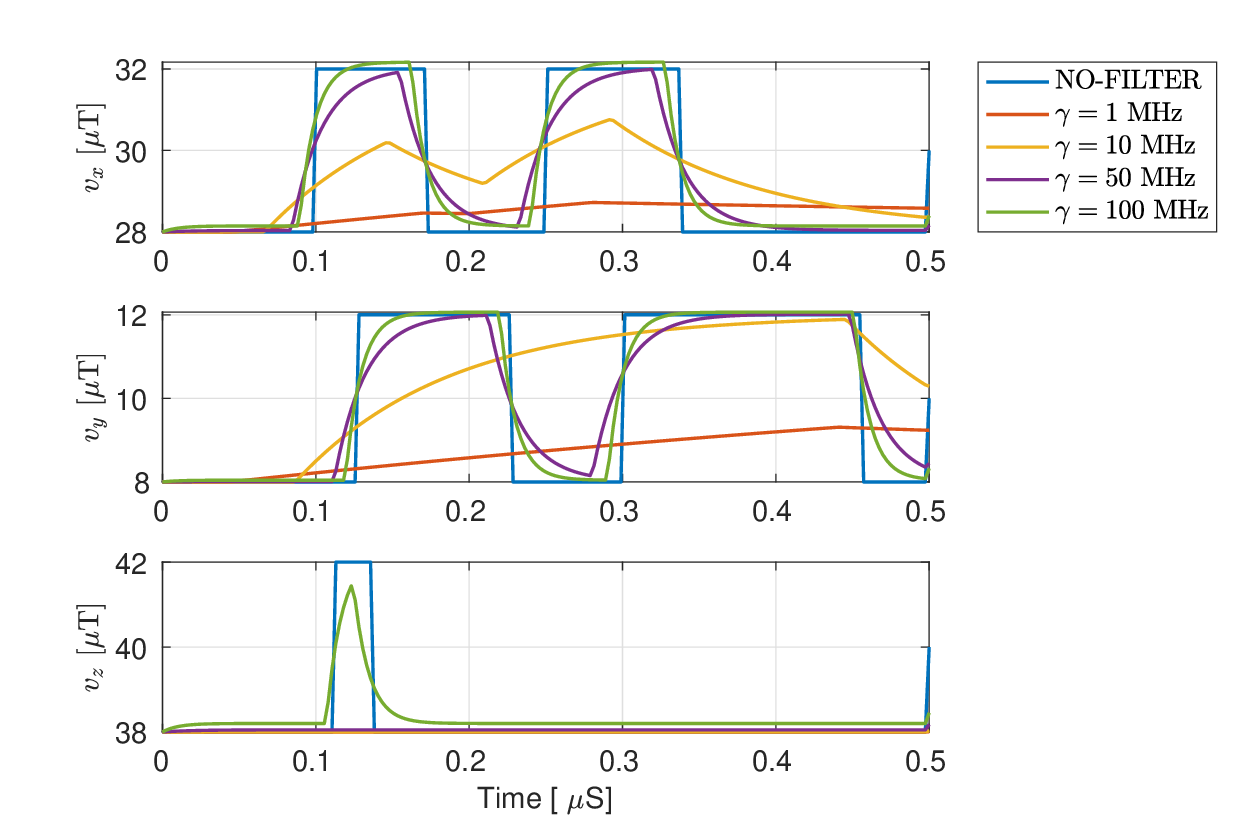}
\caption{} \label{fig:VE_1P_F04_a}
\end{subfigure}
\begin{subfigure}[b]{.45\textwidth}
\centering
\includegraphics[width=\linewidth, height=6cm]{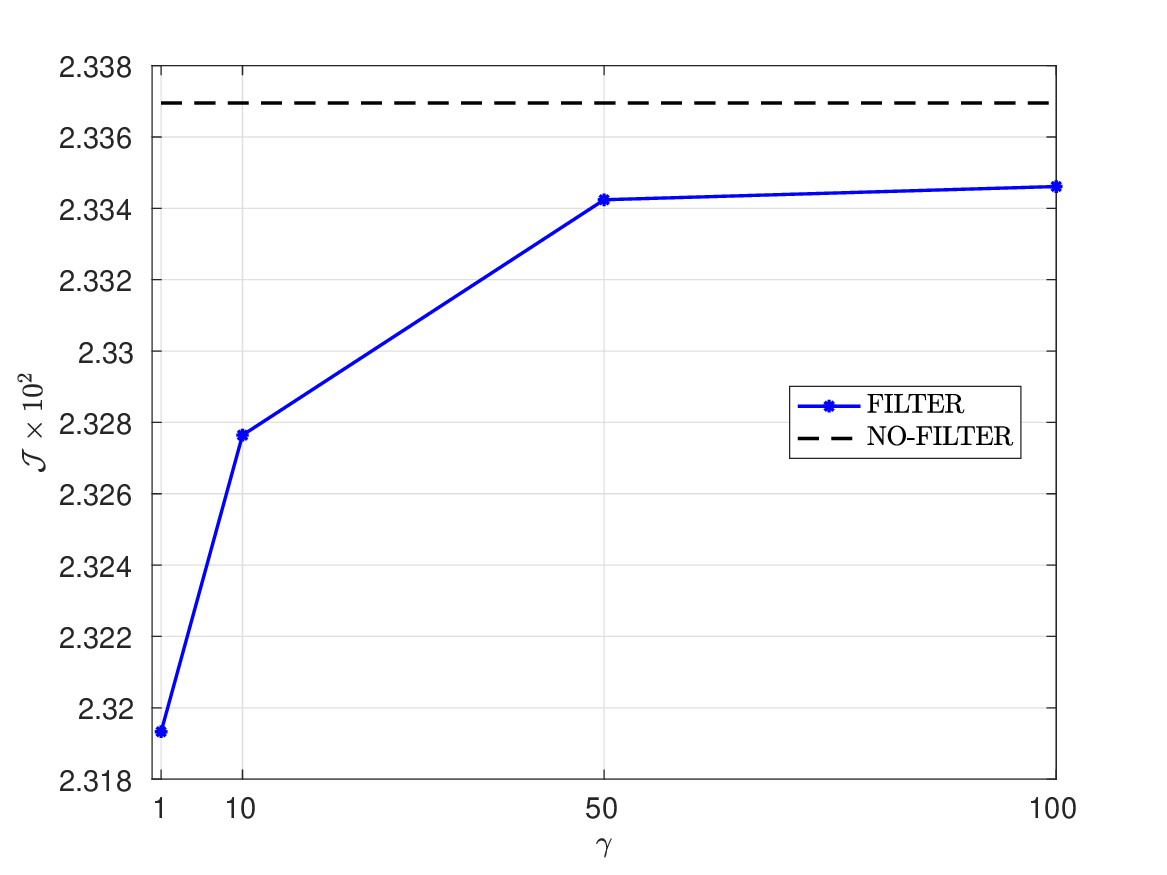}
\caption{} \label{fig:VE_1P_F04_b}
\end{subfigure}
\caption{1-Proton. (a) Optimal magnetic field coordinates in the filtered model with filtering parameter $\gamma=1,10,50,100~\mathrm{MHz}$, and optimal magnetic field for the corresponding non-filtered model. (b) Cost values for each case of filter parameter $\gamma$. Dashed line corresponds to the cost value obtained using the IPMP for the non-filtered model.}
\label{fig:VE_1P_F04}
\end{figure}

Similar results were obtained for the cases $\gamma=50~\mathrm{MHz}$ and $\gamma=100~\mathrm{MHz}$. The dynamics of the optimal magnetic field for each case $\gamma$ is displayed in Figure \ref{fig:VE_1P_F04_a}, for the filter initial magnetic field value $\mathbf{v}_0=(28,08,38)~\mathrm{\mu T}$. The corresponding cost values are displayed in Figure \ref{fig:VE_1P_F04_b}. For the particular case $\gamma=100~\mathrm{MHz}$, the quantum yield loss is (approximately) 0.1001\%.

\end{document}